\documentclass[final,1p,times]{elsarticle}

\usepackage{THGEMReview} 

\usepackage{amssymb}
\usepackage{subfig}
\usepackage{xcolor}
\usepackage{hyperref}
\usepackage{pdflscape}
\usepackage{multirow}
\usepackage{tablefootnote}

\usepackage{booktabs}
\usepackage{multirow}
\usepackage{longtable}

\hypersetup{
  colorlinks=true,
  linkcolor=blue!50!red,
  urlcolor=green!70!black
}
\usepackage{tabularray}
\usepackage[acronym]{glossaries}
\makeglossaries
\newacronym{THGEM}{THGEM}{THick Gas Electron Multiplier}
\newacronym{LEM}{LEM}{Large Electron Multiplier}
\newacronym{MWPC}{MWPC}{Multi Wire Proportional Chamber}
\newacronym{PPAC}{PPAC}{Parallel Plate Avalanche Counter}
\newacronym{RPC}{RPC}{Resistive Plate Chamber}
\newacronym{MSGC}{MSGC}{Micro Strip Gas Chamber}
\newacronym{MHSP}{MHSP}{Micro Hole \& Strip Plate}
\newacronym{MPGD}{MPGD}{Micro Pattern Gaseous Detector}
\newacronym{Micromegas}{Micromegas}{MicroMesh Gaseous Detectors}
\newacronym{PCB}{PCB}{Printed Circuit Board}
\newacronym{ToF}{ToF}{Time of Flight}
\newacronym{MM}{MM}{MicroMesh Gaseous Detectors}
\newacronym{GEM}{GEM}{Gas Electron Multiplier}
\newacronym{RMS}{RMS}{Root Mean Square}
\newacronym{GPM}{GPM}{Gaseous Photo-Multiplier}
\newacronym{CRAD}{CRAD}{CRiogenic Avalanche Detector}
\newacronym{LHM}{LHM}{Liquid Hole Multiplier}
\newacronym{LAr}{LAr}{Liquid Argon}
\newacronym{LXe}{LXe}{Liquid Xenon}
\newacronym{FWHM}{FWHM}{Full Width Half Maximum}
\newacronym{THWELL}{THWELL}{THick-WELL}
\newacronym{RWELL}{RWELL}{Resistive WELL}
\newacronym{SRWELL}{SRWELL}{Segmented Resistive WELL}
\newacronym{RPWELL}{RPWELL}{Resistive Plate WELL}
\newacronym{THCOBRA}{THCOBRA}{THick Cobra}
\newacronym{MIP}{MIP}{Minimum Ionizing Particle}
\newacronym{PE}{PE}{Primary Electron}
\newacronym{PDE}{PDE}{Photon Detection Efficiency}
\newacronym{DLC}{DLC}{Diamond Like Carbon}
\newacronym{EL}{EL}{electroluminescence}
\newacronym{IBF}{IBF}{The Ion back flow}
\newacronym{RETGEM}{RETGEM}{Resistive THGEM}
\newacronym{PMT}{PMT}{Photo Multiplier Tube}
\newacronym{SiPM}{SiPM}{Silicon Photo Multiplier}
\newacronym{UV}{UV}{Ultra Violet}
\newacronym{IR}{IR}{Infra Red}
\newacronym{VUV}{VUV}{Vacuum Ultra Violet}
\newacronym{ETE}{ETE}{Electron Transfer Efficiency}
\newacronym{M-THGEM}{M-THGEM}{Multi-layer THGEM}
\newacronym{AT}{AT}{Active Target}
\newacronym{TPC}{TPC}{Time Projection Chamber}
\newacronym{FAT-GEM}{FAT-GEM}{Field-Assisted Transparent Gaseous EL Multiplier}
\newacronym{ELCC}{ELCC}{EL light collection cell}
\newacronym{ESD}{ESD}{electrostatic dissipative}
\newacronym{DHCAL}{DHCAL}{Digital Hadronic Calorimeter}
\newacronym{RICH}{RICH}{Ring Imaging CHerenkov}
\newacronym{EDXRF}{EDXRF}{Energy Dispersive X-Ray Fluorescence}
\newacronym{PID}{PID}{Particle Identification}
\newacronym{EIC}{EIC}{Electron Ion Collider}
\newacronym{STCF}{STCF}{Super Tau-Fharm Facility}
\newacronym{COMPASS}{COMPASS}{COmmon Muon Proton Apparatus for Structure and Spectroscopy}
\newacronym{SPS}{SPS}{Super Proton Synchrotron}
\newacronym{PS}{PS}{Proton Synchrotron}
\newacronym{MAPMT}{MAPMT}{Multi Anode Photo-Multipliers Tube}
\newacronym{HV}{HV}{High Voltage}
\newacronym{ALICE}{ALICE}{A Large Ion Collider Experiment}
\newacronym{HMPID}{HMPID}{High Momentum Particle Identification Detector}
\newacronym{VHMPID}{VHMPID}{Very High Momentum Particle Identification Detector}
\newacronym{ArDM}{ArDM}{Ar Dark Matter}
\newacronym{DUNE}{DUNE}{The Deep Underground Neutrino Experiment}
\newacronym{ARIADNE}{ARIADNE}{ARgon ImAging DetectioN chambEr}
\newacronym{DARWIN}{DARWIN}{DARk matter WImp search with liquid xenoN}
\newacronym{HPTD}{HPTD}{High Transverse Momentum Trigger Detector}
\newacronym{CBM}{CBM}{Compressed Baryonic Matter}
\newacronym{FPT}{FPT}{Focal Plane Trackers}
\newacronym{NUMEN}{NUMEN}{Nuclear Matrix Elements}
\newacronym{CSR}{CSR}{Cooling Storage Ring}
\newacronym{CNC}{CNC}{Computer Numerical Control}
\newacronym{CEE}{CEE}{CSR External-target Experiment}
\newacronym{HCAL}{HCAL}{Hadronic Calorimeter}
\newacronym{CT}{CT}{Computed Tomography}
\newacronym{QE}{QE}{Quantum Efficiency}
\newacronym{FHM}{FHM}{Floating Hole Multiplier}

\usepackage{lineno}

\journal{PROGRESS IN PARTICLE AND NUCLEAR PHYSICS}

\begin{document}
\setcounter{tocdepth}{2}

\begin{frontmatter}

\title{The Thick Gas Electron Multiplier and its derivatives: physics, technologies and applications}

\author[inst1]{Shikma Bressler}
\author[inst1]{Luca Moleri}
\author[inst1]{Abhik Jash}
\author[inst1]{Andrea Tesi}
\author[inst1]{Darina Zavazieva}

\affiliation[inst1]{organization={Department of Particle physics and Astrophysics, Weizmann Institute of Science},
            addressline={Hrzl st. 234}, 
            city={Rehovot},
            postcode={7610001}, 
            country={Israel}}

\begin{abstract}
The Thick Gas Electron Multiplier (THGEM) is a robust high-gain gas-avalanche electron multiplier – a building block of a variety of radiation detectors. It can be manufactured economically by standard printed-circuit drilling and etching technology. We present a detailed review of the THGEM and its derivatives. We focus on the physics phenomena that govern their operation and performances under different operation conditions. Technological aspects associated with the production of these detectors and their current and potential applications are discussed. 
\end{abstract}




\end{frontmatter}

\tableofcontents
\printglossary[type=\acronymtype]

\section{Introduction}
\label{sec:introduction}

Gas-avalanche radiation detectors have been subject to intensive development since the 1960s. The invention of the Multi Wire Proportional Chamber (\acrshort{MWPC}) in 1968 \cite{CHARPAK1968262} has revolutionized the field of high energy physics, allowing precision detection of particles with electronic readout at an unprecedented rates. Based on early studies of Parallel Plate Avalanche Counters (\acrshort{PPAC}s) \cite{busser1965large}, the Resistive Plate Chamber (\acrshort{RPC}) \cite{santonico1981development} was proposed in 1981, offering a cost-effective solution for experiments requiring large area coverage. It was the first gaseous detector to incorporate resistive electrodes to enhance electrical stability in the presence of highly ionizing radiation at the cost of limited rate capabilities.  

In 1988, the Micro Strip Gas Chamber (\acrshort{MSGC}) \cite{oed1988position}, alternating thin anode- and broader cathode-strips printed on an insulating substrate, was the first proposed gaseous detector manufactured with micro-lithographic techniques. The small strip-to-strip pitch, of the order of a few hundred \um, resulted in significantly improved rate capabilities (up to a few \MHZMM) and much better position resolution (down to tens of \um) compared to \acrshort{MWPC}s \cite{sauli2020micro} and \acrshort{RPC}s. It marked the beginning of the Micro Pattern Gaseous Detector (\acrshort{MPGD}) era. For a recent review on \acrshort{MPGD}s, see~\cite{sauli2020micro}.

\acrshort{MPGD}s share similar operation principles with other gas-avalanche detectors. Ionization-induced primary electrons (\acrshort{PE}s) undergo charge avalanche multiplication in a region of a high electric field.  Current signals are induced on the readout electrodes by the movement of avalanche charges (electrons and ions), as described by the Shockley-Ramo theorem \cite{Shockley:1938itm, Ramo:1939}. 

However, in \acrshort{MWPC} and \acrshort{RPC} concepts, the electric field is typically determined by a single voltage difference between the anode and cathode electrodes. Thus, the production of \acrshort{PE}s, their multiplication, and signal induction occur in a single predefined volume and field configuration. In \acrshort{MPGD}s, these processes occur in dedicated and independently optimized field regions defined by the various electrodes (e.g., conversion/drift, multiplication, and the signal induction regions), providing additional flexibility in adapting \acrshort{MPGD}s to specific applications. Indeed, \acrshort{MPGD}s are used over a broad variety of applications, among which are tracking systems, photo-sensors in Cherenkov counters, standalone gaseous photo multipliers, x-ray imaging detectors, thermal and fast neutron imaging detectors, and more \cite{sauli2020micro} (see Section \ref{sec:Applications}).  

The currently leading \acrshort{MPGD} technologies are the MicroMesh Gas detectors (\acrshort{Micromegas}, \acrshort{MM}) \cite{attie2021current} and cascaded Gas Electron Multipliers (\acrshort{GEM}) \cite{sauli2016gas} conceived in 1996 and 1997, respectively. In \acrshort{MM}, the multiplication region is defined by two parallel electrodes - a thin (few \um{} thick)  mesh and a readout anode - distant typically 100 \um{} from each other. The mesh enables transferring efficiently the \acrshort{PE}s from a conversion/drift region preceding the micromesh to the multiplication region. The movement of the avalanche electrons within the narrow multiplication gap induces fast signals on the readout electrode. The \acrshort{GEM} consists of a thin (typically 50 \um{}) insulating foil metal-clad on both sides. Few tens of \um-diameter holes are etched through the foil. A voltage difference applied between the two conductive surfaces induces an intense dipole-like electric field inside the holes where charge multiplication  of radiation-induced \acrshort{PE}s drifting into the holes occurs. The avalanche electrons drift along an induction region, inducing a signal on the segmented readout anode. The operation of multiple \acrshort{GEM} foils in cascade ensures better electrical stability at high-charge avalanche amplification. The \acrshort{PE}s drift towards the holes of the first electrode in the cascade, and the resulting avalanche electrons are transferred to the next multiplication stage and so on. Once extracted from the last multiplication stage, the avalanche-electrons drift along an induction region, inducing a signal on the readout anode.  

\begin{figure}[htbp]
    \centering
    \subfloat[]{
    \includegraphics[width=0.35\linewidth]{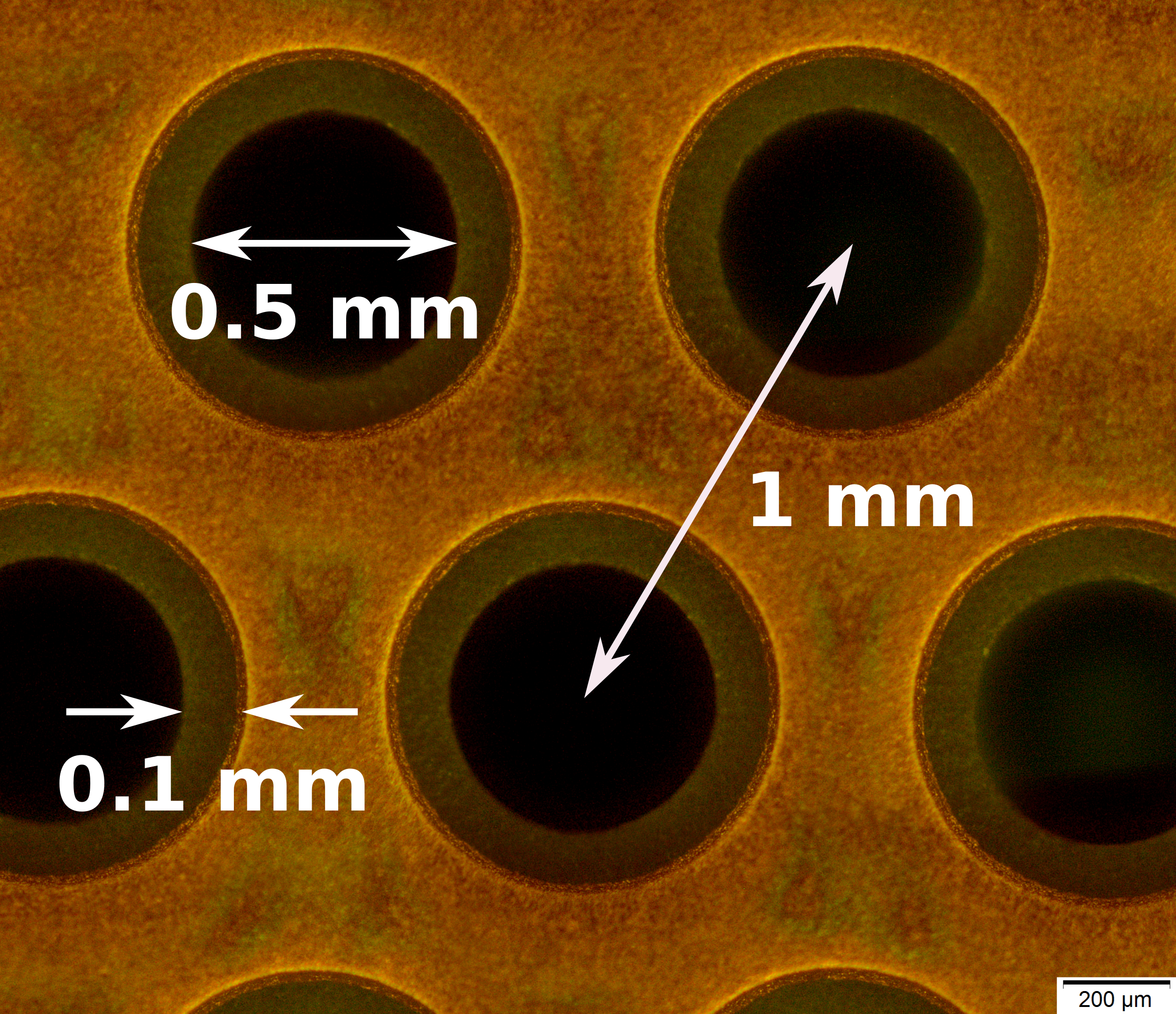}
\label{fig:THGEMElectrode}
    }
    \subfloat[]{
        \includegraphics[width=0.42\linewidth]{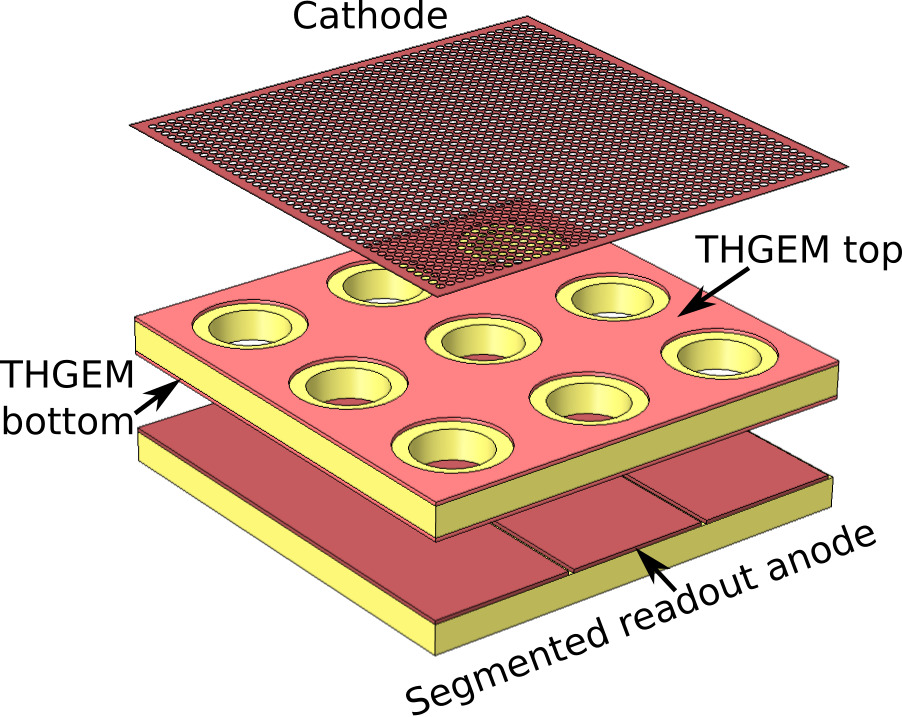}
        \label{fig:schema_3D}
    }

    \caption{\protect\subref{fig:THGEMElectrode} A \acrshort{THGEM} electrode with typical parameters. \protect\subref{fig:schema_3D} Schematic of a standard \acrshort{THGEM} configuration.}
\label{fig:THGEM_geom_schema}
\end{figure}

The THick Gas Electron Multiplier (\acrshort{THGEM}), also referred to as a Large Electron Multiplier (\acrshort{LEM}), was proposed independently by several authors between 2001 and 2004 \cite{Jeanneret:2001qxs,139,detlab_69}. It has a hole structure similar to that of a \acrshort{GEM} but with dimensions approximately ten-fold larger. A typical \acrshort{THGEM} electrode is presented in Figure \ref{fig:THGEMElectrode}; it can be manufactured economically by mechanically drilling sub-millimeter diameter holes, spaced by a $\sim$mm pitch in a fraction of a millimeter thick 2-layer PCB. To improve the electrical stability, a rim of a few tens of \um{} can be chemically etched around the holes. Being simple and robust, \acrshort{THGEM} detectors have been the subject of extensive studies and continuous development for various applications requiring radiation detection with submillimeter localization accuracy and a few ns time resolution over a large area \cite{131,detlab_16}. When coated with a photosensitive material (e.g., CsI), the \acrshort{THGEM}-electrode top face can be used as a photocathode with a relatively small dead area, making these detectors attractive gaseous photomultipliers (\acrshort{GPM}s) \cite{288,95,detlab_16}.

A standard \acrshort{THGEM}-detector configuration comprises a conversion-and-drift gap and an induction gap, followed by a readout electrode, as depicted in Figure \ref{fig:schema_3D}. In this configuration, the majority of the charge-multiplication occurs within the holes. This closed geometry limits photon feedback effects so that stable operation is achieved even in noble gases \cite{detlab_12} or highly scintillating ones like pure \cf{} \cite{detlab_1}. Under the most common conditions, detecting soft x-rays using Ar- and Ne-based gas mixtures at room temperatures, gas gains of the order of several thousand can be reached (see Section \ref{sec:maxGainRTSP}). Besides their most common operation at standard temperature and pressure, \acrshort{THGEM}-based detectors have shown good operation properties from low to high gas pressures \cite{detlab_5} and at cryogenic conditions \cite{417}. THGEM detectors preceded by proper converters were proposed for fast-neutron imaging \cite{133}. 

Electrical stability in the presence of large charge avalanches is achieved with cascaded structures (see Section \ref{sec:cascades}). For applications requiring a thinner configurations, the induction gap can be eliminated by directly coupling a \acrshort{THGEM} electrode with a metal clad on its top side only to the readout anode in a so-called THick-WELL (\acrshort{THWELL}) configuration \cite{detlab_41} (see Section \ref{sec:WELL}). Intermediate resistive layers are often deployed to mitigate the detrimental effect of discharges. To reduce secondary effects due to photon and ion feedbacks, the electrical field can be optimized by patterning the \acrshort{THGEM} electrode, e.g., in a THick COBRA (\acrshort{THCOBRA}) \cite{60} (see Section \ref{sec:Cobra}) or combining \acrshort{THGEM} with other \acrshort{MPGD} technologies in various hybrid configurations (see Section \ref{sec:Hybrids}). 

This review is structured as follows: The \acrshort{THGEM} detector, its basic operation principles, and its properties are detailed in Section \ref{sec:THGEM}. Its different \acrshort{THGEM} derivatives are presented in Section \ref{sec:DetectorDerivatives}.  Section \ref{sec:Technology} describes the technological aspects related to \acrshort{THGEM}-based detectors. The performance of the different configurations operated under various conditions are detailed in Section \ref{sec:PhysicsPerformance}. Section \ref{sec:Applications} is devoted to applications of \acrshort{THGEM}-based detectors.

\section{The THGEM}
\label{sec:THGEM}

A schematic description of a standard \acrshort{THGEM} detector configuration is presented in Figure \ref{fig:schema_3D}. The perforated electrode is located between a cathode (drift electrode) and an anode (readout electrode). Conventionally, the side facing the cathode is defined as the \acrshort{THGEM}-top and the one facing the anode, as \acrshort{THGEM}-bottom. The drift gap is defined by the cathode and the \acrshort{THGEM}-top and the induction gap by the \acrshort{THGEM}-bottom and anode. The electric-field configuration results from potentials applied to the various electrodes, in a sequence: $\Vcathode < \Vtop < \Vbottom < \Vanode$. \Vanode{} is typically kept at zero potential (to ease readout), implying that all other potentials have negative polarities.

\begin{figure}[htbp]
    \centering
    \subfloat[]{
    \includegraphics[scale=0.145]{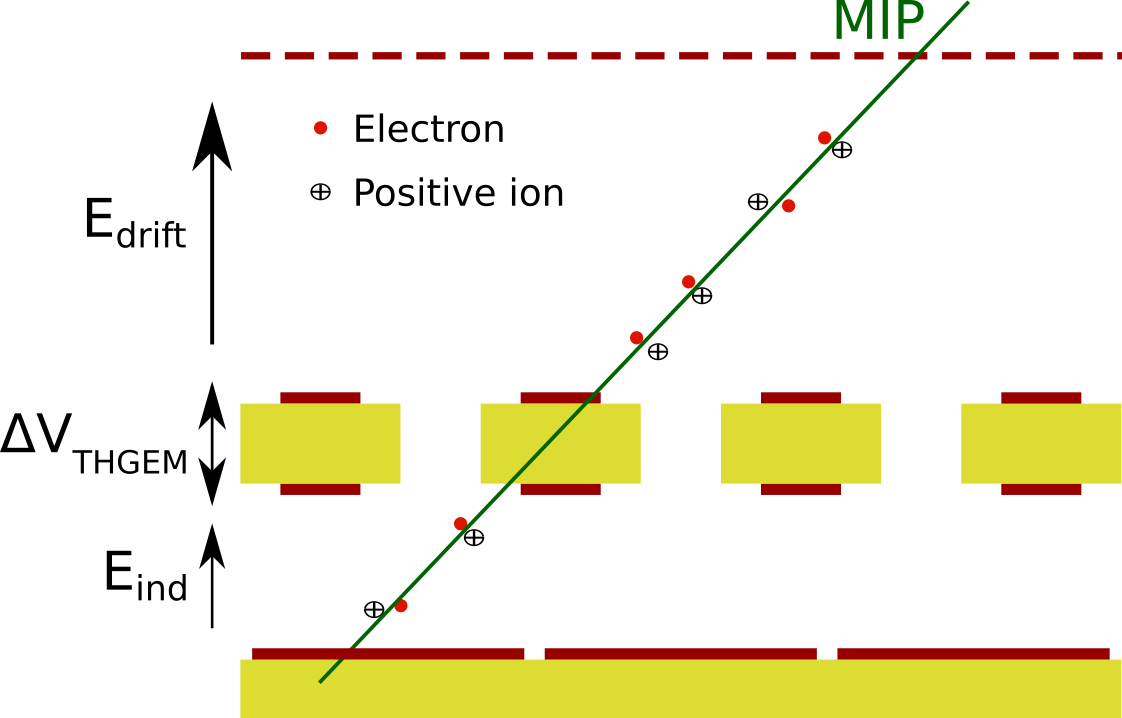}
\label{fig:THGEM_op1}
    }
    \subfloat[]{
        \includegraphics[scale=0.145]{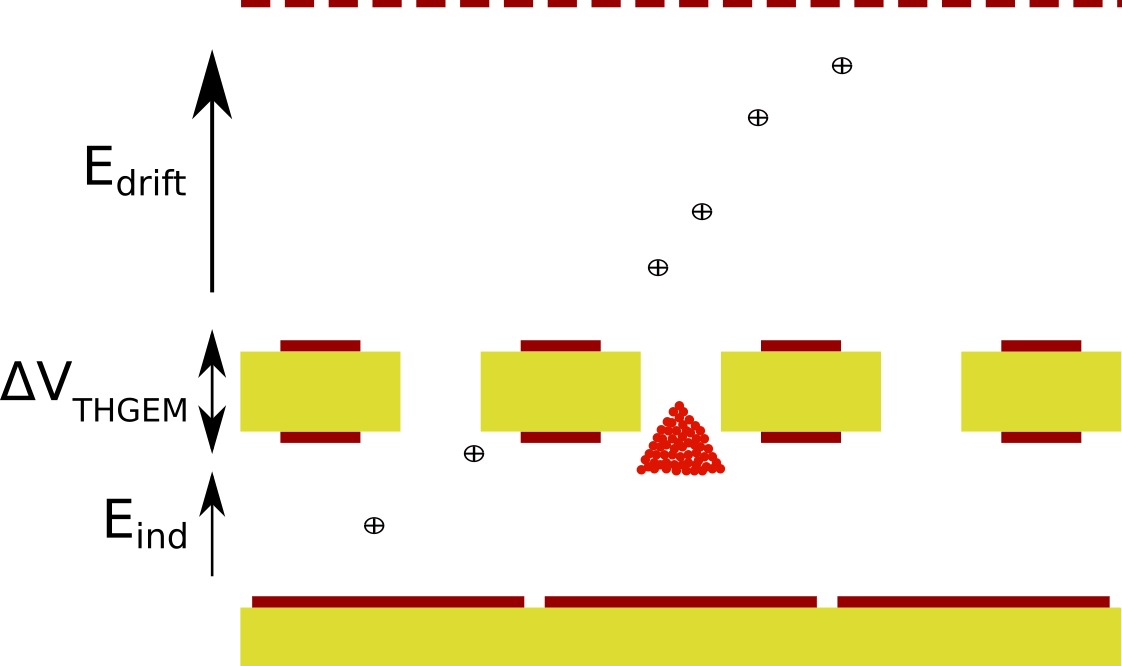}
        \label{fig:THGEM_op2}
    }
    \subfloat[]{
        \includegraphics[scale=0.145]{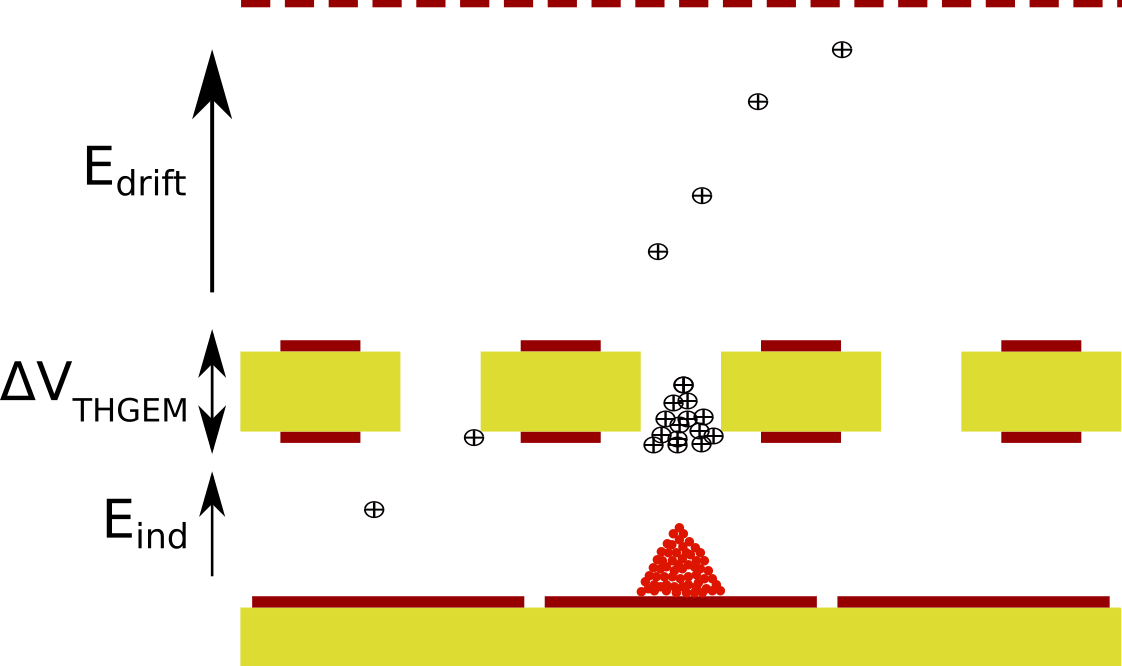}
        \label{fig:THGEM_op3}
    }
\caption{The principle of operation of a \acrshort{THGEM}: \protect\subref{fig:THGEM_op1} electron-ion pairs created by incoming radiation, \protect\subref{fig:THGEM_op2} electron avalanche in the \acrshort{THGEM} holes, \protect\subref{fig:THGEM_op3} drift of the avalanche electrons and ions to the anode and \acrshort{THGEM}-top, respectively.}
\label{fig:THGEM_op}
\end{figure}

The operating principle is illustrated in Figure \ref{fig:THGEM_op}. In this example, a \acrshort{MIP} traversing the detector creates electron-ion pairs along its trajectory (Figure \ref{fig:THGEM_op1}). While the ions drift slowly towards the cathode, the \acrshort{PE}s drift three orders of magnitude faster along the field lines into the holes, where they undergo charge-avalanche multiplication (Figure \ref{fig:THGEM_op2}). The avalanche electrons are extracted into the induction gap and drift towards the anode. The avalanche ions slowly drift within the holes toward the \acrshort{THGEM}-top (Figure \ref{fig:THGEM_op3}). The primary electron-ion pairs created in the induction gap do not contribute to the avalanche process. 

The detector's performance is determined by the geometry of the \acrshort{THGEM} electrode (substrate thickness, hole diameter, hole pattern, and rim size), the drift and induction gaps, and the operating conditions. The latter refers to the gas mixture, its temperature and pressure, the applied voltages, and the type of incoming radiation. In what follows, we discuss the basic principles related to \acrshort{THGEM} detectors using as an example a typical configuration with a 0.8 mm thick substrate perforated with 0.5 mm diameter holes (0.1 mm etched rims) drilled into an hexagonal pattern with 1 mm pitch, a drift gap of 5 mm, and an induction gap of 2 mm. The detector was operated in an $\mathrm{Ar/CO_2}$ (93:7) gas mixture.

\begin{figure}[htbp]
    \centering
    \subfloat[]{
        \includegraphics[width=0.45\textwidth, trim={0.2cm 0.0cm 0cm 0cm},clip]{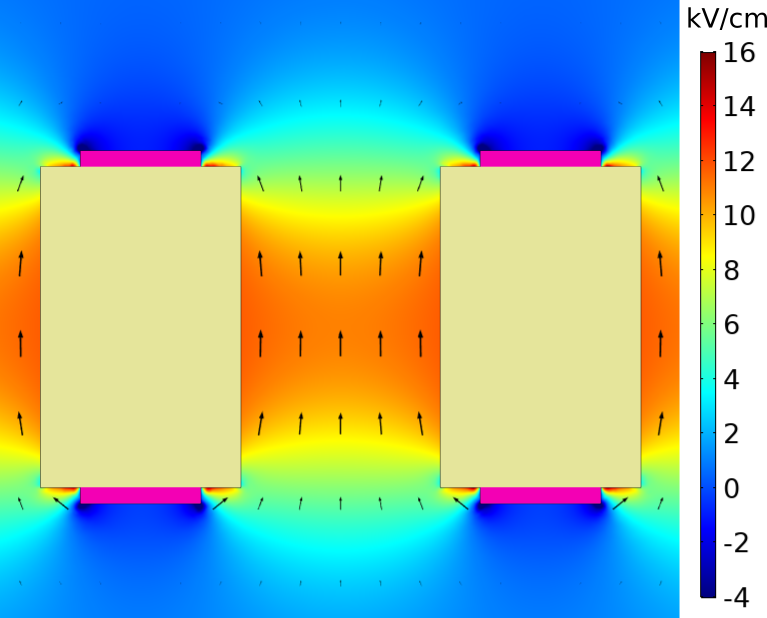}
        \label{fig:E_intensity}
    }
    \subfloat[]{
        \includegraphics[width=0.55\textwidth]{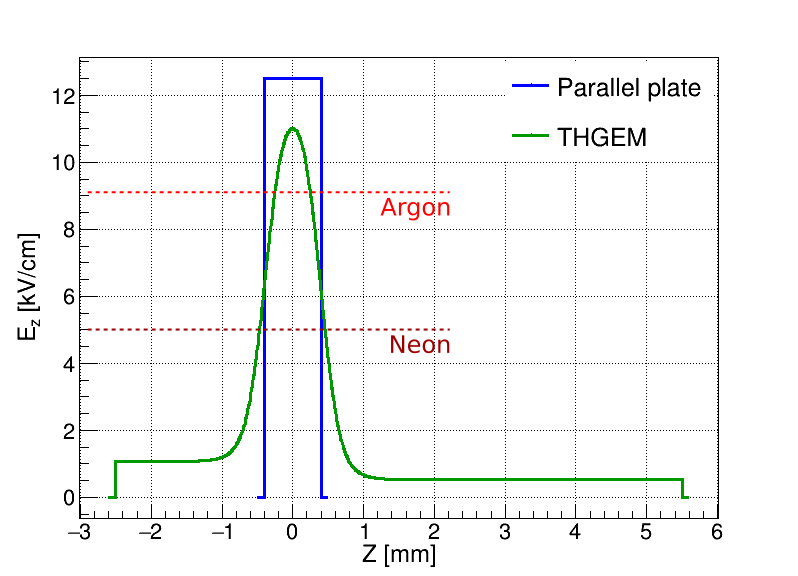}
        \label{fig:comp_THGEM_parallelplate}
    }
    \caption{Characteristic electric field of a \acrshort{THGEM} electrode (thickness = 0.8 mm, hole diameter = 0.5 mm, rim width = 0.1 mm) for \DVTHGEM{}= 1000 V, \Eind{}= 1 \kvcm{}, and \Edrift{}= 0.5 \kvcm.  \protect\subref{fig:E_intensity} A field intensity map. Arrows represent the field direction.
    \protect\subref{fig:comp_THGEM_parallelplate} The field intensity along the line through the the hole-center axis. The constant field in a parallel plate configuration at the same voltage is shown for comparison. The horizontal dashed lines represent the threshold for charge multiplication in the indicated gases.}
\label{fig:THGEMField}
\end{figure}

\paragraph{Electric field and charge-avalanche multiplication} The voltage difference between the \acrshort{THGEM}-top and bottom, $\DVTHGEM=\Vtop-\Vbottom$ results in an intense (tens of \kvcm{} at the holes center) dipole-like electric field within the holes, where most of the charge-avalanche multiplication occurs.  Studies have shown that this field is maximal for an aspect ratio of one between the thickness of the electrode and the diameter of the hole \cite{detlab_4}. The dependence of the field strength on different electrode parameters was simulated in \cite{408}. The drift field, \Edrift, is determined by $\Vcathode - \Vtop$, while the induction field, \Eind, is determined by $\Vbottom - \Vanode$. The field map and profile in the center of the hole in a typical configuration are shown in Figure \ref{fig:E_intensity} and Figure \ref{fig:comp_THGEM_parallelplate}, respectively, for typical values of \DVTHGEM{}= 1000 V, \Eind{}= 1 \kvcm{}, and \Edrift{}= 0.5 \kvcm. For comparison, the uniform field in a 0.8 mm thick parallel plate detector biased at 1000 V is shown to be higher than the \acrshort{THGEM} field at the hole's center. 

Typically, the maximum field value of a \acrshort{THGEM} detector is obtained at the hole's center at half the electrode's thickness \cite{detlab_4}; however, the high-field region above the multiplication threshold (10–15 \kvcm{}, depending on the specific gas) could extend beyond the holes, at the rim vicinity. For taking advantage of the multiplier’s “closed geometry” (e.g. compared to the “open” one of \acrshort{MWPC}s, \acrshort{PPAC}, \acrshort{RPC}, \acrshort{MM}, etc.) in reducing of the probability of avalanche-photons initiating secondary avalanches away from the incoming particle trajectory (so-called photon feedback \cite{detlab_18}), it is beneficial to reduce the field intensity outside the hole. Simulations have shown that the latter is inversely proportional to the electrode thickness and increases with hole diameter \cite{detlab_12,detlab_14}. Due to their closed geometry, limiting photon feedback effects, \acrshort{THGEM} detectors can be stably operated in mildly-quenched gas mixtures and even in pure, noble gases \cite{detlab_12} or in other scintillating ones, e.g., in $\mathrm{CF_4}$ \cite{detlab_4,83}. On the other hand, large \Eind{} \cite{Coimbra:2013jpa,Bressler:2013qsa}, and \Edrift{} \cite{205,499} may also be applied to extend the high field region and increase the total amplification, but at the cost of enhanced secondary effects.

\begin{figure}[htbp]
    \centering
    \includegraphics[width=0.7\textwidth, trim={0.2cm 0.0cm 2cm 1cm},clip]{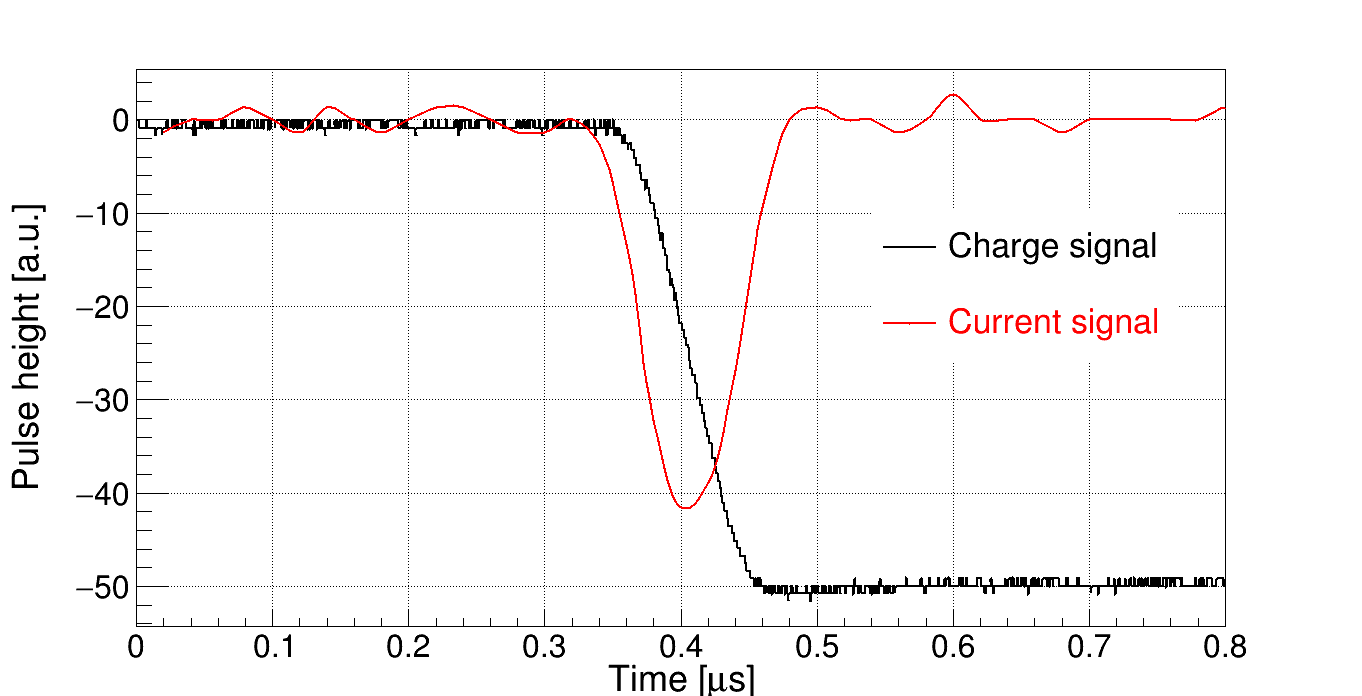}
    \caption{X-ray induced charge and current signals recorded from the anode of a typical \acrshort{THGEM} configuration (Figure \ref{fig:THGEM_geom_schema}) operated in \arco{} (93:7). A 0.8 mm thick \acrshort{THGEM} electrode perforated with 0.5 mm diameter holes (0.1 mm rims) drilled into an hexagonal pattern with 1 mm pitch was used for the measurement.}
    \label{fig:THGEMAnodeSignal}
\end{figure}

\paragraph{Signal shape} 
According to the Shockley-Ramo theorem \cite{Shockley:1938itm, Ramo:1939}, currents are induced on all electrodes by the movement of charges, i.e., electrons and ions in gas. The signal intensity is proportional to their velocity. An example of an x-ray signal recorded with a charge-sensitive preamplifier\footnote{By design, the output signals of charge sensitive preamplifiers are inverted with respect to the input ones. The original polarity of the input signal is restored in all presented figures.} from the anode in a standard \acrshort{THGEM} configuration (in Figure \ref{fig:THGEM_geom_schema}), operated in \arco{} (93:7), is shown in Figure \ref{fig:THGEMAnodeSignal}. Since the amplifier acts as a current integrator, the signal amplitude represents the total induced charge as a function of time. The corresponding current signal, obtained by differentiating the charge one, is overlaid. The \acrshort{THGEM}-bottom electrode shields the anode from charges moving within the hole. Thus, fast-rising (a few tens of nanoseconds) signals of negative polarity are induced on the anode mainly by the movement of electrons towards the anode in the induction gap \cite{detlab_61}. 

\begin{figure}[htbp]
    \centering
    \subfloat[]{
        \includegraphics[width=0.5\textwidth]{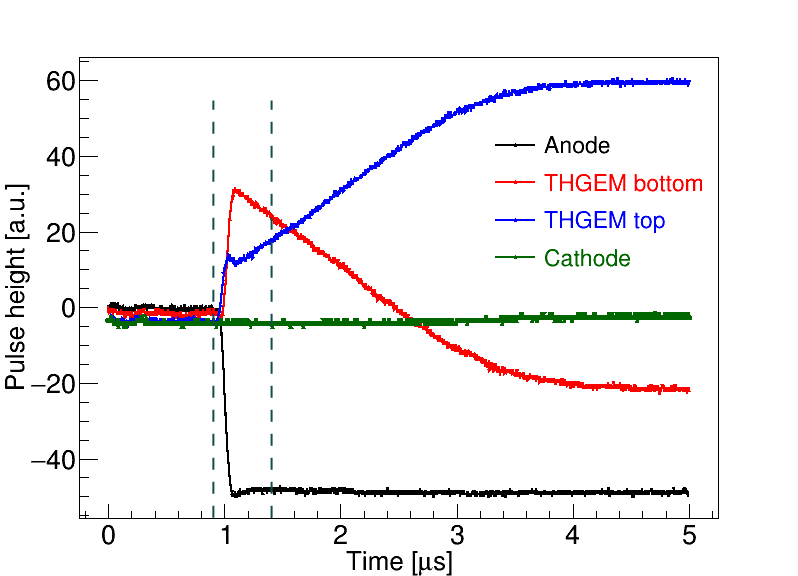}
        \label{pulses_allElectrodes}
    }
    \subfloat[]{
        \includegraphics[width=0.5\textwidth]{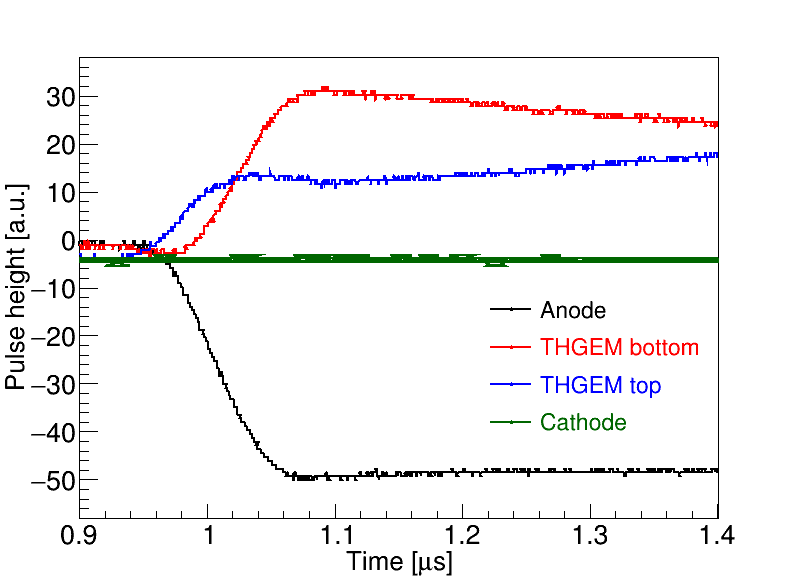}
        \label{pulses_allElectrodes_zoom}
    }
    \caption{\protect\subref{pulses_allElectrodes} Typical x-ray charge signals recorded on the electrodes of the detector configuration of Figure \ref{fig:THGEM_geom_schema} operated in \arco{} (93:7). A 0.8 mm thick \acrshort{THGEM} electrode perforated with 0.5 mm diameter holes (0.1 mm rims) drilled into an hexagonal pattern with 1 mm pitch was used for the measurement. \protect\subref{pulses_allElectrodes_zoom} An expanded view of the fast component of the signals (region marked by the dashed lines in \protect\subref{pulses_allElectrodes}).}
\label{fig:THGEMSignals}
\end{figure}

Figure \ref{fig:THGEMSignals} depicts charge signals recorded from the anode, the \acrshort{THGEM}-top and bottom, and the cathode. The signal induced on the \acrshort{THGEM}-bottom is characterized by a fast-rise component of positive polarity due to the avalanche electrons drifting away from it towards the anode and a slow-rise negative component due to the positive ions drifting towards the \acrshort{THGEM}-top. The slow component is a few \us{} long, consistent with the typical drift-time of ions in gas. The exact calculation of the latter depends on the composition of the ions and their dynamics \cite{kalkan2015cluster}.  

The signal induced on the \acrshort{THGEM}-top has a fast rise component of positive polarity, similar to that of the \acrshort{THGEM}-bottom but with a smaller amplitude since the \acrshort{THGEM}-top is located farther away from the drifting electrons. Additionally, the bottom electrode partially shields the \acrshort{THGEM}-top from the electrons moving in the induction gap; moreover, it has a slow positive component induced by the ions drifting in its direction. The small signal on the cathode is induced only by ion movements; thus, it has a slow rise and positive polarity.

\begin{figure}[htbp]
\centering
\includegraphics[width=0.65\textwidth, trim={0.0cm 0.0cm 0cm 1.0cm},clip]{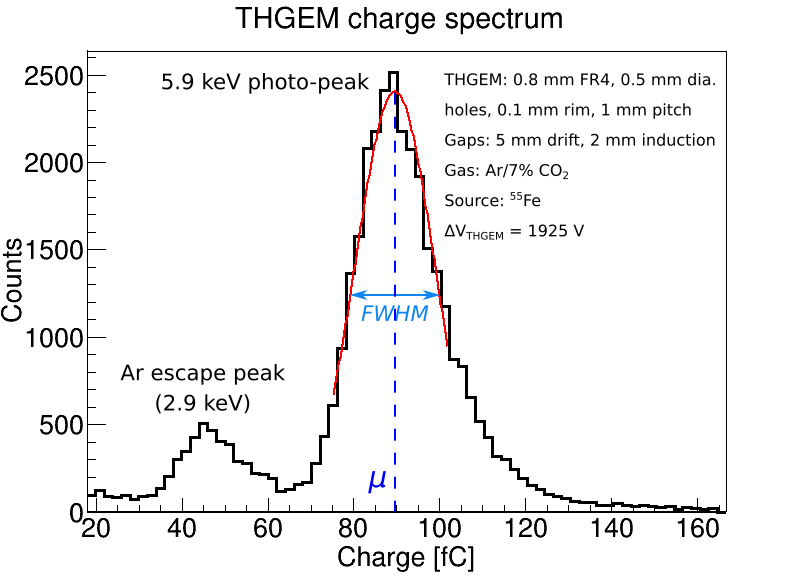}
\caption{Typical \acrshort{THGEM} charge spectrum measured with the detector configuration of Figure \ref{fig:THGEM_geom_schema} operated in $\mathrm{Ar/CO_{2}}$ (93:7) gas mixture.
}
\label{fig:THGEMFeSpectrum}
\end{figure}

\paragraph{Charge spectrum}
The induced charge-spectrum is the histogram of charge signals. Its shape depends on the irradiation source and the detector properties. As an example, a \fe{} x-ray spectrum measured in \arco{} (93:7) is shown in Figure \ref{fig:THGEMFeSpectrum}. The characteristic distribution consists of a characteristic photo-peak and an "escape" one. The representative avalanche charge, $\mu$, is the mean value of the photo-peak Gaussian. 
Instead, the characteristic spectrum of \acrshort{MIP}s has a broad Landau distribution \cite{detlab_26}. The representative avalanche charge, in this case, is the most probable value of the distribution. 

\paragraph{Energy resolution} The energy resolution is normally defined as $E_{res} = \frac{FWHM}{\mu}$\footnote{Some authors also define $E_{res} =\frac{\sigma_E}{E}$}, where $FWHM$ and $\mu$ are the full-width at half-maximum and the mean value of a Gaussian-like spectrum measured (e.g., in Figure \ref{fig:THGEMFeSpectrum}) with a soft x-rays, respectively. In standard THGEM configurations, the energy resolution is typically of the order of $\mathrm{20-30\%}$ \cite{detlab_4,detlab_12,detlab_14}; it depends on the gas, fluctuations in the number of \acrshort{PE}s and in the number of electrons reaching the multiplication region, the electric field configuration, statistical avalanche fluctuations, electronic noise, and detector uniformity. The energy resolution can be optimized by tuning the gain, drift, and induction fields, as well as using specific gaseous mixtures \cite{detlab_12,499}. When measured as a function of the detector gain, the energy resolution often has a minimum value \cite{detlab_73}. 

\paragraph {Detector gain}
Under the assumption of 100\% collection efficiency of the \acrshort{PE}s into the multiplication region, the absolute detector gain, \absG, is defined as the ratio between the final number of electrons (after multiplication), $N^{final}_e$, and the number of \pes, $N_{\pe}$:
\begin{equation}
    \absG = \frac{N^{final}_e}{N_{\pe}}
\label{eq:absG}
\end{equation}

The effective gain, \effG, is defined as the total charge measured by the readout system over the number of \acrshort{PE}s. It could be significantly different from the absolute gain for several reasons. First, the collection efficiency of the \acrshort{PE}s 
could be limited by losses to the \acrshort{THGEM}-top. Likewise, the electron transfer efficiency (\acrshort{ETE}) - the fraction of electrons transferred through the hole into the induction region - could be limited by electron losses to the \acrshort{THGEM}-bottom and in the holes' walls. Second, the charge measured by the readout electronics depends on their response (e.g., the effect of the ballistic deficit in \cite{detlab_61}). In addition, different effective gains could be measured on different electrodes.

In a standard \acrshort{THGEM} configuration, electron collection and \acrshort{ETE} values of nearly 100\% can be reached at relatively low fields \cite{detlab_1,403,205,500}. Thus, by using optimized readout electronics, the effective gain measured from the anode could be similar to the absolute one. 

Typically, the effective gain is estimated either from the DC current supplied to an electrode at high-rate detector irradiation or from the charge spectrum. When estimated from the DC current \cite{detlab_4}, $\effG = \frac{I}{N_{\pe}\cdot r_{source} \cdot q_e}$. Here, $I$ is the measured current, $N_{\pe}$ is the number of \acrshort{PE}s, $r_{source}$ is the source rate, and $q_e$ is the electron charge. When estimated from the induced current signals, $\effG = \frac{\mu}{N_{\pe} \cdot q_e}$, where $\mu$ is the representative value of the charge spectrum. 

\begin{figure}[htbp]
    \centering
    \subfloat[]{
        \includegraphics[width=0.515\textwidth]{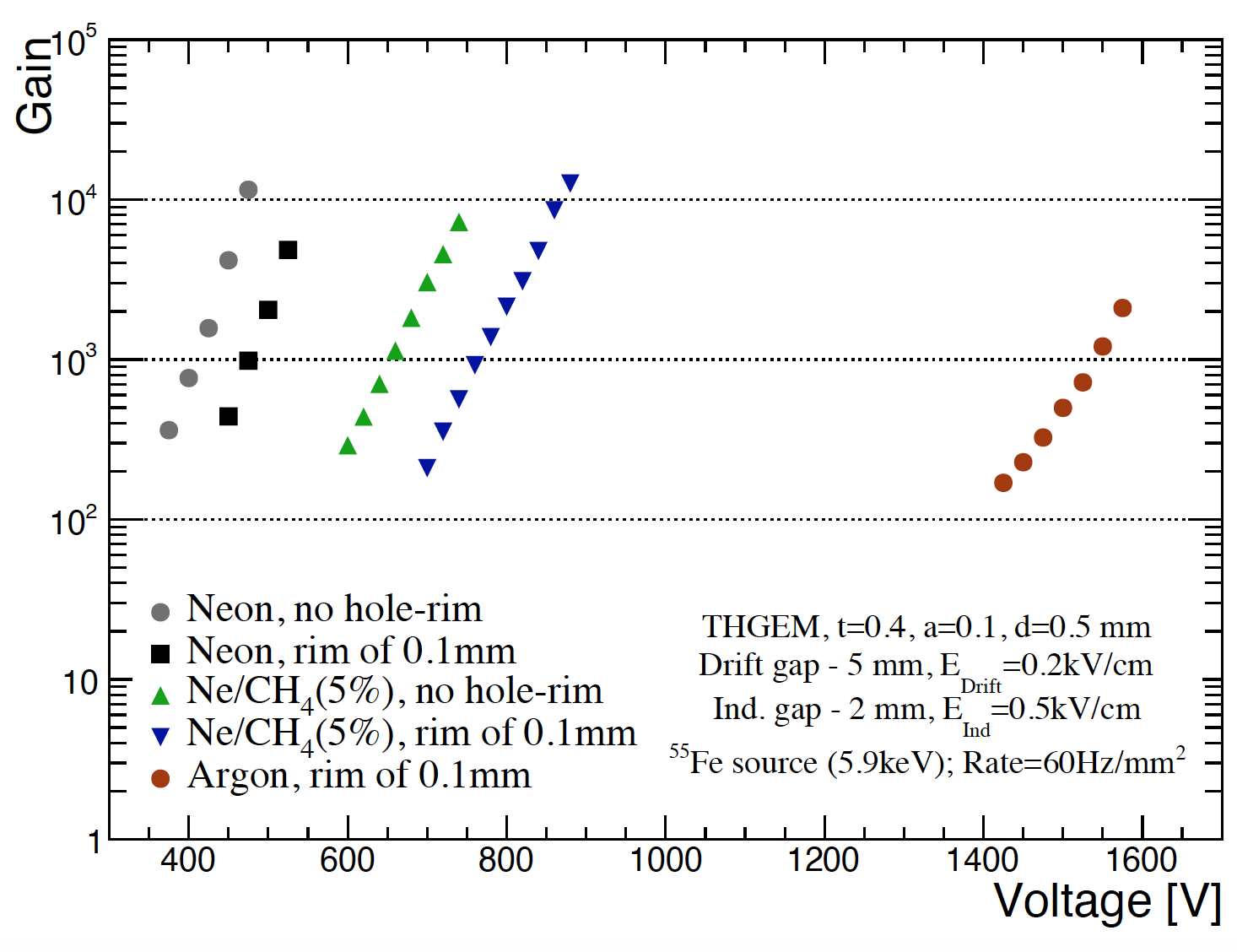}
        \label{fig:THGEMGain}
    }
    \subfloat[]{
        \includegraphics[width=0.45\textwidth, trim={0.0cm 0.1cm 0cm 0cm},clip]{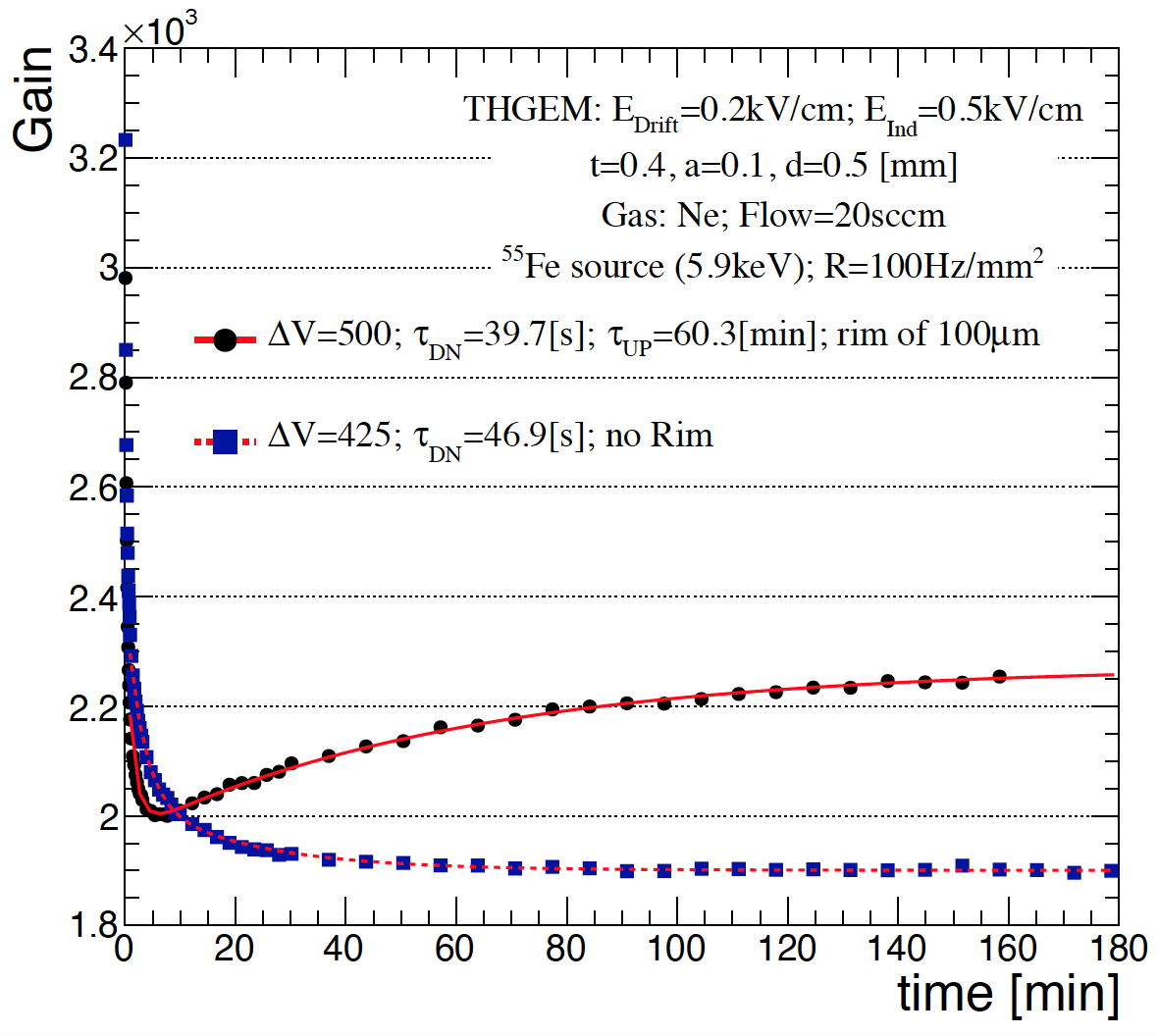}
    \label{fig:THGEMGainStabilization}
    }
    \caption{\protect\subref{fig:THGEMGain} Typical \acrshort{THGEM} gain curves measured with \fe{} x-rays, operated in different gas mixtures at standard temperature and pressure with \Eind{} = 0.5 \kvcm{} and \Edrift{}= 0.2 \kvcm{}. \protect\subref{fig:THGEMGainStabilization} A typical \acrshort{THGEM} gain stabilization curve. Figures from \cite{detlab_55}.}
\label{fig:gain}
\end{figure}

Typical gain curves (gain as a function of \DVTHGEM) measured with 0.4 mm thick \acrshort{THGEM}, operated in different gas mixtures with \Eind{}= 0.5 \kvcm{} and \Edrift{}= 0.2 \kvcm{}, are shown in Figure \ref{fig:THGEMGain}. In agreement with the Townsend theory (see for example Chapter 6 in \cite{knoll2010radiation}), an exponential trend is observed for all gases and geometries investigated. For all configurations, the maximum achievable gain, \maxG, is determined by the onset of electrical instabilities.  

\paragraph{Detection efficiency}
The detection efficiency is defined as the number of detected particles over the number of incoming ones. It is determined by the probability that the incoming particles induce primary ionization, $\varepsilon_{\pes{}}$, the collection efficiency of the \acrshort{PE}s into the multiplication region, $\varepsilon_{col}$, and the signal-to-noise separation, $\varepsilon_{S/N}$, for each avalanche gain. 
The detection efficiency of \acrshort{MIP}s and x-rays as a function of the amplification often reaches a plateau. Its value is dominated by the probability that the particles interact with the gas and induce \acrshort{PE}s in the drift gap. 

Instead, for UV photons, the single-photon detection efficiency (\acrshort{PDE}), $\varepsilon_{\gamma}$, is given by:
\begin{equation*}
    \varepsilon_{\gamma} = A \times QE \times \varepsilon_{ext} \times \varepsilon_{col} \times \varepsilon_e
\end{equation*}
Here, $A$ and \acrshort{QE} are the effective area and the quantum efficiency (in vacuum) of the photocathode, respectively. $\varepsilon_{ext}$ is the extraction efficiency of a photoelectron, dictated by its probability of backscattering on gas molecules and recombination. The resulting effective quantum efficiency, $QE \times \varepsilon_{ext}$, depends on the gas and on the electric field at the photocathode surface \cite{215}.  $\varepsilon_{col}$ is the collection efficiency of the extracted electron into the amplification region and $\varepsilon_e$ is the efficiency of detecting a single electron pulse above noise. 

High \acrshort{QE} values are obtained with electrodes coated with CsI photocathodes \cite{breskin1996csi}. Spurious signals were observed in measurements with intense irradiation, which were attributed to ion or photon feedback effects \cite{detlab_18, 214}. These could affect \acrshort{PDE} measurements.

\paragraph{Rate capability} 
The rate capability of a standard \acrshort{THGEM} is determined by the ions' drift time along the \acrshort{THGEM} holes and by the avalanche size - the larger the avalanche, the lower the rate in which gain drop is observed. Different authors reported different dependency of the gain on the incoming particle's rate \cite{detlab_4,detlab_18,detlab_41}, e.g., in a typical configuration, with an avalanche size of $10^4$ electrons, a constant gain is measured up to an incoming particle's rate of $\sim$$\mathrm{10^5}$, after which it drops abruptly. 

\paragraph{Gain stabilization}
A drawback of having an exposed insulator surface in the proximity of the multiplication region is the so-called charging-up effect - the accumulation of positive and negative charges on the insulator surface; it causes time variations of the field and, thus, of the gain. Charging-up effects have been studied in detail in simulations and dedicated experiments \cite{27,detlab_55,detlab_56,70,38,detlab_54,63,205,526}. A typical gain stabilization curve is shown in Figure \ref{fig:THGEMGainStabilization}, taken from \cite{detlab_55}.

The gain stabilizes at an equilibrium condition in which no more charges reach the insulator. Thus, the stabilization time depends on the field configuration, detector gain, and incoming radiation type – namely the avalanche size and charge density within a hole. The presence of etched rims around the holes allows higher gain values to be reached \cite{detlab_1,detlab_40,16} at the expense of slowly charging up the insulator. This adds a slow component to the stabilization process. The charging-up trend is strongly dependent on the structure of the hole \cite{476} and on the electrode substrate \cite{476,27}. It was demonstrated that the drift field influences 
the charging-up effect \cite{detlab_55} as well. The charging-down mechanisms responsible for slow evacuation of charges from the insulator were not studied as thoroughly.

Together with uncertainties related to gas purity \cite{detlab_14,detlab_20}, charging-up effects are a major source of uncertainty in gain measurements. In such studies, these can be reduced following a fixed procedure to "initialize" the electrode and stabilize the gain \cite{detlab_54}. Charging-up effects can also be mitigated by coating the insulators with a nanometer-thick layer of high resistivity, such as Diamond Like Carbon (\acrshort{DLC}) \cite{261}.

\paragraph{Position resolution} 
\acrshort{THGEM} detectors have a position resolution of a few hundred \um. The position resolution depends on the type of radiation (extended or point like), signal-to-noise ratio, and type of readout. However, studies have shown that it is 
mostly limited by the pitch of the holes into which the primary electrons are focused \cite{detlab_51}. Nonetheless, precision better than the holes' pitch is reached \cite{detlab_51,detlab_6}. 

\paragraph{Time resolution}
It was shown that a time resolution of the order of a few \ns{} could be reached \cite{detlab_72,105}. The physics governing the time resolution of THGEM detectors, such as its dependency on the signal shape, gas mixture, signal-to-noise ratio, readout electronics, etc. was not studied in detail. 

\paragraph{Electrical instabilities}
The occurrence of discharges is a limiting factor of all gaseous detectors, and, typically, the onset of discharges defines their dynamic range. In the following discussion, we assume high-quality \acrshort{THGEM} electrodes (see, e.g., \cite{118} in Section \ref{sec:MaterialsAndProduction}) with no production imperfections, such as sharp edges that cause localized instabilities.  

The sequence of events leading to a discharge is initiated when the avalanche size exceeds a critical charge limit ($10^{6}$ - $10^{7}$ electron-ion pairs), the so-called Raether limit \cite{von1965electron}. The resulting local electric field becomes large enough to induce a transition of the avalanche to a forward–backward propagating streamer, a well-studied process in gas-avalanche detectors (see for example \cite{von1965electron, Fonte:1991yt, Fonte:1997tw, Peskov:1997ti, Battistoni:1983wk, Taylor:1989uy}). Due to the small distance between the electrodes in most \acrshort{MPGD} configurations, the streamer is likely to form an electrical connection between neighboring electrodes of different potentials, consequently discharging part of the energy stored in the capacitor defined by this two-electrode system. 
In \acrshort{THGEM} detectors, the addition of etched rims around the holes allows for a stable operation at higher gain values \cite{detlab_1,detlab_40}; they smoothen the sharp conductive edges and increase the distance between the two electrodes. 

Although \acrshort{THGEM} electrodes are robust against discharges, the latter can still induce dead time, charge up the substrate,  and damage the readout electronics. The discharge can also propagate through the induction gap and cause a delayed secondary discharge to the anode. This effect depends on the intensity of the induction field and the \acrshort{THGEM} clad material \cite{469,505} (whereas the probability of primary discharge does not).

Different methods have been developed over the years to solve or mitigate the discharge problem. One common approach is to employ cascaded structures \cite{Breskin:1979sa, Bressan:1998uu, Charles:2011zz} to reduce the charge density at each stage. The use of gases with large diffusion coefficients to spread the \acrshort{PE}s over several holes has also been suggested \cite{detlab_14,68}. 

Dividing the detector into smaller segments does not allow larger gain values. Instead, it reduces the area affected by a discharge and the corresponding discharge energy stored in the two-electrode capacitor \cite{Charles:2011zz}. In recent years, the most common approach to mitigate discharges in \acrshort{MPGD}s (including structures based on \acrshort{THGEM}) is embedding resistive electrodes in the detector assembly \cite{Dixit:2003qg, Alexopoulos:2011zz, DiMauro:2007ncz, 1, ALICEITS:2012ojb, Fonte:2010aw, Bencivenni:2014exa}. This has two roles: (i) protect the readout electronics by decoupling it from the energy released in the discharge, and (ii) quenching the discharge energy. The long clearance time of the charges from the resistive electrode results in a local reduction of the electric field and self-extinction of the discharge.

\paragraph{Aging} Severe aging-related effects have not been reported for \acrshort{THGEM} detectors. However, many repeated discharges (millions) occuring at the same point, could damage the \acrshort{THGEM} electrode \cite{469}. 

\paragraph{Light yield}
Various applications make use of electroluminescence (\acrshort{EL}) produced by inelastic collisions of electrons with gas molecules in the multiplier’s high-field region \cite{BRESKIN1989457} (for \acrshort{THGEM}-related works see for example \cite{413,detlab_34} and Sections \ref{sec: light readout}, \ref{ref:DoalPhaseNobleLiquidTPC}, \ref{sec:non invasive imaging} and \ref{sec:ARIADNE}). The effect could be enhanced by a small avalanche multiplication. The light yield is defined as the number of \acrshort{EL} photons emitted per single electron. Thicker electrodes could be advantageous in this respect due to the longer drift length in the gases, in particular in configurations in which the \acrshort{EL} photons are not lost inside the holes (e.g., \cite{248}).

\paragraph{Ion back-flow} 
The Ion back-flow (\acrshort{IBF}) is defined by the percentage of avalanche ions reaching the \acrshort{THGEM}-top electrode. In a standard configuration, the \acrshort{IBF} could be as high as 100\% and become a limiting factor in various applications (see, e.g., Section \ref{sec:RICH}). Different methods have been developed over the years to reduce the \acrshort{IBF}. One common approach is to employ cascaded misaligned structures \cite{19}.

\section{THGEM Derivatives}
\label{sec:DetectorDerivatives}

Various detector configurations have been derived from the \acrshort{THGEM} concept. Some of them attempt to resolve known limitations, such as electrical instability, \acrshort{IBF}, and more. Some derivatives focus on extending the detector's dynamic range, while others are focus at meeting the requirements of specific applications.

In what follows, we describe present derivatives. We discuss the main motivation for their development, explain how the proposed concept addresses the problem it was meant to resolve, and present their performance. 

\subsection{Cascaded THGEM}
\label{sec:cascades}

Given their high electron transfer efficiency, \acrshort{THGEM} electrodes can be efficiently cascaded \cite{detlab_1,215}. Compared to a single-stage \acrshort{THGEM} configuration (1THGEM), in a cascade structure (e.g., a double, 2THGEM, and a triple, 3THGEM), higher effective gain is reached. The total amplification in each stage is smaller, and the charge in the last amplification stage, with the largest charge density, is divided between more holes \cite{detlab_34}. Consequently, cascaded structures have a broader dynamic range \cite{detlab_12} and can achieve higher gains under stable operating conditions.
This was demonstrated, e.g., in \cite{detlab_18}; gains in the order of $\mathrm{10^3-10^4}$ were reached with a 1THGEM detector operating in $\mathrm{Ne/CH_4}$ (95:5) with soft x-rays, while 2THGEM and 3THGEM detectors yielded gains of several $\mathrm{10^5}$. Further, staggering \acrshort{THGEM} electrodes in a cascade configuration were found effective in reducing \acrshort{IBF} at the expense of some gain loss. More details and specific operating conditions are provided in Section \ref{sec:PhysicsPerformance}.

\subsection{Resistive THGEM - RETGEM}
\label{sec:RETGEM}

Electrical instabilities and gaseous breakdown leading to occasional discharges have been the main limitation in operating \acrshort{THGEM} detectors over broad dynamic ranges. Attempts to reduce their occurrence and mitigate their destructive effects have been an integral part of these detectors' development. One of the first proposed methods was to incorporate resistive electrodes in their assembly, leading to the development of the resistive THGEM (\acrshort{RETGEM}) \cite{285}. The resistive material could replace the conductive one \cite{1,155,2,157,249} or be placed in direct contact with it \cite{285,163,224,7,45,4}. Related technological aspects (coating methods, materials, etc.) are detailed in Section \ref{sec:Technology}. As regular \acrshort{THGEM}s, \acrshort{RETGEM}s can be coated with photosensitive coatings for Ultra Violet (\acrshort{UV}) light detection \cite{155,4} and cascaded (see, e.g., \cite{2,262}, among many others). 

The extent to which the discharge is quenched depends on the effective resistance along the charge path, either to the ground or to power supply. 
As long as no current is flowing through the resistive electrodes, the amplification field of the \acrshort{RETGEM} is similar to that of a standard \acrshort{THGEM}. Under these conditions, the avalanche formation in a \acrshort{RETGEM} follows the same principles as in a regular \acrshort{THGEM}. Yet, at high irradiation rates or large gains, when current is flowing through the resistive materials, voltage drops across the resistive electrodes result in lower amplification fields. Thus, lower gain occurs for the same operating voltage. Minimizing the current flow along the surface and speeding up the charge evacuation from large-area resistive electrodes enable stable operations at a wide range of irradiation rates and mitigate cross-talk effects. This is achieved by segmenting the resistive surfaces with conductive grids or lines for local charge evacuation \cite{157,249}, which can also be used to enable imaging capabilities \cite{45,4}.

The signal shape measured from the \acrshort{RETGEM} electrodes could be different from that measured in a regular \acrshort{THGEM} under all operating conditions. An additional slow component is typically measured due to slow evacuation of charges (electrons or ions) from the resistive layer.   

In \cite{285,1,162,163,157,45,249}, some results for the gain of single- (1RETGEM) and double-RETGEM (2RETGEM) configurations are presented. Different resistive electrodes were operated in pure form and mixtures of Ne and Ar, in He mixtures and in air, including measurements in Ar at cryogenic temperatures \cite{287}. An energy resolution in the range of 20-40\% was reached in various Ar-based gas mixtures using 5.9 and 8 \kev{} x-ray photons \cite{259}. 

Studies with \acrshort{DLC}-coated \acrshort{RETGEM} (See section \ref{sec:Technology}) showed that when using materials with a resistivity of 100-600 \mohmsquare, the discharge energy was quenched while no gain drop was observed at irradiation rates up to approximately $\mathrm{10^3}$ \HZMM{} \cite{249}. At irradiation rates of a few $\mathrm{10^4}$ \HZMM, a 30\% drop was observed. This can be mitigated using electrodes with conducting lines for fast grounding (e.g., G-RETGEM, M-RETGEM, and S-RETGEM), for which no gain drop is observed up to rates of $\sim$$\mathrm{10^4}$ \HZMM{} \cite{157,249}.

\subsection{Thick-Cobra}
\label{sec:Cobra}

\begin{figure}[htbp]
    \centering
    \subfloat[]{
        \includegraphics[width=0.25\textwidth]{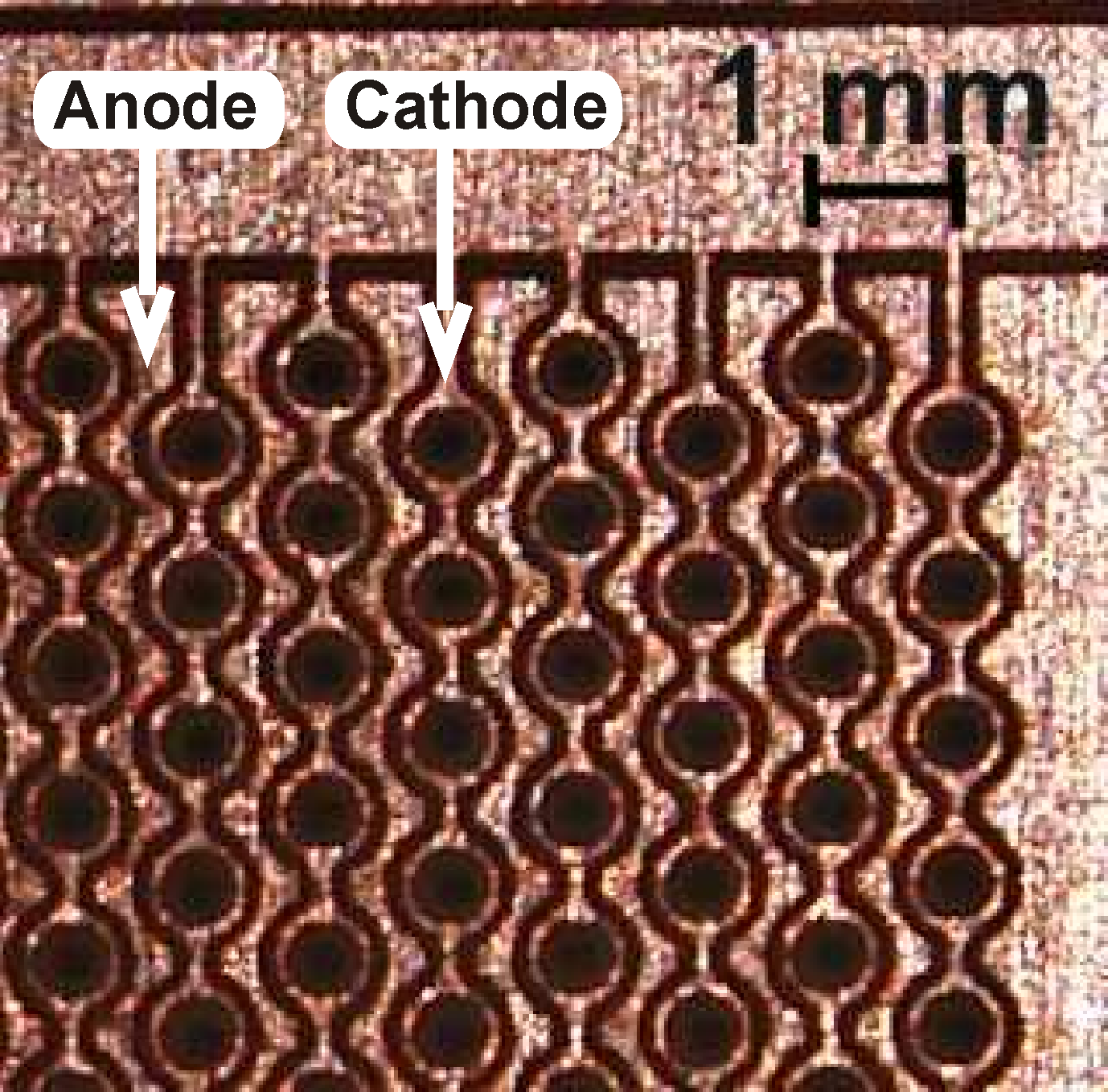}
        \label{fig:COBRA}
    }
    \subfloat[]{
        \includegraphics[scale=0.09]{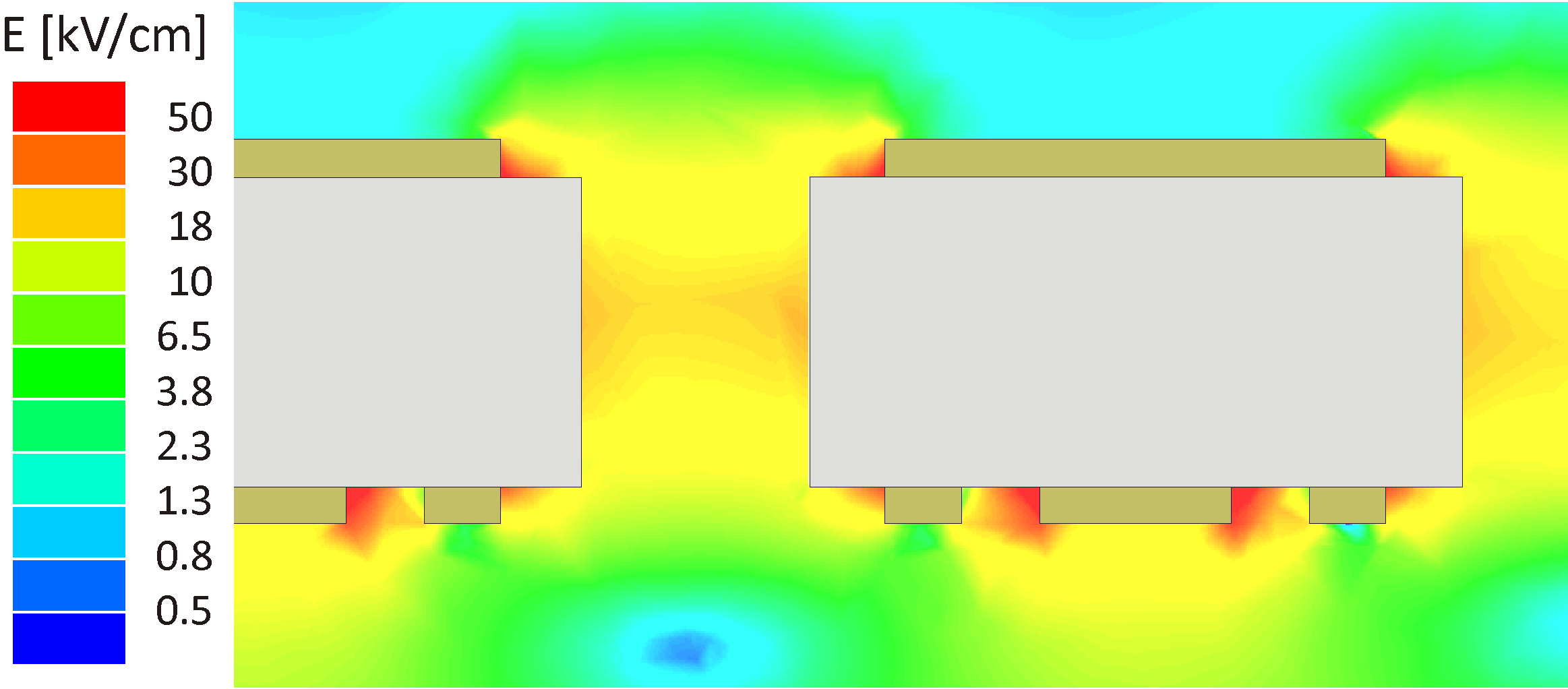}
        
    \label{fig:THCOBRAField}
    }
    \caption{\protect\subref{fig:COBRA} A \acrshort{THCOBRA} electrode. \protect\subref{fig:THCOBRAField} The electric field in a \acrshort{THCOBRA} detector configuration. Figures are taken from \cite{51}.}
\label{fig:THCOBRA}
\end{figure}

The Thick-Cobra (\acrshort{THCOBRA}) electrode \cite{51} depicted in Figure \ref{fig:COBRA} combines the concepts of \acrshort{THGEM} and the micro-hole \& strip plate \cite{veloso2000proposed} (\acrshort{MHSP}). One side of a \acrshort{THGEM} electrode (the cobra side) is patterned with lines of interconnected circles surrounding the holes and wavy conducting strips between them. The former is often biased as a cathode and the latter as an anode.  

A typical electric field in a \acrshort{THCOBRA} detector configuration is presented in Figure \ref{fig:THCOBRAField}. The two regions of high field, at the middle of the hole and close to the strips, cause a charge avalanche formation in two phases. The two amplification phases allow higher gain with a single element while operating the \acrshort{THCOBRA} at relatively low voltages. Furthermore, relative to standard \acrshort{THGEM}, the two-phase amplification provides flexibility in tuning the fields, which has been found useful in mitigating \acrshort{IBF} effects \cite{301,119}. The anode-cathode strip pattern is used to deflect the ions, preventing them from returning to the top electrode through the holes. The main disadvantage of the \acrshort{THCOBRA} is a significant slow ion component when reading the signal from the anode strips. On the other hand, optical readout of photons from the strip-avalanche, could be beneficial.

The basic properties of the \acrshort{THCOBRA} detector configuration were intensively studied in \cite{51}. When measuring 22 \kev{} x-ray photons 
in Ar and $\mathrm{Ar/CH_4}$ (90:10) gas mixtures,
effective gains up to $\sim$10$^5$ were reached with a single electrode. A similar gain was reached with single photoelectrons in 1.7 bar \neon. An energy resolution of the order of 20\% was measured with 8 \kev{} x-rays in  $\mathrm{Ar/CH_4}$ (90:10) at an effective gain of $\mathrm{10^4}$. In the same gas mixture, a Polya-like spectrum (rather than an exponential one) was demonstrated with single photoelectrons at a gain of $\mathrm{2\times 10^5}$, paving the way towards high detection efficiencies \cite{51}.

An \acrshort{IBF} below 1\% was measured operating at a single- and double-stage \acrshort{THCOBRA} (1THCOBRA and 2THCOBRA) configuration \cite{119,17}.
Further reduction to the level of 0.1-0.5\% could be reached using a cascade structure also combining standard \acrshort{GEM} electrodes \cite{119}. Measurements of single \acrshort{UV} photons were conducted with a triple-stage configuration comprising two standard \acrshort{THGEM} electrodes and one \acrshort{THCOBRA} electrode (2THGEM+THCOBRA). Operation in a \nech{} gas mixture using the \acrshort{THCOBRA} as the third amplification stage with an effective gain greater than $\mathrm{10^6}$ reduced the \acrshort{IBF} from 30\% to 20\% \cite{191}. By operating the \acrshort{THCOBRA} in a flipped reverse mode as a second multiplication stage in pure Ne, the \acrshort{IBF} could be lowered to 5\% without losing \acrshort{PDE} \cite{60}. 
In \cite{301}, a single \acrshort{THCOBRA} with CsI coating was coupled to a scintillation region to obtain a zero \acrshort{IBF} in pure \ar.

A 2D-\acrshort{THCOBRA} configuration has one cobra electrode and one strip-patterned electrode rotated at 90 degrees with respect to the cobra lines. It was proposed as a position-sensitive Vacuum \acrshort{UV} (\acrshort{VUV}) gaseous photo multiplier in \cite{191} and studied in \cite{252,136,135}.






\subsection{Thick WELL configurations}
\label{sec:WELL}

\begin{figure}[htbp]
    \centering
    \subfloat[]{
        \includegraphics[width=0.4\linewidth]{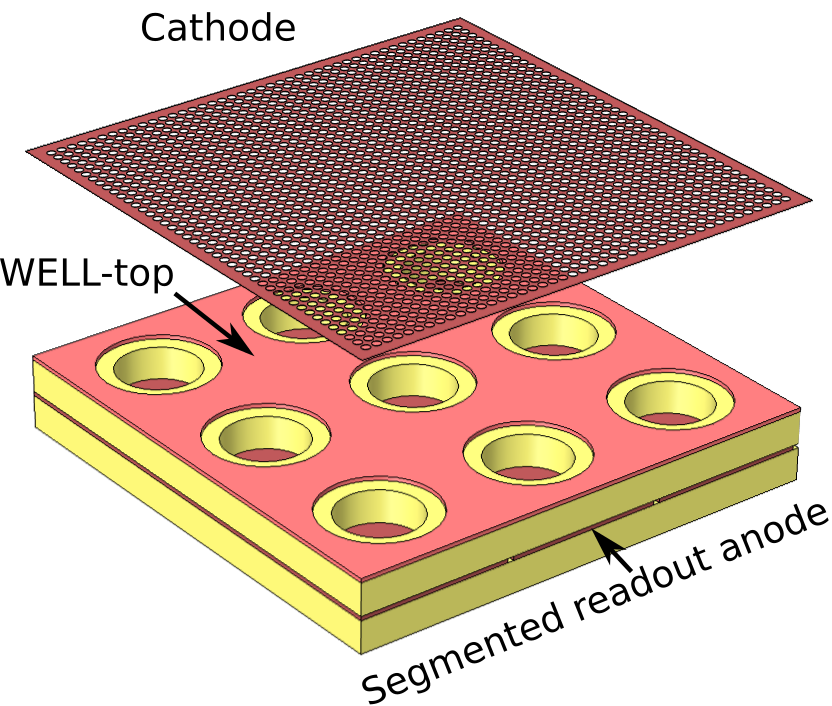}
        \label{fig:WELL_schema_3D}
    }
    \subfloat[]{
        \includegraphics[width=0.55\linewidth]{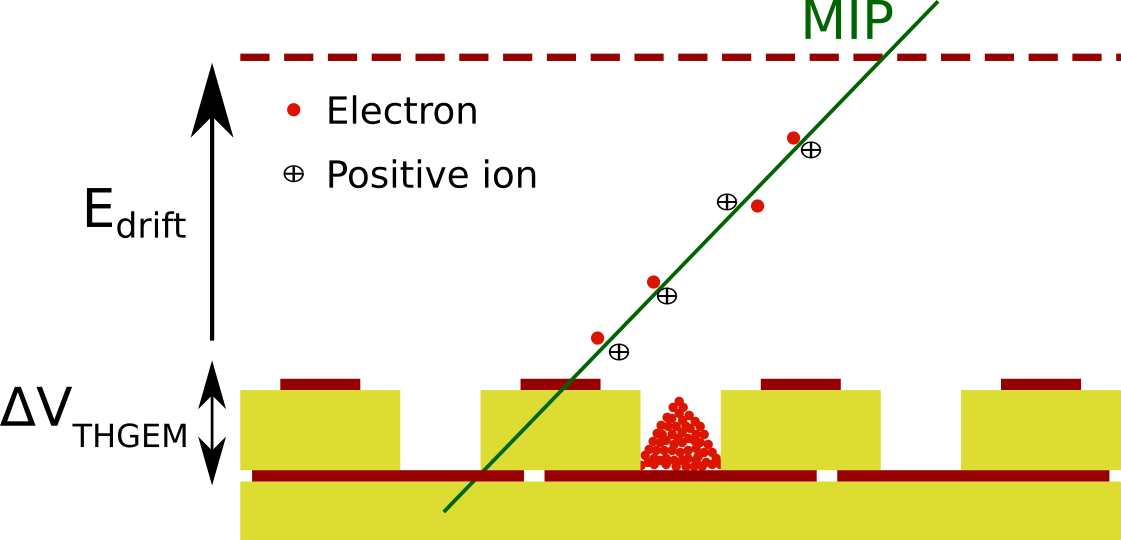}
        \label{fig:WELL_schema_2D}
    }
    \caption{The \acrshort{THWELL} detector configuration in \protect\subref{fig:WELL_schema_3D} 3D and \protect\subref{fig:WELL_schema_2D} 2D, with its principle of operation.}
\label{fig:THWELL_geom_schema}
\end{figure}

The Thick-WELL (\acrshort{THWELL}) detector configuration presented in Figure \ref{fig:THWELL_geom_schema} 
was first suggested in \cite{detlab_26}. It resembles the WELL detector \cite{bellazzini1999well} but with expanded dimensions and is similar to some extent to the CAT presented in \cite{bartol1996cat}. It consists of a single-sided \acrshort{THGEM} electrode (copper-clad on its top side only) coupled directly 
to a readout anode. Thus, the amplification field is defined by the voltage difference between the \acrshort{THGEM}-top and the anode. 

\begin{figure}[htbp]
    \centering
    \includegraphics[width=0.7\textwidth]{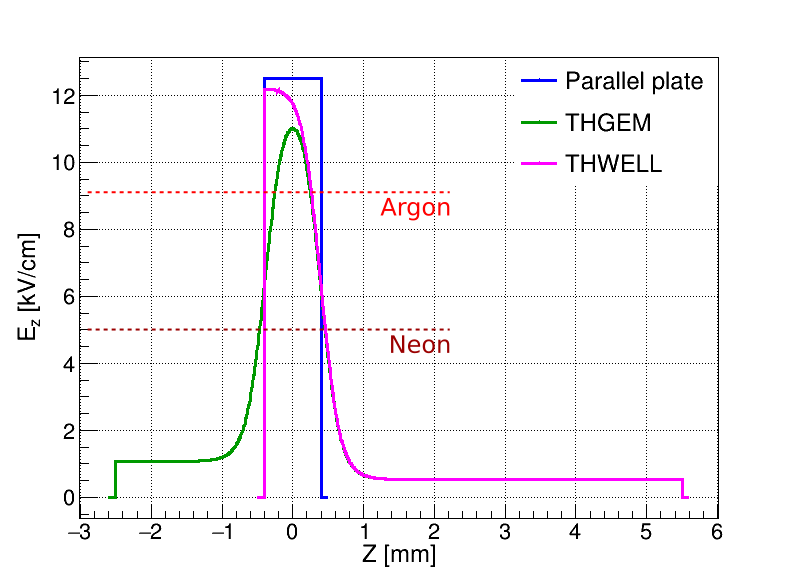}
    \caption{The electric field along a line crossing the center of a hole in a \acrshort{THWELL} and \acrshort{THGEM} configuration with a 0.8 mm thick electrode and 0.5 mm diameter holes (100 \um{} rims) operated at 1 kV. The constant field in a parallel plate configuration at the same voltage is shown for comparison. The horizontal dashed lines represent the threshold for charge multiplication in the illustrated gases.}
\label{fig:WELLField}
\end{figure}

As demonstrated in Figure \ref{fig:WELLField}, for similar geometries (0.8 mm thick, 0.5 mm hole diameter, 1 mm pitch, 0.1 mm rim), the electric field within a \acrshort{THWELL} detector hole is higher than the one in a \acrshort{THGEM} detector operated at the same voltage (here 1 kV), yielding higher gains values \cite{detlab_26}. 

The signal induced on the anode is characterized by a fast-rise avalanche-electron component followed by a slow avalanche-ion one, both of negative polarity. The latter has a typical time consistent with the ion drift along the hole. The signal induced on the top electrode is similar but of positive polarity. 

\begin{figure}[htbp]
    \centering
    \subfloat[]{
        \includegraphics[height=2.3cm]{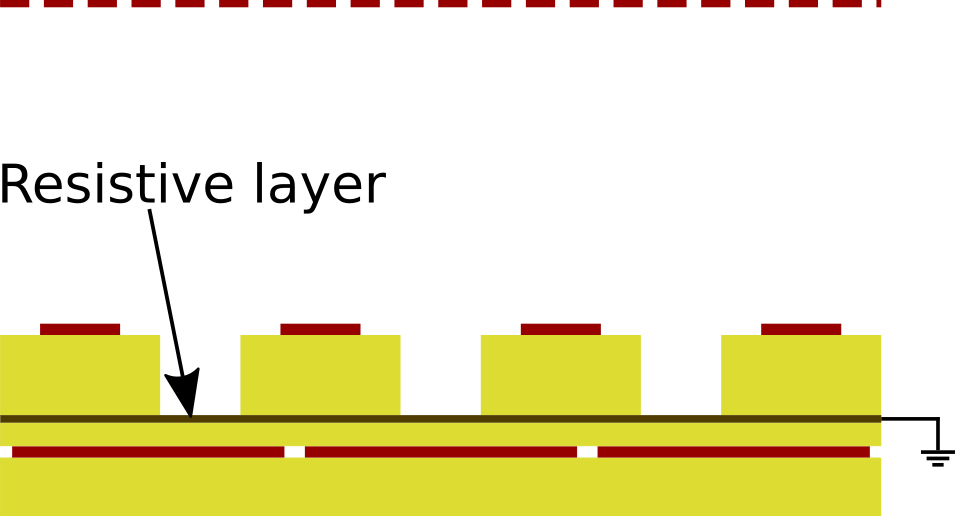}
        \label{fig:RWELL_schema}
    }
    \hfill
    \subfloat[]{
        \includegraphics[height=2.3cm]{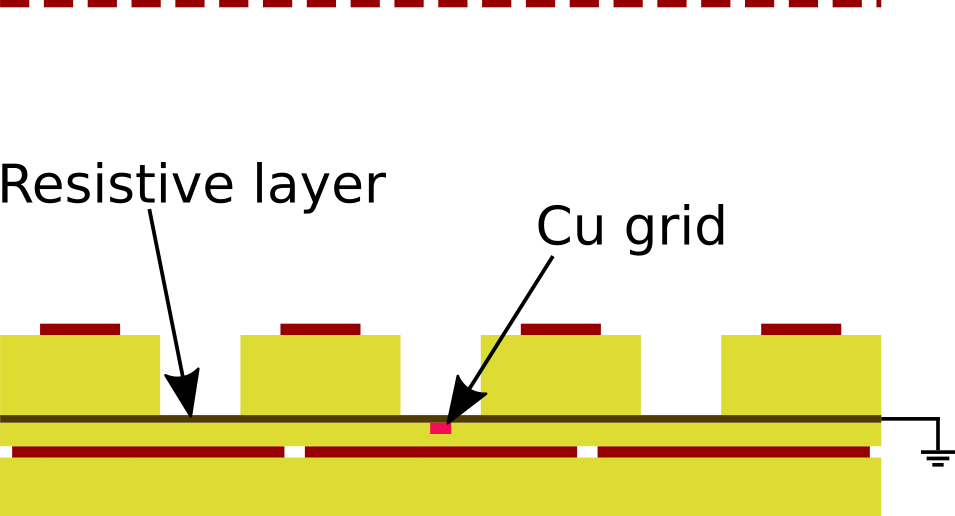}
        \label{fig:SRWELL_schema}
    }
    \hfill
    \subfloat[]{
        \includegraphics[height=2.3cm]{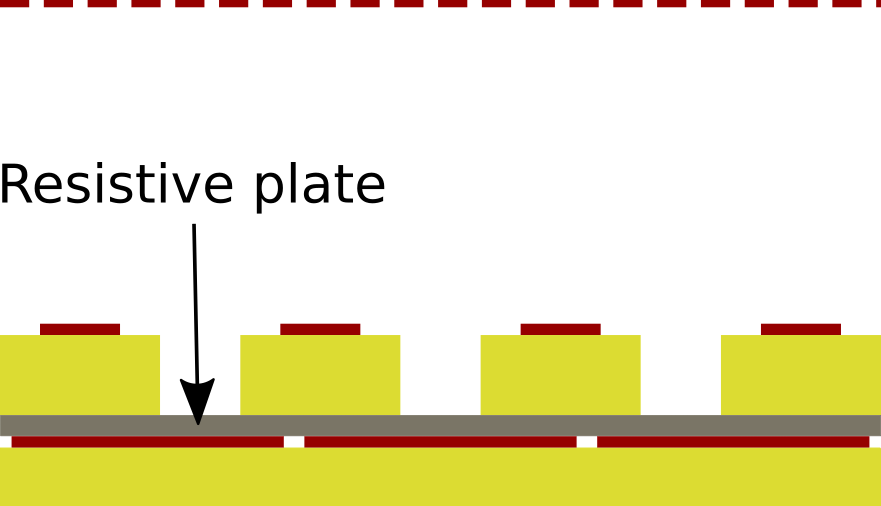}
        \label{fig:RPWELL_schema}
    }
    \caption{Resistive \acrshort{THWELL} configurations: \protect\subref{fig:RWELL_schema} \acrshort{RWELL}: a single-sided \acrshort{THGEM} coupled to the readout anode through a thin resistive layer deposited on an insulator sheet. \protect\subref{fig:SRWELL_schema} \acrshort{SRWELL}: similar to the \acrshort{RWELL} but with the resistive layer segmented by a conductive grid, minimizing the charge sharing between neighboring readout elements. \protect\subref{fig:RPWELL_schema} \acrshort{RPWELL}: a single-sided \acrshort{THGEM} electrode coupled to a readout anode through a plate of high bulk resistivity.}
\label{fig:ResistiveWELLConfigurations}
\end{figure}

All the energy released in a \acrshort{THWELL} discharge reaches the anode. 
This could damage the multiplier, its anode, and the readout electronics and impose significant dead-time effects. Various resistive configurations were proposed in an attempt to mitigate these effects: the resistive-WELL (\acrshort{RWELL}) \cite{detlab_41} depicted in Figure \ref{fig:RWELL_schema}, the segmented resistive-WELL (\acrshort{SRWELL}) \cite{detlab_33,detlab_41} shown in Figure \ref{fig:SRWELL_schema}, 
and the resistive-plate WELL (\acrshort{RPWELL}) \cite{detlab_30} presented in Figure  \ref{fig:RPWELL_schema}.

\begin{figure}[htbp]
    \centering
    \subfloat[]{
        \includegraphics[width=0.5\textwidth]{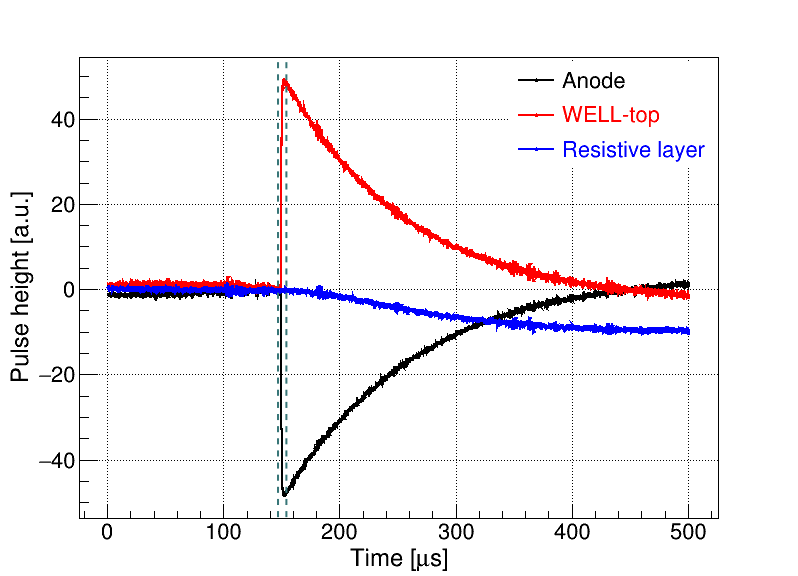}
        \label{pulses_allElectrodes_RWELL}
    }
    \subfloat[]{
        \includegraphics[width=0.5\textwidth]{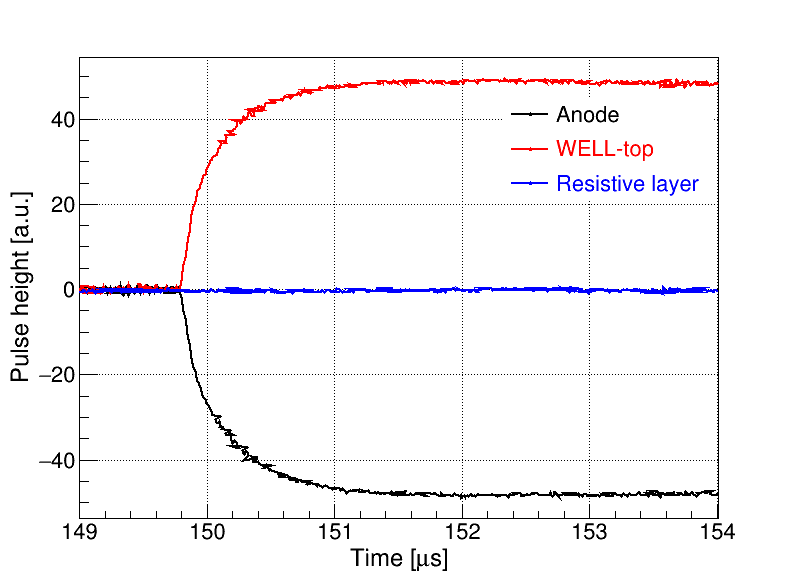}
        \label{pulses_allElectrodes_RWELL_zoom}
    }
    \caption{\protect\subref{pulses_allElectrodes_RWELL} Signal measured on the different \acrshort{RWELL} electrodes (0.8 mm thickness, 0.5 mm hole diameter, 0.1 mm rim and 1 mm pitch) operated in $\mathrm{Ar/CO_2}$ (93:7). \protect\subref{pulses_allElectrodes_RWELL_zoom} an expanded view of the same pulses for the region marked by the dashed lines in \protect\subref{pulses_allElectrodes_RWELL}.}
\label{fig:RWELLSignals}
\end{figure}

In an \acrshort{RWELL} configuration \cite{detlab_41,detlab_29} (Figure \ref{fig:RWELL_schema}), the WELL electrode is coupled to a resistive anode; a resistive layer is deposited on an insulating sheet in contact with a metalic readout electrode underneath. 
Typical \acrshort{RWELL} signals induced on the anode and the WELL-top electrode are similar to those of a \acrshort{THWELL}, as displayed in Figure \ref{fig:RWELLSignals} for a 0.8 mm thick electrode perforated with 0.5 mm diameter holes, with 0.1 mm rim and 1 mm pitch operated in $\mathrm{Ar/CO_2}$ (93:7). 
In an \acrshort{RWELL}, an additional slow signal is recorded on the resistive layer itself. The timescale of this signal is in accordance with the spread of the charge throughout the resistive layer and could cause significant cross-talk between neighboring readout channels \cite{detlab_41}.

In the \acrshort{SRWELL} configuration \cite{detlab_41} (Figure \ref{fig:SRWELL_schema}), the charge spreading across 
the resistive layer stops before inducing a signal on neighboring readout channels. This is achieved by adding a conductive grid below the resistive layer, matching the geometry of the readout channels. The holes of the \acrshort{SRWELL} are drilled with matching geometries and ’blind’ copper strips above the grid lines designed to prevent discharges in holes situated directly above the metal grid lines \cite{detlab_33}. Further, the grid allows for rapid clearance of the electrons diffusing over its surface, improving the rate capability at the cost of less discharge energy quenching \cite{detlab_41}. The grid-to-ground impedance plays a role in determining the transparency of the resistive layer and filtering out the signal's long tail induced by the ion movement in the drift region \cite{206}.\\

In the \acrshort{RPWELL} configuration \cite{detlab_30} (Figure \ref{fig:RPWELL_schema}), the WELL electrode is coupled to the readout anode through a plate of high bulk resistivity ($\mathrm{10^9-10^{12}}$ \ohmcm)\footnote{This idea was briefly mentioned in \cite{285}, and a similar concept with reverse fields was discussed in \cite{298} for an ion detector.}. Compared to an \acrshort{RWELL}, in the this configuration, the charges typically flow through a path of higher resistivity values, resulting in superior discharge quenching. In addition, the lateral charge spread on the readout anode is reduced since the accumulated charges are transported through the layer (as opposed to transversely across its surface in the \acrshort{RWELL}). The pulse shape induced on the anode is similar to that recorded in the other WELL configurations. Gain saturation is observed at high avalanche charge, which could be attributed to the self-avalanche saturation mechanism \cite{detlab_30} or to detector instability \cite{detlab_78}. \\

The induced signals in the various \acrshort{THWELL} configurations were studied experimentally and with MC simulations \cite{detlab_29,detlab_41,detlab_61}. For sufficiently large resistivity values (in \acrshort{RWELL} and \acrshort{RPWELL}), when the charge movement across and within the resistive material is longer than the typical time of the avalanche formation, the signal shape is not affected by the presence of the resistive material \cite{riegler2016electric}.
	
The effective gain of the \acrshort{RWELL} configuration was shown to be affected by the presence of the resistive anode \cite{detlab_78} when using an insulating substrate with a thickness of the same order of magnitude as the WELL electrode. It is attributed to the reduced weighting field \cite{riegler2002induced}, lowering the amplitude of the signal, despite having similar charge amplification.

\begin{figure}[htbp]
	\centering
	\includegraphics[width=0.7\textwidth]{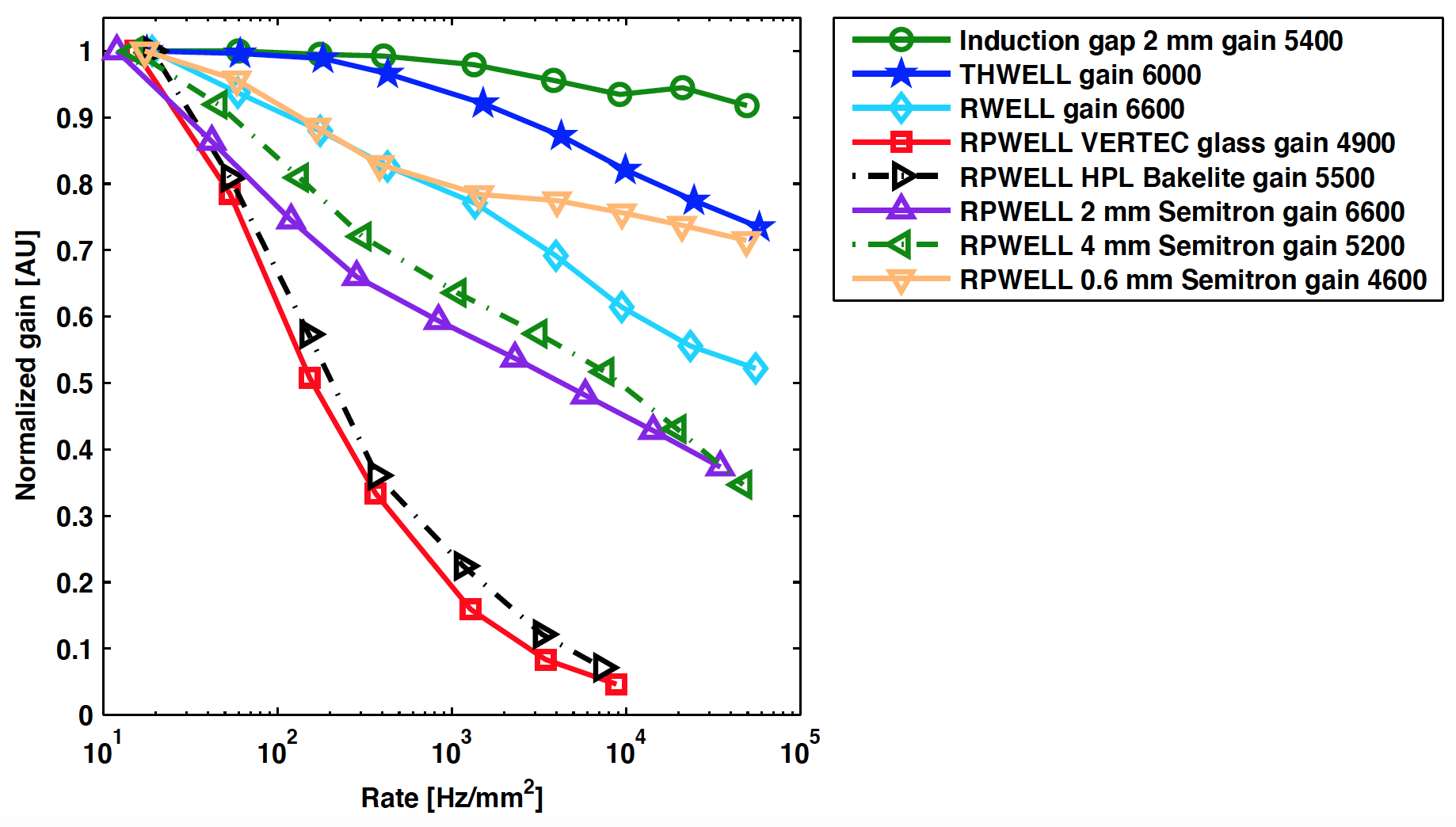}
	\caption{The gain as a function of irradiation rate for different WELL configurations. Figure reproduced from \cite{detlab_30}.}
	\label{fig:WELLGainVsRate}
\end{figure}

A maximal achievable gain of the order of a few $\mathrm{10^4}$ was measured with 
\acrshort{THWELL}, \acrshort{RWELL}, \acrshort{SRWELL}, and \acrshort{RPWELL} detectors (see \cite{detlab_29} and references therein), operated in a $\mathrm{Ne/CH_4}$ (95:5) gas mixture
with soft x-rays. A maximum gain of $\mathrm{8\times 10^3}$ was measured with an \acrshort{RWELL} irradiated with 8 \kev{} x-rays in $\mathrm{Ar/iC_4H_{10}}$ (95:5) \cite{65}.  Like most detectors incorporating resistive materials, current flow across the resistive material results in voltage and, consequently, gain drops. The gain drop is more pronounced for configurations with larger resistivity values (see, for example, \cite{detlab_30}). Figure \ref{fig:WELLGainVsRate} demonstrates the dependency of the gain on the irradiation rate for the various configurations.

An energy resolution of the order of 20\% was measured with all WELL configurations \cite{detlab_29,detlab_30}. The position resolution was measured with an RPWELL detector and modeled using a dedicated MC simulation module to be smaller than the hole pitch \cite{detlab_51}. 

The dynamic range of the WELL-like configurations was studied in \cite{detlab_26, detlab_30, detlab_40, detlab_18, detlab_78}. Two different measurements were conducted; one using a fixed number of \acrshort{PE}s and varying the gain, and the other fixing the gain and using a pre-amplification stage to vary the number of \acrshort{PE}s \cite{detlab_40}. In both cases, spikes in the current supplied to the electrodes appeared in the \acrshort{THWELL}, \acrshort{RWELL}, and \acrshort{SRWELL} configurations, indicating discharges. No current spikes were observed in the \acrshort{RPWELL} configuration. However, discharges were identified by the presence of large-magnitude signals induced on the readout anode \cite{detlab_78}. 

\subsection{M-THGEM} 
\label{sec:mthgem}

The Multi-layer THGEM (\acrshort{M-THGEM}) proposed in \cite{313} and studied in \cite{102} consists of a multi-layer PCB; a stack of copper-clad FR4 layers mechanically drilled in a \acrshort{THGEM}-like geometry. Unlike \acrshort{THGEM}, operated in a cascade structure (with transfer gaps), the closed geometry provides efficient containment of the avalanche-induced \acrshort{EL}  light, thereby significantly reducing photon feedback effects. This is a key factor in obtaining a stable operation with significant charge multiplication in pure gases at low pressures. Similar to a cascaded \acrshort{THGEM} structure, the avalanche forms over several multiplication stages (but without spreading over several holes), potentially allowing reaching higher gains.


A double \acrshort{M-THGEM} (2M-THGEM) configuration was tested in low-pressure He-based gas mixtures \cite{102}. Operated in a symmetric field configuration in $\mathrm{He/CO_2}$ (90:10) and in pure He (contamination <0.1\%), \acrshort{UV} photons could be stably measured with a gain in the order of $\mathrm{10^5}$ at pressures of 150-760 Torr. The small difference in maximum achievable gain between the quenched and unquenched operation gases suggested that the photon feedback was indeed reduced. A gain of $\mathrm{\sim}$$10^6$ could be reached under similar conditions by cascading two 2M-THGEM elements with a transfer gap between them. 

A triple M-THGEM (3M-THGEM) configuration was used as the readout element in an active-target (\acrshort{AT}) time projection chamber (\acrshort{TPC}) and tested with 5.5 MeV alpha particles at different pressures \cite{102}. The lower field on the cathode side allowed reaching higher gain (of the order of $\mathrm{10^6}$). This is attributed to better photon feedback suppression and reduced \acrshort{IBF}. \acrshort{IBF} could be further reduced using an MM-THGEM configuration, namely a 3M-THGEM that includes fine meshes as inner electrode layers, similar to those used in \acrshort{MM} detectors \cite{82}. Maximum achievable gains of $\mathrm{10^4}$ and $\mathrm{10^5}$ could be reached with \acrshort{UV} photons when operated at standard temperature and pressure in $\mathrm{Ar/CH_4}$ (90:10) and $\mathrm{He/CO_2}$ (90:10), respectively. 

\subsection{Exotics}
\label{sec:exotics}

Proposed to enhance the light yield for \acrshort{EL} with an optical readout in noble gases, the Field-Assisted Transparent Gaseous \acrshort{EL} Multiplier (\acrshort{FAT-GEM}) is a \acrshort{THGEM} made of light transparent materials (see Section \ref{sec:MaterialsAndProduction}). A high light yield is obtained with thick ($\sim$5 mm) electrodes with large hole diameters (>2 mm) and operation below the multiplication threshold. 

A \acrshort{FAT-GEM} made of 5 mm PMME with a thin meshed electrode with 2 mm diameter holes and 5 mm pitch was studied with 5.9 \kev{} x-rays in Xe at a 2-10 bar pressure range \cite{248}. Several photons per \acrshort{PE} were measured in all configurations, outperforming mesh-based structures. Furthermore, an energy resolution of 25-30\% was recorded. \acrshort{FAT-GEM} made of a wavelength-shifting material (polyethylene naphthalate) was studied as a solution for reading scintillation light in dual phase \acrshort{TPC}s emitting 128 (\acrshort{LAr}) and 178 (\acrshort{LXe}) nm \acrshort{VUV} light \cite{481}.  \\

The \acrshort{EL} light collection cell (\acrshort{ELCC}) \cite{488} is an alternative approach employing PTFE as a substrate, which is \acrshort{UV} reflective rather than transparent (see Section~\ref{sec:MaterialsAndProduction}) with a mesh as one of the electrodes. An energy resolution of $\mathrm{\sim}$4\% was measured with a 5 mm thick electrode with 4 mm diameter holes at 7.5 mm pitch (compatible with electron diffusion in 1 m Xe) using 122 \kev{} x-ray at 4 bar Xe.\\

Similar to the CAT detector \cite{bartol1996cat}, the wall-less THGEM was proposed in an attempt to suppress charging up and mitigate discharge effects. It is a \acrshort{THGEM} detector in which the perforated top and bottom electrodes are separated by a gas gap. The electrodes can be made of plain thick metal \cite{307,306} or supported by an insulating layer \cite{209,219}. A wall-less \acrshort{THGEM} detector made of 0.2 mm Polyamide film copper clad with 0.8 mm diameter holes with 0.5 mm gas gap was operated in an $\mathrm{Ar/CO_2}$ (80:20) gas mixture. Large current signals were recorded without developing into streamers or discharges \cite{219}. In \cite{209,161}, wall-less \acrshort{THGEM} electrodes were used to image alpha events.\\

The TIP-HOLE detector is an \acrshort{M-THGEM} in a WELL configuration with an additional needle anode at its center \cite{467}. A field-shaping ring can be added around the needle. The field lines concentrated at the needle tip give rise to a high local field, which enables a point-like avalanche. This can allow obtaining Townsend multiplication in configurations in which particularly high field values are required, for example, in heavy hydrocarbon vapors. Having a closed geometry, the \acrshort{M-THGEM} serves mainly to limit photon-feedback effects. Operated in an $\mathrm{Ar/CH_4}$ (90:10) gas mixture at pressure ranges of 130-760 Torr, it can record 5.5 MeV alpha particles at a total charge avalanche of $\mathrm{10^6}$-$\mathrm{10^7}$ electrons.

\subsection{Hybrids}
\label{sec:Hybrids}

Numerous \acrshort{THGEM}-based configurations operated in a cascade structure with other \acrshort{MPGD}s or wires were studied in an attempt to benefit from the advantages of the different concepts. 
A 2THGEM+\acrshort{THWELL} detector was studied in \cite{detlab_13,109,110}, demonstrating a higher maximal achievable gain relative to 1THGEM or \acrshort{THWELL} detectors. A cascade of \acrshort{THGEM}+\acrshort{MWPC} was studied in \cite{120,31,33}, aiming at \acrshort{UV} photon detection with \acrshort{MIP} suppression. Using a 2THGEM+\acrshort{MM} configuration, with a CsI-coated first THGEM electrode, a high \acrshort{PDE} at low \acrshort{IBF} values were reached \cite{105}. \acrshort{MM} detectors were also combined with RETGEMs \cite{163}. A constant gain of $\mathrm{\sim}$$10^4$ up to an irradiation rate of several $\mathrm{10^4 ~Hz/mm^2}$ could be reached using a \acrshort{DLC}-based \acrshort{RETGEM}+\acrshort{RWELL} with surface resistivity in the order of 100 $\mathrm{M\Omega/\square}$, with 8 keV x-rays, in an $\mathrm{Ar/iC_4H_{10}}$ (95:5) gas mixture \cite{249}. A cascade structure of \acrshort{RETGEM}+microdot-microstrip was studied in \cite{220,207}.

\section{Technology}
\label{sec:Technology}

Governed by advances in industrial technologies and developments in the field of material science, progress has also been made in the field of \acrshort{MPGD}s. In this section, we discuss the main technological aspects relating to the \acrshort{THGEM} detector and its derivatives. After describing in detail the production technique of a standard FR4-based \acrshort{THGEM} electrode, we discuss production using other insulating substrates. Techniques used to coat the electrode with \acrshort{UV}-sensitive and resistive materials are also presented, followed by a discussion of resistive plates. Finally, we discuss the challenges and progress made in an attempt to scale up the detectors' sizes.

\subsection{Standard THGEM electrode production}
\label{sec:stdTHGEMProduction}

A standard \acrshort{THGEM} electrode consists of a perforated 2-layer (copper) FR4-\acrshort{PCB} board in which the holes are drilled by a \acrshort{CNC} machine \cite{detlab_4,49,Rui_THGEM_cleaning}. The detector quality and, in particular, its ability to sustain high voltages depends on the quality of the holes; the presence of glass fibers sticking out from their walls or sharp copper edges around the holes are known instability sources. A single defected hole is sufficient to jeopardize the performance of the entire detector. To avoid sharp copper edges, rims are often chemically etched around the holes. These should be concentric to ensure a uniform field across the electrode. Frequent replacement of the \acrshort{CNC} drills is essential to obtain smooth holes.

\begin{figure}[htbp]
    \centering
    \includegraphics[width=0.8\textwidth]{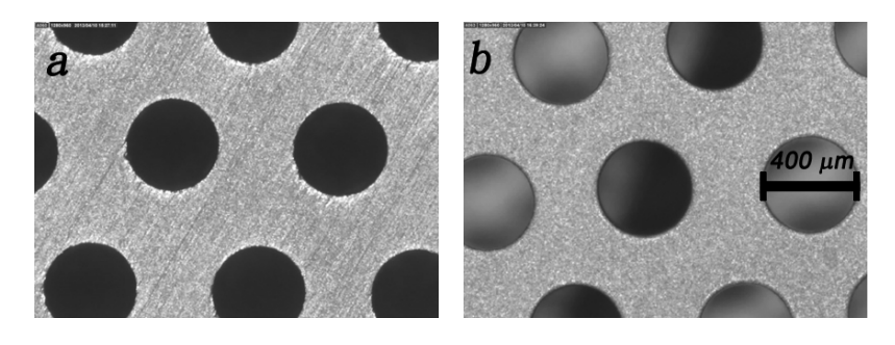}
    \caption{Left: a \acrshort{THGEM} surface acquired via a digital microscope just after the production. Hole-edge defects and irregularities can be clearly observed. Right: the same \acrshort{THGEM} layer after applying the surface polishing treatment procedure. Figure reporoduced from \cite{118}.}
    \label{fig:THGEMpolish}
\end{figure}

Two post-treatment techniques that greatly improve the electrode quality have been developed. The most common one encompasses multiple “washing” cycles of the drilled electrode in an ultrasonic bath with different chemicals, followed by oven drying \cite{Rui_THGEM_cleaning}. The other method is based on long-term polishing of the electrode with pumice powder and cleaning \cite{118,99}. An example of a \acrshort{THGEM} electrode surface before and after the polishing process is provided in Figure \ref{fig:THGEMpolish}. This technique was used to manufacture good-quality electrodes with rim-less holes (beneficial in terms of charging up, as discussed above).

\subsection{Materials and production techniques}
\label{sec:MaterialsAndProduction}

FR4 with copper clad is the most common raw material used in the \acrshort{PCB} industry. However, other materials may be utilized for various applications requiring particular properties  such as mechanical strength, chemical stability, radio-purity (in rare-event searches), compatibility with high-purity gases (outgasing), large-area scalability and more. While mechanical drilling is typically deployed, holes may also be formed using other techniques.

\subsubsection{Insulating substrates} 
Various substrate materials were tested.  One example is Kevlar, with good mechanical and chemical-resistivity properties \cite{detlab_4}; e.g. it may be important when using gases containing aggressive elements like Fluorine. PVC substrate simplifies the post-treatment procedure after drilling \cite{335}. Ceramic substrates made of pure elements should lead to low intrinsic radioactivity; however, the addition of some glass fibers to improve mechanical strength affects the radio-purity  \cite{116}. Ceramics typically have low neutron scattering due to their low hydrogen content and outgassing rate. These are important features for some applications \cite{116,124,29,270,293,296,499,506,518,517}. Compared to FR4, ceramics are more expensive, brittle, and difficult to produce in large sizes. Another radio-pure THGEM substrate is Polyimide (Cirlex), investigated in \cite{Gai:2007dj}.

Glass made \acrshort{THGEM} (glass-\acrshort{THGEM}) \cite{476} might be useful in applications requiring low outgassing, such as in sealed \acrshort{GPM} detectors \cite{235}. Large-area electrodes can also be made of fused silica, allowing for radio-pure detectors conception \cite{476}. As an alternative, Kapton \cite{112}, PTFE \cite{112,312,458,166} and PMMA \cite{196,248} were also tested. 

Being a suitable \acrshort{UV} light reflector, PTFE can be used to achieve better \acrshort{EL}-light collection from the holes \cite{488}. In addition to being radiopure, PMMA is transparent to visible light \cite{248}. A more sophisticated solution for light collection utilizes PEN as a transparent, wavelength-shifting material coated with PEDOT:PSS transparent electrodes \cite{481}. It may become useful in noble-gas detectors, shifting UV-photos to visible ones.

\subsubsection{Conductors} 
Copper is the most common conductor used as an electrode in the \acrshort{PCB} industry. For coating with CsI photocathode, the copper is usually Au-coated to avoid chemical reaction. Various other conducting or resistive materials can also be used \cite{469}. In some cases, the electrodes are applied to the substrate by surface coatings \cite{481,476}, which can be feeble against discharges. The destructive effect of discharges can be mitigated by adding guard rings around each hole \cite{201}. Thin metal films or grids (e.g. Ni, Cr) transparent to light have also been considered \cite{476}. 

\subsubsection{Hole drilling} 
In addition to standard mechanical drilling, holes can be produced with different techniques, depending on the substrate and conductive materials. Sandblasting on glass electrodes results in conical or double conical holes \cite{235,476}, whose shapes can also be polygonal \cite{476}. Glass-\acrshort{THGEM} can be also produced by laser drilling \cite{204}; holes can be chemically etched in photosensitive etchable glasses \cite{158,230,244,245,283,251}. In this case, the shape of the holes could be non-cylindrical. Similar results can be obtained by etching PMMA with deep x-ray lithography \cite{196,197,158,309,223}. 
In low-temperature co-fired ceramic substrates without glass fibers, holes can be produced by punching \cite{293}.

\subsubsection{3D printing and similar techniques} 
A promising technology is the production of \acrshort{THGEM} electrodes in 3D printing. Recent attempts \cite{73} have shown that this front requires further improvement of both the materials and processes used; the quality of the \acrshort{THGEM} holes and the concentricity of the hole' rims were far from optimal. Production of \acrshort{THGEM} electrodes from pure epoxy with a silver conductor showed excellent precision. However, the technique is expensive, and the quality of the substrate degrades with time \cite{renous2017thesis}.

\subsection{Surface coatings and resistive materials}
\label{Sec:SurfaceCoatings}

Coating either the substrate or the conductor surface of the \acrshort{THGEM} electrode allows tuning their properties; coating with photosensitive materials enables the efficient detection of \acrshort{UV} photons. Charging-up effects can be mitigated by coating the insulator with materials of high resistivity \cite{261,18,249}. The replacement of conductors with resistive materials allows for quenching of the energy released in discharges. Examples are the \acrshort{RETGEM}, as discussed in Section \ref{sec:RETGEM}, and \acrshort{RWELL} and \acrshort{RPWELL} discussed in Section \ref{sec:WELL}. 

\subsubsection{UV-sensitive coatings}

CsI is the standard \acrshort{UV}-photocathode material used for coating electrodes of \acrshort{GPM}s. It is known to be stable in pure gases and has rather high \acrshort{QE} values in the \acrshort{VUV} spectral range  \cite{breskin1996csi,288,detlab_24}. Gold-plated electrodes are usually used as the basis for CsI coating. They are free from oxidation effects or reactions with CsI that can compromise the quality of the photocathode \cite{breskin1996csi,93}. 
Diamond Photocathodes made by various procedures, e.g. CVD \cite{Amos_CVD}
or nano-diamond coatings \cite{79,88,528} have lower \acrshort{QE} but higher stability and resistant to humidity compared to CsI. The latter technique was employed and tested with \acrshort{THGEM} detectors.

\subsubsection{Resistive coatings}
\label{sec:ResCoating}

Different resistive materials, e.g., resistive Kapton \cite{162}, and resistive coating techniques, such as screen printing of resistive paste \cite{2}, graphite paint \cite{detlab_26}, and \acrshort{DLC} evaporation \cite{65,249}, were used in various \acrshort{THGEM} derivatives. In recent years, \acrshort{DLC} coating has attracted increasing attention; it is relatively easy to deploy, its resistivity can be tuned, and it can easily cover large areas. Unlike resistive Kapton, studies have demonstrated its robustness, as it can sustain $\mathrm{\sim}$$10^6$ discharges without being damaged \cite{314}.
The resistive coating can be deposited on the metal surface \cite{285,45,4} or can substitute the metal \cite{1,2,249}, in which case different metallic structures can be implemented for the charges' evacuation \cite{162,157}. 

\subsubsection{Resistive plates}
\label{sec:ResPlates}

The use of materials of high bulk resistivity, in the range of $\mathrm{10^8}$-$10^{12}$ $\mathrm{\Omega \cdot cm}$, is unique to the \acrshort{RPWELL} structure (see Section \ref{sec:WELL}). 
In the industry, such materials are usually defined as bad insulators or electrostatic dissipative (\acrshort{ESD}) ones. When employed in an \acrshort{RPWELL} detector, sub mm-thick plates with beneficial mechanical properties and resilience to aging and radiation are typically required. This narrows the list of possibilities to a handful of materials. The materials tested are VERTEC 400 glass and Bakelite \cite{detlab_30}, Semitron \acrshort{ESD} 225 \cite{detlab_30, detlab_49,detlab_48,detlab_53,detlab_62,detlab_63, detlab_68}, Fe-doped glass \cite{detlab_51,detlab_54,detlab_78}, and ceramics \cite{detlab_59}.
Among those, \acrshort{ESD} plastic and Fe-doped glass have been shown to perform well also under relatively high particle fluxes. Semitron \acrshort{ESD} 225 is commercially available only in rather thick plates, which necessitates massive machining for deployment in standard \acrshort{RPWELL} assemblies. Like most plastics, it is soft and degrades with time. The Fe-doped glass is stiff and easily produced in thin sheets with good precision. On the other hand, it is seldom found in the industry. 

Employing resistive plates at low-temperatures, e.g. in noble liquid detectors,  poses an additional challenge. Most materials of high bulk resistivity exhibit an exponential increase in resistivity with decreasing temperatures. Thus, materials of appropriate bulk resistivity at room temperature are likely to reach extremely high resistivity values at low temperatures, becoming effectively insulators. A possible solution is to tune the resistivity of the material at room temperature so that it develops the required resistivity at low temperatures. This concept was demonstrated with a ceramic resistive plate at a \acrshort{LXe} temperature \cite{detlab_59}.

\subsection{Upscaling}
\label{sec:LargeArea}

The size of the largest THGEM electrodes currently produced and operated is $\mathrm{\sim}$0.2-0.25 $\mathrm{m^2}$ \cite{87,detlab_63,249}. The main challenge in producing high-quality large-area \acrshort{THGEM} electrodes is to achieve adequate thickness uniformity. Non-uniform thicknesses affect the local detector gain and energy resolution, as well as the overall discharge probability and maximum achievable gain \cite{detlab_63}. Due to their production technique, the thickness tolerance of standard FR4 plates is in the order of a few tens of \%. Depending on the production technique and desired plate thickness, better precision, in the order of a few \%, can be reached \cite{118}. Strict preselection of the raw material is often required \cite{105}.

Better precision can be achieved more easily using a glass substrate \cite{476}. Fairly large areas of $\mathrm{30\times30 ~cm^2}$ and 20 cm diameters have been achieved with photosensitive etchable glass \cite{242,387,251} and sand-blasted borosilicate glass \cite{476}, respectively. 

Guaranteeing the quality of millions of holes, where an imperfection in any of them may cause electrical instabilities in the entire detector, is another challenge. Different quality inspection methods can be used. In \cite{83}, an optical system is used to measure the pitch and eccentricity of the holes and rims. 

For \acrshort{THGEM} derivatives incorporating resistive electrodes, more challenges could arise. For example, in \acrshort{RWELL} and \acrshort{RPWELL} detectors, the \acrshort{THGEM} electrode has to be pressed against the resistive plate across its entire surface. Different solutions were employed in an attempt to minimize additional dead areas \cite{detlab_48,detlab_63}.

\section{Performance under different operation conditions}
\label{sec:PhysicsPerformance}

The underlying physical processes governing the operation of \acrshort{THGEM}-based detectors (see Section \ref{sec:THGEM}) and their performances were studied in dedicated R\&D projects and in the context of specific applications (see Section \ref{sec:Applications}). In this section, we provide a detailed summary of the performance of \acrshort{THGEM} and \acrshort{THGEM}-based detector configurations operated with various radiation sources under different conditions.

\subsection{Standard temperature and pressure}
\label{sec:perfSTP}

Significant effort has been made to characterize the performance of various \acrshort{THGEM}-based detector configurations, primarily at standard temperature and pressure. A summary of studies performed under these conditions is given in Table \ref{tab:STTP} in Appendix \ref{sec: summary tables}. A systematic study of the effect of the \acrshort{THGEM} electrode geometry in terms of thickness, hole diameter, pitch and rim sizes on its operation in various gas mixtures and in pure Ne, Ar, Xe can be found in \cite{detlab_12, detlab_14, detlab_4,27,58}. 
Drift field optimization with respect to the gain and electron collection efficiency can be found in \cite{215,205,20}, while detailed experimental and simulation studies of the electron collection and transfer efficiency in different gases are available in \cite{detlab_4,detlab_14,26,215,125,205,221}.

The effect of adding impurities or quencher gases to the noble gases on the performance of the \acrshort{THGEM} detector was studied in \cite{detlab_14}. Throughout the avalanche multiplication, impurities and quenchers limit both photon and ion feedback effects. The former is achieved by reducing the probability of photon emission, absorbing \acrshort{UV} photons emitted in the avalanche \cite{review_2}, or wavelength shifting them \cite{detlab_14}. The latter is by reducing the extraction of secondary electrons from ion feedback via a charge transfer process between the gas species \cite{kalkan2015cluster,detlab_20}.  An additional significant effect on the gain in certain gas mixtures is the Penning transfer. It is demonstrated in \cite{103} by reproducing THGEM gain curves for three different gas mixtures using a Monte Carlo simulation.

By far, the majority of the literature focuses on gain measurements with different detector configurations.  Information is also provided on energy resolution, spatial and time resolution, detection efficiency, \acrshort{IBF}, rate capabilities, discharge probabilities, and light yields. These are summarized below for different radiation sources. 


\subsubsection{Maximum achievable gain}
\label{sec:maxGainRTSP}

The Maximum achievable gain is determined by the onset of discharges, mostly attributed to crossing a critical charge known as the Raether limit \cite{von1965electron}. Thus, it is strongly dependent on the radiation source and in particular, on the number of \acrshort{PE}s undergoing avalanche multiplication in a single hole within a short time period. 

There is a difference between single electrons, point-ionization events (e.g. x-ray induced photoelectrons or neutron-induced low-range charged particles) and extended-ionization events, e.g. \acrshort{MIP}S. Photocathodes are used to convert \acrshort{UV} photons into single photoelectrons. These are extracted and collected into a single hole. X-ray photons convert in the drift gap, leaving a small cloud of primary ionization; for soft x-rays, the cloud size is typically a few hundred microns. In such cases, the majority of \acrshort{PE}s are shared between a few holes; this depends on the lateral and transverse diffusion of the electrons along their drift in the gas, on the hole diameter and the pitch \cite{detlab_34}. At most, a few hundred of \acrshort{PE}s are multiplied within a single hole. 

\acrshort{MIP}s leave ionization clusters along their trajectory in the gas. The first \acrshort{PE}s to reach the holes are those that were deposited in gas in their proximity, while the rest drift towards the holes for a duration depending on the drift length and drift velocity: usually tens of nanoseconds (tracking detector) to tens of microseconds (\acrshort{TPC}). Their time of arrival to a hole depends on the track inclination. Dense ionization clusters are created along the trajectory of highly ionizing particles (alpha particles, heavy ions), in particular at the end of the track (Bragg peak). Thus, an order of a few thousands of \acrshort{PE}s are multiplied within a hole over very short time periods (ns-to-tens of ns).

Any comparison of the maximum achievable gain between different configurations should be taken with a grain of salt since the result could be strongly dependent on the detector geometry. Here, the gain values are presented for the different irradiation sources and separately for the different gas mixtures.

\subsubsection*{Soft x-ray photons}

In Ne (grade not reported), maximal gains at the order of several $\mathrm{10^4}$ were measured with a 1THGEM configuration \cite{detlab_14,detlab_55}; lower values (a few $\mathrm{10^3}$) were reached without rims \cite{detlab_55}). The maximal achievable gain increased to $\sim$$\mathrm{10^5}$-$\mathrm{10^6}$ using 2THGEM configurations \cite{detlab_16,detlab_14}. Similar performances were recorded with resistive single-element multipliers: a maximal gain of several $\mathrm{10^4}$ to $\mathrm{10^5}$ was reached with \acrshort{RETGEM}, M-RETGEM, and G-RETGEM \cite{4,2,45,487,157}. Gains of the order of $\mathrm{10^5}$ were also reached with thick (2.4 mm) \acrshort{RETGEM} and Kapton-\acrshort{RETGEM} \cite{156}.  The gain measured with 2RETGEM configurations ranged between $\mathrm{10^5}$-${10^6}$ \cite{156,285,2,45}. A hybrid detector consisting of \acrshort{RETGEM} proceeded by resistive-\acrshort{MSGC}-like electrode reached a gain of $\mathrm{10^4}$ \cite{220,207}. 

Operating in $\mathrm{Ne/CH_4}$ (95:5), gains in the order of several $\mathrm{10^3}$-${10^4}$ were measured with 1THGEMs \cite{detlab_14,detlab_18,132,detlab_41,124,316,112}. An exception is found in \cite{116}, reaching a gain of $\mathrm{10^5}$. Higher gains of the orders of $\mathrm{10^4}$-${10^5}$ and several $\mathrm{10^5}$ were measured with 2THGEM  \cite{detlab_14,detlab_18,173,124,199,112} and 3THGEM \cite{detlab_18} configurations, respectively. A gain of several $\mathrm{10^5}$ was reached using 1THGEM with additional multiplication in the induction gap \cite{detlab_31}. \acrshort{RETGEM} configurations yielded a gain slightly below $\mathrm{10^4}$ \cite{13}. Gains in the range of $\mathrm{10^4}$-${10^5}$ were measured with \acrshort{THWELL}, \acrshort{RWELL}, and \acrshort{RPWELL} configurations, with the highest gains measured for the \acrshort{RPWELL} \cite{detlab_30,detlab_41,detlab_78,detlab_26}. A similar gain, slightly lower than $\mathrm{10^5}$, was measured with a \acrshort{THCOBRA} detector \cite{194,252}. A hybrid \acrshort{THGEM}+\acrshort{MM} configuration yielded a gain of $\mathrm{10^6}$ \cite{199}. More details on the effect of different $\mathrm{CH_4}$ concentrations in Ne are available in \cite{detlab_14,detlab_18,87}.

Measurements in $\mathrm{Ne/CO_2}$ (90:10) were carried out with hybrid \acrshort{RETGEM}+Resistive \acrshort{MSGC} \cite{220} and \acrshort{RETGEM}+Resistive \acrshort{THCOBRA} \cite{207}. Gains of the order of $\mathrm{10^4}$ were reached with both detector configurations. 

When operated in $\mathrm{Ne/iC_4H_{10}}$ (90:10), gains of $\mathrm{5 \times 10^3}$ and $\mathrm{10^4}$ were reached with 1THGEM and 2THGEM configurations, respectively \cite{230}. 

The use of $\mathrm{Ne/CF_4}$ (95:5) and (90:10) was studied in the context of \acrshort{RICH} detectors (see Section \ref{sec:RICH}) for achieving high electron extraction efficiency.  A gain of $\sim$$\mathrm{10^4}$ was reached with a 1THGEM configuration \cite{detlab_18,detlab_73}. A higher gain of $\sim$$\mathrm{10^5}$ was reported for glass-\acrshort{THGEM} \cite{283} and FR4-\acrshort{THGEM}, operated with additional multiplication in the induction gap \cite{detlab_31}. The gain measured with a 2THGEM configuration was in the order of $\mathrm{10^5}$ \cite{detlab_18,detlab_73,230,265}. The 3THGEM configuration yielded a gain of several $\mathrm{10^5}$ up to $\mathrm{10^6}$ \cite{470,detlab_18}.

In Ar (grade not reported), a gain of the order of $\mathrm{10^3}$ was reached with a 1THGEM \cite{132,316,112,166,57}. A higher gain, in the order of several $\mathrm{10^4}$ was measured with a 2THGEM configuration. This gain was reduced to a few $\mathrm{10^3}$ when the Ar was purified (though the purity level was not reported). The gain measured in Ar with a single (1, 2, and 4 mm thick) \acrshort{RETGEM} reached $\mathrm{10^4}$ and $\mathrm{10^5}$, while the gain measured with a 2RETGEM configuration it exceeded $\mathrm{10^6}$ \cite{1,156,45,4,287,285,162,2}. Gains in the range of $\mathrm{10^3}$-${10^5}$ were reported with \acrshort{THCOBRA} configurations \cite{51,301}, where the lower value was measured in a configuration optimized for reduced \acrshort{IBF}. 

Many measurements were conducted in $\mathrm{Ar/CH_4}$ mixtures. Gains in the range of $\mathrm{10^3}$-${10^4}$ were measured with a 1THGEM in a (90:10) mixture \cite{196,278,293}, while in a (95:5) mixture, the gain was slightly higher than $\mathrm{10^4}$ \cite{124,detlab_14}. Glass-\acrshort{THGEM} operated in a (90:10) mixture reached a slightly higher gain of the order of $\mathrm{5 \times 10^4}$ \cite{254,292,242,201,387,254,283}. Operated in a (90:10) mixture, 2THGEM and 3THGEM configurations yielded similar gains, larger than $\mathrm{10^4}$ \cite{317}, also with different hole-rim sizes \cite{16}. A gain of $\sim$$\mathrm{7 \times 10^5}$ was obtained with a 2THGEM configuration operated in a (95:5) gas mixture \cite{124}; a somewhat lower gain was reported in \cite{detlab_6}. A gain of $\mathrm{10^5}$ was reached with a \acrshort{THCOBRA} detector operated in a (90:10) mixture \cite{51}, while a gain of several $\mathrm{10^4}$ was reached with \acrshort{RETGEM} configurations \cite{278,16}. A slightly higher gain, of the order of $\mathrm{10^5}$, was reported with a 2RETGEM configuration \cite{156,285}.

A gain of several $\mathrm{10^5}$ was measured in an $\mathrm{Ar/CH_4}$ (80:20) with a 1THGEM operated with additional multiplication in the induction gap \cite{detlab_66} and with 2RETGEM configurations \cite{156,285}. Lower gains were measured in an $\mathrm{Ar/CH_4}$ (70:30) mixture with configurations of 1THGEM \cite{293,112}, single glass-\acrshort{THGEM} \cite{245}, and 2THGEM \cite{112,245,detlab_6}. Measurements' results with higher concentrations of $\mathrm{CH_4}$ can be found in \cite{131,27}. 

Detector configurations were characterized in $\mathrm{Ar/CO_2}$ mixtures as well. In a (90:10) mixture, a gain of approximately 700 was measured with a 1THGEM \cite{123}. Higher gains of several $\mathrm{10^4}$ and $\mathrm{10^3}$-${10^4}$ were measured with single glass- \cite{241} and ceramic- \cite{266,296,112,499} \acrshort{THGEM}, respectively. The gain reached by a 1THGEM did not change much when adding more $\mathrm{CO_2}$ (up to a (70:30) mixture) \cite{117,123,124,499}. PTFE- and ceramic-\acrshort{THGEM} operated in an (80:20) mixture yielded a gain of several $\mathrm{10^3}$ \cite{112,116,499}. An FR4-\acrshort{THGEM} could reach a gain of $\mathrm{10^4}$ \cite{116}. A 2THGEM configuration operated in a (70:30) mixture yielded gains of $\sim$$5 \times 10^3$-${10^4}$ \cite{detlab_6,112,196}. In measurements of 2THGEM made of PTFE and Kapton in (80:20) mixtures, gains of the order of several $\mathrm{10^4}$ were reached \cite{112}. 2THGEM operated in an (80:20) mixture yielded a gain of $\sim$$\mathrm{5 \times 10^4}$ \cite{112,124}. \acrshort{RETGEM} configurations operated in (80:20) and (90:10) mixtures yielded gains of  $\mathrm{10^4}$ \cite{156} and $\mathrm{10^4}$-${10^5}$ \cite{1,285}, respectively. A somewhat lower gain, $\mathrm{10^3}$, was recorded in a (95:5) mixture \cite{45,157}. Operated at a (70:30) mixture, a gain of several $\mathrm{10^3}$ was measured with a \acrshort{THCOBRA} configuration \cite{119}, while a gain of the order of several $\mathrm{10^4}$ was measured with an \acrshort{RWELL} detector at a (90:10) mixture \cite{206}. 

A mixture of $\mathrm{Ar/CO_2/CH_4}$ (89:10:1) was tested with \acrshort{THGEM} electrodes of 0.5 mm and 1 mm thickness with hole diameters in the 0.3-0.6 mm range. The highest gain of $\mathrm{10^4}$ was achieved with the smallest holes for both thicknesses \cite{58}. 

Measurements with $\mathrm{Ar/iC_{4}H_{10}}$ were carried out with different ratios between (98:2) and (90:10). 1THGEM detectors yielded gains in the range of several $\mathrm{10^3}$ up to $\mathrm{10^4}$ in all mixtures, where slightly higher gains could be reached with lower $\mathrm{iC_4H_{10}}$ concentrations \cite{132,173,166,124,181,112,316,261,116}. A similar gain was measured with \acrshort{M-THGEM} \cite{258} and \acrshort{RETGEM} \cite{249} configurations, while a higher gain, in the order of $\mathrm{10^5}$, was reported with Glass-\acrshort{THGEM} \cite{204}. A gain of several $\mathrm{10^4}$ was measured with 2THGEM detectors for the various mixtures \cite{173,124,262,267,112}. Operated at a (95:5) mixture, a gain in the order of $\mathrm{10^4}$ was obtained with \acrshort{DLC}-\acrshort{RETGEM} and \acrshort{RWELL} detectors \cite{65,249}.

\subsubsection*{UV photons}

Though investigated in many gases, for UV detectors, those containing large concentrations of $\mathrm{CH_4}$ or $\mathrm{CF_4}$ are mostly adequate for photoelectron extraction \cite{breskin1996csi}. Further, not all the studies were carried out with, e.g., CsI photocathode deposited on the multiplier or on the first multiplier in a cascade, essential for reaching high \acrshort{PDE}.

Operated in Ne (grade not reported), depending on the rim size, gain values from several $\mathrm{10^3}$ (40 \um{} rim) to $\mathrm{10^5}$ (120 \um{} rim) were reached, while the gain measured with a 2THGEM configuration reached $\mathrm{10^7}$ \cite{detlab_14}. A full photoelectron collection efficiency was measured with the latter. S-RETGEM and 2RETGEM configurations reached gains of $\mathrm{10^5}$ \cite{4} and $\sim$$\mathrm{10^6}$ \cite{216}, respectively. For the latter, a full collection efficiency of the extracted photoelectron to the hole was measured \cite{4,216}.

A gain of the order of $\mathrm{10^6}$ was reached with a 1THGEM configuration operated in $\mathrm{Ne/CF_4}$ (90:10) and (95:5) gas mixtures, with a slightly lower gain measured with the latter. An order of 75\% extraction efficiency was measured from the CsI-coated electrode, followed by a full photoelectron collection efficiency into the hole \cite{detlab_21,detlab_18,215}. The lower gain measured with a (95:5) mixture could be recovered with a 3THGEM configuration \cite{detlab_24}, while a gain of the order of $\mathrm{10^5}$ was measured with an \acrshort{RPWELL} detector \cite{detlab_68}. A 2THGEM detector operated with $\mathrm{Ne/CF_4}$ (90:10) in sealed mode yielded a gain in the order of $\mathrm{10^5}$ \cite{235}. Generally, lower gains were measured in $\mathrm{Ne/CH_4}$ gas mixtures; less than $\mathrm{10^5}$ was measured with 1THGEM \cite{detlab_21,detlab_14} and several $\mathrm{10^5}$ with a 3THGEM \cite{detlab_24} in (95:5) and (90:10) mixtures. At higher $\mathrm{CH_4}$ concentrations of (77:23), gains in the range of $\mathrm{10^5}$-${10^6}$ were reported with 1THGEM \cite{detlab_21,detlab_14} and RETGEM configurations \cite{287}. The hybrid 2THGEM+\acrshort{THCOBRA} configuration operated in the (95:5) mixture yielded a gain in the order of $\mathrm{10^6}$ \cite{191}, while a hybrid \acrshort{THGEM}+\acrshort{RPWELL} reached a similar gain when operated in (98:2) to (85:15) mixtures \cite{detlab_68}. Measurements were carried out also with various \acrshort{RETGEM} configurations in $\mathrm{Ne/iC_{4}H_{10}}$ \cite{235}, Ne/air, Ne/EF, and air/EF gas mixtures \cite{7}.

In Ar (grade not reported), gains of several $\mathrm{10^4}$ and $\mathrm{10^5}$ were reached with 1THGEM \cite{4} and 2RETGEM \cite{216} configurations, respectively. Operated in $\mathrm{Ar/CH_4}$ (95:5), gains of the order of $\mathrm{10^5}$ \cite{detlab_69,detlab_1,detlab_4} and $\mathrm{10^6}$ \cite{detlab_14} were reported with a 1THGEM configurations. A higher gain of $\mathrm{10^7}$ was measured with 2THGEM \cite{detlab_1,detlab_4}. Measurements were also carried out with \acrshort{RETGEM} \cite{4} and hybrid \acrshort{THGEM}+\acrshort{RPWELL} configurations \cite{detlab_68}. At higher concentrations of $\mathrm{CH_4}$, gains of $\mathrm{10^4}$-${10^5}$ were measured with \acrshort{THCOBRA} \cite{51}, 2RETGEM \cite{216}, 1THGEM+\acrshort{MM} \cite{231} and MM-THGEM \cite{82} configurations. Operated in an $\mathrm{Ar/CO_2}$ (70:30) gas mixture, gains in the order of $\mathrm{10^5}$ and $\mathrm{10^7}$ were measured with 1THGEM and 2THGEM configurations, respectively \cite{detlab_1,detlab_4}, as well as with a hybrid consisting of a 2THGEM+\acrshort{MM} \cite{105}. A gain in the order of $\mathrm{10^5}$ was measured using 3THGEM in (50:50) \cite{228} and (70:30) \cite{186} $\mathrm{Ar/CO_2}$ mixtures. A lower gain of several $\mathrm{10^3}$, was measured with various configurations at lower $\mathrm{CO_2}$ concentrations \cite{4}. 

Operated in $\mathrm{CF_4}$ and $\mathrm{CH_4}$, gains in the order of $\mathrm{10^4}$ were reached with a 1THGEM configuration \cite{detlab_1,detlab_4,detlab_18}. Higher gain values were measured with various configurations in He (grade unknown); $\sim$$\mathrm{10^5}$ with a \acrshort{THWELL}, several $\mathrm{10^6}$ with a hybrid \acrshort{THGEM}+\acrshort{THWELL}, and several $\mathrm{10^7}$ with a hybrid 2THGEM+\acrshort{THWELL} \cite{110}. A gain of the order of $\mathrm{10^6}$ was also reached with a 2THGEM+\acrshort{MM} operated in He \cite{108}. Gains in the range $\mathrm{10^5}$-${10^6}$ were obtained with 2M-THGEM and 3M-THGEM detectors. Similar values were obtained with 2M-THGEM and 3M-THGEM, operated in a $\mathrm{He/CO_2}$ (90:10) mixture \cite{102}. 1THGEM and 2THGEM detectors operated in $\mathrm{He/CO_2}$ (70:30) resulted in 10$^3$ and 5$\times$10$^3$ respectively \cite{512}. 
MM-THGEM and a cascade MM-THGEM+\acrshort{THWELL} operated in the same mixture yielded gains of several $\mathrm{10^5}$ and $\mathrm{10^6}$, respectively \cite{82}. Gain values in the range of $\mathrm{10^4}$-${10^5}$ were reached with a 1THGEM operated in (95:5) to (60:40) $\mathrm{He/CH_4}$ and $\mathrm{He/CF_4}$ mixtures \cite{detlab_75}.  

Using single electrons, gain values in the order of $\mathrm{10^4}$ were measured with a 1THGEM in $\mathrm{CH_4}$ and $\mathrm{CF_4}$ gases. Higher gain values, of the order of $\mathrm{10^5}$ and $\mathrm{10^7}$, were measured in $\mathrm{Ar/CH_4}$ (95:5) and $\mathrm{Ar/CO_2}$ (70:30) using 1THGEM and 2THGEM configurations, respectively \cite{detlab_1, 288}. Gains of $\mathrm{10^5}$-${10^6}$ were obtained with a 1THGEM operated in Ne mixtures ($\mathrm{Ne/CH_4}$ and $\mathrm{Ne/CF_4}$) \cite{detlab_21,detlab18,470}. 

Operated in $\mathrm{CF_4}$, a stable gain of $\mathrm{10^4}$ was measured at event rates up to $\sim$$\mathrm{10^7 Hz/mm^2}$ with 1THGEM \cite{detlab18}. Examples of 1THGEM gain curves measured in various gas mixtures are presented in Fig. \ref{fig:THGEMGain}.

\subsubsection*{Alpha particles}

Operated in Ne (grade not reported) gains of approximately 400 and 500 were measured with 1THGEM and 2THGEM configurations, respectively. After further purifying the Ne, the gain measured with the two configurations dropped by an order of magnitude \cite{detlab_20}. \acrshort{RETGEM} and Kapton-\acrshort{RETGEM} operated in non-pure Ne, yielded gains of 10-100 \cite{2,157} and 100 \cite{1,156}, respectively.  

Ceramic-1THGEM operated in sealed mode in He/CO$_2$ (90:10), Ne/CO$_2$ (90:10) and Ar/CO$_2$ (90:10), resulted in a maximum gain of about 200, 100 and 70, respectively \cite{517}.

A \acrshort{RETGEM} operated in Ar (grade not reported) yielded gain values of several tens \cite{2,157, detlab_20}, which was reduced by an order of magnitude after purification \cite{detlab_20}. 

\acrshort{THCOBRA} operated in Kr (grade not reported) yielded a gain of several $\mathrm{10^4}$, an energy resolution in the order of 35\%, and a rate capability up to $\sim$$\mathrm{10^5}$ \HZMM{} \cite{136}. A gain of several $\mathrm{10^3}$ was recorded operating 2M-THGEM in  He \cite{102}, while a gain of at most 20 was measured with a 1RETGEM in a $\mathrm{He/Ar/CH_4}$ (86.8:12.5:0.7) mixture \cite{157}.


\subsubsection{Energy resolution}
The energy resolution of gaseous detectors is typically measured with soft x-rays, as detailed in Section \ref{sec:THGEM}). It depends on the multiplier, gas, gain, rate etc and often has a minimum value when measured as a function of gain (see e.g., \cite{detlab_73}). In most THGEM-like configurations discussed above, operated in various gas mixtures, the resolution was in the range of 20-30\% \acrshort{FWHM}. A few examples of the best published values are detailed here. About $\mathrm{20\%}$ was measured with 1THGEM and 2THGEM configurations and with a glass-\acrshort{THGEM} operated in various Ar-based mixtures \cite{59,112,245,283,293,123,252,132,499}. About $\mathrm{25\%}$ was measured with 1THGEM and 2THGEM operated in $\mathrm{Ne/CF_4}$ \cite{detlab_73}, also with some multiplication in the induction gap \cite{detlab_31,detlab_66}. \acrshort{THWELL}, \acrshort{RWELL}, and \acrshort{RPWELL} configurations operated in $\mathrm{Ne/CH_4}$ (95:5) yielded 20-25\% \cite{detlab_30}. A \acrshort{RETGEM} operated in $\mathrm{Ar/iC_{4}H_{10}}$ (90:10), yielded about 30\% \cite{13}. The energy resolution was also measured with \acrshort{THCOBRA} detectors, e.g., \cite{301,51,252,136}. In He/CO$_2$ (70:30), a resolution of about 30\% and 40\% was reported for 1THGEM and 2THGEM, respectively\footnote{Values for FWHM/$\mu$; calculated using $\sigma$/$\mu$ in the paper.}. 
It was demonstrated in \cite{388} that an energy resolution of 25\% \acrshort{FWHM} could be achieved with a 1THGEM, also by reading out the \acrshort{EL} light emitted by the avalanche in a Ar/CF$_4$ (90:10) gas mixture.

\subsubsection{Detection efficiency}

\subsubsection*{MIPs}

Measurements with various \acrshort{THGEM}-based detector configurations were carried out in muon and high-rate pion beams, mostly in the context of Digital Hadronic Calorimetry (\acrshort{DHCAL}, Section \ref{sec:AppCalorimetry}). Operated with $\mathrm{Ne/CH_4}$ (95:5), detection efficiencies greater than 98\% were reached with 1THGEM and 2THGEM \cite{detlab_26}, \acrshort{THWELL}, \acrshort{RWELL}, \acrshort{SRWELL}, hybrid \acrshort{THGEM}+\acrshort{RWELL}, and \acrshort{RPWELL} \cite{detlab_32,detlab_33,detlab_48,detlab_49,detlab_53,detlab_26} configurations. Similar results were obtained with the \acrshort{RPWELL} configuration operated in $\mathrm{Ar/CH_4}$ (95:5) and $\mathrm{Ar/CO_2}$ (93:7) mixtures \cite{detlab_48,detlab_53}. 

Efficiency measurements were performed with 1THGEM configurations in several gas mixtures; detection efficiency values of 80-90\% and 93-99\% were measured for pions and protons, respectively, in $\mathrm{Ar/iC_{4}H_{10}}$ (97:3) \cite{115}. The respective values in $\mathrm{Ne/CH_{4}}$ (95:5) were 80-90\% and 91-97\% were measured \cite{193}.

\subsubsection*{UV photons}

As discussed in Section \ref{sec:THGEM}, the \acrshort{PDE} of \acrshort{THGEM} detectors, depends on the the photocathode’s \acrshort{QE}, the extraction efficiency of photoelectrons into gas, their focusing into the holes, and on the avalanche-gain – dictating the fraction of signals detected above noise \cite{288}.

Studies were conducted to characterize the performance of \acrshort{THGEM}-based \acrshort{UV}-photon detectors and optimize them for different applications (see, e.g., Section \ref{sec:RICH}). It was shown that in gas mixtures such as $\mathrm{Ne/CH_4}$ (95:5) and (90:10), $\mathrm{Ne/CF_4}$ (95:5) and (90:10), and $\mathrm{CH_4}$ and $\mathrm{CF_4}$, the extraction efficiency ($\varepsilon_{ext}$) increased rapidly as a function of the photoelectron extraction field  (reversed drift field, \Vcathode > \Vtop) at the photocathode surface, reaching a plateau above 0.5-1 \kvcm. The best $\varepsilon_{ext}$, in the order of 85\%, was obtained with $\mathrm{CH_4}$ \cite{detlab_21}. Measurements in $\mathrm{Ar/CH_4}$ (30:70) revealed that the product of the extraction and collection efficiency ($\varepsilon_{ext} \times \varepsilon_{col}$) was optimal at a drift field of 0.2 \kvcm{} with minor dependency on the multiplication field. A hole diameter of 300 \um{} and a pitch of 800 \um{} were found optimal with respect to the area coverage and $\varepsilon_{ext} \times \varepsilon_{col}$ \cite{38}.

Using Ne mixtures, absolute \acrshort{PDE} in the range of 12–14$\%$ was estimated for 170 nm \acrshort{UV} photons at gains greater than $\mathrm{10^5}$, which could increase to $\sim$20$\%$ with an optimized hole geometry \cite{detlab_21}. Large CsI-coated detectors (hybrid 2THGEM+\acrshort{MM}) were operated efficiently in experiments \cite{66}.

\subsubsection{IBF}

A challenge in gas-avalanche detectors is the reduction of the \acrshort{IBF} value, with minimal losses of electrons. The latter is particularly critical for keeping high \acrshort{PDE} values of single \acrshort{UV}-photon detectors (e.g., in \acrshort{RICH}) but also in getting good energy resolutions. Thus, \acrshort{IBF} should be considered relative to the electron-detection efficiency. 

\acrshort{IBF} is best suppressed in misaligned cascade structures \cite{286}. A 2THGEM detector operated in $\mathrm{Ar/CO_2}$ (90:10) resulted in 100\% \acrshort{IBF}, which could be lowered to 60\% by misaligning the holes of the two stages with a mild loss of electrons. Operated in an $\mathrm{Ar/CH_4}$ (70:30), 3\% \acrshort{IBF} was measured with a 3THGEM configuration but at about 50\% gain loss \cite{19,186}. An \acrshort{IBF} greater than 10\% was recorded with a 3THGEM configuration operated in an $\mathrm{Ar/CO_2}$ (70:30) mixture. This was reduced to 1-3\% by misaligning the holes of the three electrodes albeit at the cost of significant loss of electron detection efficiency \cite{19,114}. MM-THGEM operated in $\mathrm{He/CO_2}$ (90:10) yielded \acrshort{IBF} values in the range of 1.5-2\% \cite{82}. 5\% \acrshort{IBF} was recorded with a hybrid \acrshort{THGEM}+\acrshort{THCOBRA}+\acrshort{THGEM} operated in pure Ne keeping high electron detection efficiency \cite{60}. 20\% \acrshort{IBF} was obtained with a hybrid 2THGEM+\acrshort{THCOBRA} detector operated in $\mathrm{Ne/CH_4}$ (95:5) \cite{191}. Very low \acrshort{IBF} values, in the range 0.1-0.5\%, were recorded with a hybrid \acrshort{GEM}+2THCOBRA in $\mathrm{Ne/CO_2}$ (90:10) \cite{119}. Exploiting the inherent low \acrshort{IBF} of \acrshort{MM} detectors, the large area hybrid 2THGEM+\acrshort{MM} configuration deployed in the \acrshort{COMPASS}-\acrshort{RICH} experiment yielded \acrshort{IBF} at the 3\% level with sufficient single-electron detection efficiency \cite{66}. In \cite{500} a Monte Carlo simulation shows that a 4THGEM configuration with staggered holes could reach \acrshort{IBF} as low as 0.2\% in Ar/iC$_4$H$_{10}$ (90:10).

\subsubsection{Spatial resolution} 

Comparison of the spatial resolution of different detectors should be done with care. The measured value could depend on the readout scheme and readout electronics. 

Using a soft x-ray, spatial resolution in the order of 0.7 mm \acrshort{FWHM} was measured with a 2THGEM structure in a $\mathrm{Ne/CH_4}$ (95:5) gas  mixture \cite{detlab_6}. A somewhat poorer resolution, in the order of 2.5 mm \acrshort{FWHM}, was measured in the same gas mixture with a \acrshort{THCOBRA} configuration \cite{300,194}.  

A spatial resolution in the order of 100-300 \um{} \acrshort{FWHM} was measured with a hybrid 2THGEM+\acrshort{THCOBRA} configuration operated with a \acrshort{UV} source in a $\mathrm{Ne/CH_4}$ (95:5) gas mixture \cite{26,191}. The expected resolution of a 1THGEM operated in $\mathrm{Ar/CO_2}$ (70:30) was found in a simulation study to be 150-300 \um{} \acrshort{RMS}, depending on the drift and induction fields \cite{339}.  

Operated with a $\mathrm{Ne/CH_4}$ (95:5) gas mixture in a muon beam, a spatial resolution of approximately 200 \um{} \acrshort{RMS} was measured with an \acrshort{RPWELL} detector. Simulation studies have shown that the resolution limitation stems from the holes' pitch \cite{detlab_51}. Spatial resolution studies with alpha particles are presented in \cite{294}. Similar results were obtained by measuring cosmic rays with a 3THGEM detector operated in $\mathrm{Ar/CO_2}$ (90:10) \cite{25}.

\subsubsection{Time resolution} 

A 2THGEM detector with semitransparent- or reflective-photocathode operated in an $\mathrm{Ar/CH_4}$ (95:5) mixture yielded a time resolution of 8-10 ns \acrshort{RMS} when operated with single \acrshort{UV} photons and relativistic electrons. Using photoelectron pulses, a time resolution of 0.5-1 ns \acrshort{RMS} was reached, depending on the number of photons in the pulse \cite{detlab_72}. Using a \acrshort{UV} source, a time resolution in the order of 10 ns was also measured with a 3THGEM configuration operated with an $\mathrm{Ar/CH_4}$ (50:50) mixture \cite{131}. A hybrid 2THGEM+\acrshort{MM} configuration operated in $\mathrm{Ar/CH_4}$ (70:30) yielded a 7 ns \acrshort{RMS} time resolution when irradiated with single \acrshort{UV} photons \cite{105}. 
 
\subsubsection{Light yield}
\label{sec: light readout}

\acrshort{THGEM} is particularly suitable for light readouts due to the electrode thickness; it enables considerable electron drift length in the high field region, resulting in a significant light yield. 
Light yields in the order of $\mathrm{7\times 10^4}$ and $\mathrm{1.5\times 10^4}$ photons per primary electron (emitted in 4$\pi$) were measured with a 1THGEM operated in pure Xe and Ar, respectively \cite{240}. Using $\mathrm{Ar/CF_4}$ (90:10), a light yield of $\mathrm{9\times 10^4}$ photons per \kev{} was detected \cite{251,388}. Preliminary Monte Carlo simulations of \acrshort{EL} in pure Neon were also performed \cite{80}.

\subsubsection{Rate dependence}

The gain variation with irradiation rate is strongly dependent on the detector configuration. When using configurations without resistive materials, the main dependency stems from the drift time of the ions in the multiplication region, several \us{} in most gas mixtures used. 
Thus, in these configurations, gain drops were evident at irradiation rates in the order of a few \MHZMM{} \cite{detlab_4,detlab_18}. In configurations that incorporate resistive materials, gain drops are mainly due to the current across the resistive material causing a voltage drop. More details can be found in \cite{detlab_14,detlab_49,detlab_18,detlab_26,detlab_48,detlab_53,detlab_41,21,249,252,136}. In \acrshort{THWELL}-like structures, gain drops are observed at all rates, and the greater the resistivity, the greater the drop \cite{detlab_30}.
Effects were observed at irradiation rates of \KHZMM{} in \acrshort{DLC}-coated \acrshort{THGEM}s \cite{249}.

\subsubsection{Discharge probability} 

Several efforts were aimed at characterizing the discharge probability and magnitude of different \acrshort{THGEM} detector configurations \cite{detlab_40, detlab_78,detlab_26, detlab_56, detlab_33,210,486}. In most cases, discharges were defined as visible spikes in the current supplied to one or more of the electrodes. Using this definition, discharge probabilities in the order of $\mathrm{10^{-5}}$ and $\mathrm{10^{-4}}$ were measured with a 1THGEM in muon and high rate pion beams, respectively. The detector was operated in $\mathrm{Ne/CH_4}$ (95:5) at a gain of several $\mathrm{10^3}$ \cite{detlab_26}. Under similar operation conditions, a 2THGEM structure showed a discharge probability of less than $\mathrm{2\times 10^{-6}}$ for both muons and pions \cite{detlab_26}. Similar measurements conducted with \acrshort{SRWELL} resulted in a discharge probability of $\mathrm{10^{-6}}$ in the muon beam and $\sim$$\mathrm{10^{-5}}$ in the high rate pion beam \cite{detlab_33}. This was reduced to approximately $\mathrm{10^{-7}}$ for muons and $\mathrm{10^{-6}}$ for pions when using cascaded \acrshort{THGEM}+\acrshort{SRWELL} \cite{detlab_33}. 

Laboratory studies using x-rays demonstrated that the discharge probability as a function of the number of \acrshort{PE}s at a fixed gain and a function of gain at fixed number of \acrshort{PE}s is similar for \acrshort{THWELL} and \acrshort{RWELL} structures \cite{detlab_40,detlab_78}, but the energy released in a discharge of an \acrshort{RWELL} was measured to be an order of magnitude lower \cite{detlab_78,detlab_41}. 

No current spikes were observed with highly resistive configurations like the \acrshort{RPWELL} \cite{detlab_53}. However, it was shown in \cite{detlab_78} that in such configurations, low-intensity electrical instabilities are also present and induce large signals on the readout electrodes. The onset of discharges was consistent with the crossing of the Raether limit \cite{von1965electron} ($\mathrm{10^6-10^7}$ electrons) in most measurements \cite{detlab_78}. 

In \cite{68,486}, measurement and modeling studies of the discharge probability of a 1THGEM were performed in different Ne and Ar mixtures using alpha particles. It was demonstrated that charge density inside the hole is the main factor influencing discharge probability. Additionally, operation in Ne-based mixtures is less prone to discharges compared to Ar-based ones because of their different atomic number. Moreover, the amount of quenchers does not directly relate to reduced discharge probability. Rather, it is the final charge density that matters most.

\subsection{Room temperature and low pressure}
\label{sec:RTLP}

Several applications involve detecting tracks from heavily ionizing radiation, for which low gas pressures, down to sub-Torr values, are required. Examples
are ion spectrometry (Section \ref{sec: focal plane tracker}), \acrshort{AT}-\acrshort{TPC}s detecting heavy products of nuclear reactions (Section \ref{sec: AT TPC}), and detection of track-structure of very low-energy ions for nanodosimetry studies (Section \ref{sec:medicalApps}). 

\acrshort{THGEM}-based detectors might be attractive for these applications. Compared to operation in standard pressure, at low pressures, the electron mean free path increases significantly. Therefore, thicker electrodes with larger hole diameters are preferable. Indeed, gain values similar to those obtained at standard pressure are reached \cite{detlab_5}, outperforming thinner \acrshort{THGEM}s and \acrshort{GEM}s by several orders of magnitude \cite{detlab_5,410}.

\subsubsection*{Molecular gases} 
Operations down to very low-pressure values, below 50 Torr, were carried out in $\mathrm{iC_{4}H_{10}}$ \cite{detlab_5,detlab_69}, $\mathrm{CF_{4}}$ \cite{410,83}, $\mathrm{SF_{6}}$ \cite{101,137,76,512}, $\mathrm{CH_{4}/iC_{4}H_{10}}$ \cite{74}, $\mathrm{CO_2}$ \cite{97}, and $\mathrm{C_3H_8/CO_2/N_2}$ \cite{177,304,72}. 

1THGEM detectors with various electrode geometries were operated in low-pressure $\mathrm{iC_{4}H_{10}}$, irradiated with single \acrshort{UV} photons. Gain values above $\mathrm{10^5}$ were reached for pressure values of 1-50 Torr, with a peak in the order of $\mathrm{10^6}$ at $\sim$10-20 Torr \cite{detlab_69,detlab_5}. A gain of $\mathrm{10^7}$ was reached with a very thick electrode of 2.2 mm with 1 mm-diameter holes, operated at 10 Torr \cite{detlab_5}. At 0.5 Torr, the maximum achievable gain was of several $\mathrm{10^3}$ \cite{detlab_5}. Similar results at 0.5, 1, and 10 Torr were obtained with a 2THGEM structure \cite{detlab_5}. Lower gains of the order of $\mathrm{10^3}$ and 500, were obtained with a 1THGEM operated in  $\mathrm{iC_{4}H_{10}}$ at 50 and 180 Torr, respectively \cite{67,71}. In \cite{503} a 3M-THGEM operated with alpha particles in iC$_4$H$_{10}$ at 7.5, 15 and 22.5 Torr (10, 20 and 30 mbar) demonstrated a maximum gain of 3$\times$10$^4$, 10$^4$ and 4$\times$10$^3$, respectively. In all cases the \acrshort{IBF} was between 10-20\%. 

In 50, 35, and 25 Torr $\mathrm{CF_{4}}$, gains higher than $\mathrm{10^5}$ were measured with 1THGEM, 2THGEM, and 3THGEM configurations, respectively. A 40\% energy resolution was measured with the 2THGEM configuration. The three detector configurations were irradiated with a \fe{} source \cite{410,83}.

$\mathrm{SF_{6}}$ is an important gas for negative ion \acrshort{TPC}s (Section \ref{sec:IonTPC}). Having high electro-negative properties, \acrshort{PE}s attach to the gas molecules to form negative ions. The latter drift to the amplification region where the electrons are stripped from the ion and undergo charge avalanche multiplication. Measurements with a \fe{} source in 30 and 40 Torr $\mathrm{SF_{6}}$ with 1THGEM detectors, resulted in a gain of several $\mathrm{10^3}$\cite{101,512} and energy resolutions of 20-40\% \cite{101}. A hybrid \acrshort{THGEM}+\acrshort{MWPC} configuration operated at 20 Torr $\mathrm{SF_{6}}$ was irradiated with an alpha source, yielding a gain of approximately 2500 \cite{137,76}. 

Measurements in a tissue-equivalent gas $\mathrm{C_3H_8/CO_2/N_2}$ (55:39.6:5.4) (Section \ref{sec:medicalApps}) were performed. At 100 Torr, with an alpha source, several $\mathrm{10^2}$ and $\mathrm{10^3}$ were obtained with 1THGEM and 2THGEM configurations, respectively \cite{72}. A gain of several $\mathrm{10^2}$ was obtained with a 1THGEM at $\sim$50 Torr \cite{177}. A ceramic-\acrshort{THGEM} was tested in the range of 50-300 Torr, resulting in gains ranging between $\sim$10-500 \cite{304}. 

Results were also reported for different \acrshort{M-THGEM} structures. A gain of the order of $\mathrm{10^4}$ was obtained with a 2M-THGEM detector operated with alpha particles at 40 Torr  $\mathrm{CF_{4}/iC_{4}H_{10}}$ (80:20) \cite{74}.
2M-THGEM (3M-THGEM) detectors operated at low-pressure $\mathrm{CO_2}$ yielded gains ranging from $\mathrm{10^6}$ at 50 Torr to $\mathrm{10^3}$ ($\mathrm{10^4}$) at 450 Torr. An energy resolution in the order of 5\% was measured with a hybrid 2THGEM+\acrshort{MM} configuration at 50 Torr $\mathrm{CO_2}$ using alpha particles \cite{97}.

Measurements in $\mathrm{H_2}$ \cite{109,108}, $\mathrm{H_2/iC_4H_{10}}$ \cite{108}, and $\mathrm{D_2}$ \cite{160,180,109} are particularly relevant for \acrshort{AT}-\acrshort{TPC} applications (Section \ref{sec: AT TPC}). Several measurements were carried out with a \acrshort{UV} source. A single \acrshort{THWELL} configuration operated in $\mathrm{H_2}$ and $\mathrm{D_2}$ reached gains in the order of 1000 at 100 Torr and $\mathrm{10^4}$ at 450 Torr \cite{109}. A slightly higher gain of the order of 2000 was measured with a cascade \acrshort{THGEM}+\acrshort{THWELL} configuration at 100 Torr $\mathrm{H_2}$. An \acrshort{IBF} below 5\% was also demonstrated in this configuration. A hybrid 2THGEM+\acrshort{MM} configuration operated at 200 and 300 Torr reached gains of approximately 800 and 300, respectively with alpha particles and heavy nuclear fragments \cite{108}. Mixing the $\mathrm{H_2}$ with 2\% $\mathrm{iC_4H_{10}}$ increased the gain significantly, to $\sim$$\mathrm{10^4}$ at 200 Torr with a 2THGEM configuration irradiated with alpha particles. 2THGEM and 3THGEM configurations operated at 130-400 Torr $\mathrm{D_2}$ yielded gains of the order of $\mathrm{10^3}$ \cite{160} and $\mathrm{10^4}$ \cite{180}, respectively with alpha particles.

\subsubsection*{Noble gases} 
Measurements in pure and mixtures of noble gases in the range of 50-750 Torr were conducted in He \cite{110,108,102}, $\mathrm{He/CO_2}$ \cite{277,11,280,102,189,512}, Ne \cite{485}, Ne/H$_2$ \cite{485}, $\mathrm{Ar/CO_2}$ \cite{detlab_5,462,282,529}, $\mathrm{Ar/CH_4}$ \cite{72,529}, Kr \cite{310}, Xe, and Ar/Xe \cite{detlab_12}.

Further, He gas is particularly relevant for \acrshort{AT}-\acrshort{TPC} applications (Section \ref{sec: AT TPC}). A gain of several $\mathrm{10^3}$-${10^4}$ was reached with a 1THGEM when a $\mathrm{He/CO_2}$ (90:10) mixture was operated at 120 Torr (0.16 bar), using an alpha source \cite{277}. Gain variations associated with charging-up of the rims were also demonstrated in these conditions \cite{11}. The measured energy resolution was approximately 5\%, using a transverse ion beam source \cite{189}. A 1THGEM measuring \fe{} x-rays in 380 Torr and 570 Torr He/CO$_2$ (70:30) yielded a gain of about 2$\times$10$^3$ \cite{512}. The energy resolution was in the range of 40-50\%\footnote{Values for FWHM/$\mu$; calculated using $\sigma$/$\mu$ in the paper.}.

\begin{figure}
\centering
\includegraphics[width=0.9\textwidth]{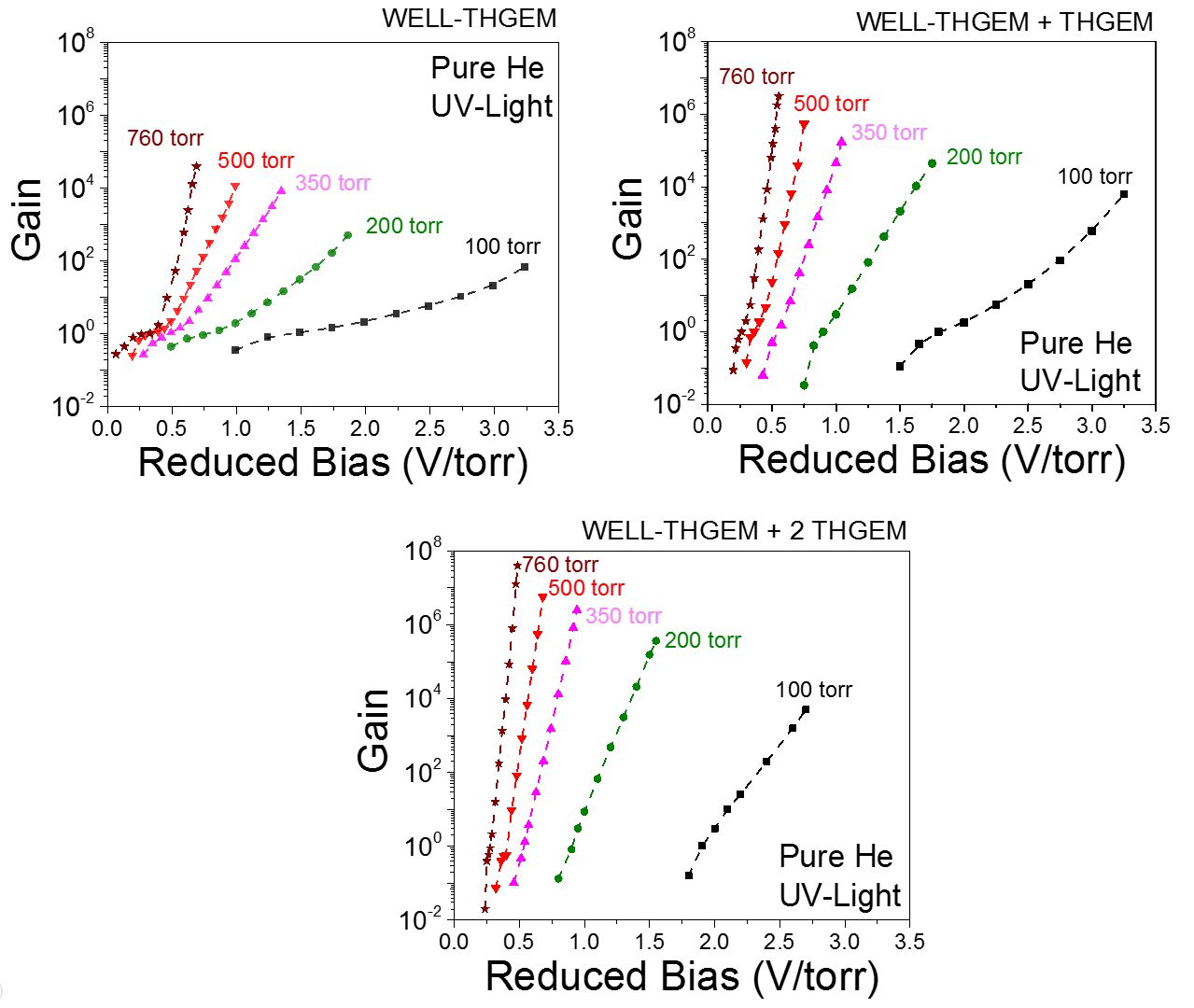}
    \caption{Effective gain curves for single photoelectrons  in 100–760 torr He for different THGEM-based detector configurations. Figure obtained from \cite{110}.}
    \label{fig:gain lowP He}
\end{figure}

Studies in He at 100-600 Torr were carried out with various \acrshort{THWELL} \cite{110} and \acrshort{M-THGEM} \cite{102} configurations. Irradiated with a single-\acrshort{UV} source, gain values ranging from 80 at 100 Torr to $\mathrm{10^4}$ at 500 Torr were reached with a \acrshort{THWELL} detector. Higher gain values were measured with cascaded \acrshort{THGEM}+\acrshort{THWELL} and 2THGEM+\acrshort{THWELL}, operated under the same conditions. For the former, the gain ranged from $\mathrm{10^4}$ at 100 Torr to $\mathrm{8 \times 10^5}$ at 500 Torr, while for the latter, from $\mathrm{10^4}$ at 100 Torr to $\mathrm{2 \times 10^6}$ at 500 Torr. The corresponding gain curves are presented in Figure \ref{fig:gain lowP He}. Over a similar pressure range and using \acrshort{UV} photons, gains of the order of $\mathrm{10^4}$-${10^5}$ and $\mathrm{10^5}$-${10^6}$ were reached with 2M-THGEM and a cascade of two 2M-THGEM configurations, respectively \cite{102}. A hybrid 2THGEM+\acrshort{MM} irradiated with \acrshort{UV} photons reached gains of $\mathrm{10^6}$ at 200 Torr and $\mathrm{3 \times 10^6}$ at 500 Torr, although the gain added by the \acrshort{MM} was negligible \cite{108}. The \acrshort{M-THGEM} configurations were also tested with a low-pressure $\mathrm{He/CO_2}$ (90:10). Irradiated with \acrshort{UV} photons, a gain of several $\mathrm{10^5}$ was measured with a single 2M-THGEM operated in pressures in the range 150-760 Torr. Higher gains of about $\mathrm{10^6}$ could be reached with a cascade of double 2M-THGEM configurations \cite{102}. 3M-THGEM operated with alpha source yielded a gain in the range of a few $10^3$ to a few $10^3$ when operate in pure He and $\mathrm{He/CO_2}$ (90:10) \cite{102}. Operated in He, an energy resolution of approximately 2.5\% was measured by irradiating a hybrid 2THGEM+\acrshort{THWELL} configuration with alpha particles \cite{110}. 

The maximum gain of a 2M-THGEM detector in 350 Torr Ne was measured to be about 200 using a alpha source. The same detector could reach an order of magnitude higher gain of a few 10$^3$ when operated in 150-300 Torr Ne/H$_2$ (98:2) and (95:5), due to the Penning energy-transfer occurring in this mixture \cite{485,511}. These results were supported by Monte Carlo simulations.

The gain measured with a \fe{} source with a 1THGEM operated in $\sim$180 Torr $\mathrm{Ar/CO_2}$ (70:30) mixture was of the order of $\mathrm{10^5}$ \cite{462,282}. At pressures of 1, 5, and 30 Torr, the maximum gains were measured to be $\mathrm{10^2}$, $\mathrm{10^4}$, and $\mathrm{10^5}$, respectively \cite{detlab_5}. A single glass-\acrshort{THGEM} operated in $\mathrm{Ar/CO_2}$ (90:10) at 200 Torr could measure alpha particles at a gain of 60 \cite{529}.
In studies of negative ion \acrshort{TPC}s (Section \ref{sec:IonTPC}), using a 1\acrshort{THGEM} in $\mathrm{Ar/CO_2/O_2}$ (66:30:4), the gain at 180 Torr was slightly less than $\mathrm{10^5}$ \cite{462}. 

In 100 Torr $\mathrm{Ar/CH_4}$ (90:10), a gain of $\mathrm{2 \times 10^4}$ was reached measuring a \fe{} source with a 1\acrshort{THGEM} \cite{72}. In the same gas mixture, a single glass-\acrshort{THGEM} operated in the range 200-700 Torr could measure alpha particles at a gain of about 100 \cite{529}. The reported energy resolution was 3\% to 7\%, depending on gain and gas pressure.

The gains measured with a 2THGEM configuration operated in $\mathrm{Ar/iC_4H_{10}}$ (97:3) at 380 and 570 Torr (0.5 and 0.75 atm, respectively) were of the order of $\mathrm{10^4}$ and $\mathrm{5 \times 10^3}$, respectively \cite{267}.
A gain of several $\mathrm{10^4}$ was measured with 1THGEM and 2THGEM configurations, operated at 375 Torr (0.5 bar) Xe gas \cite{detlab_12}. Similar gains were obtained with different \acrshort{THGEM} geometries. A 2THGEM configuration operated in a 150 Torr (0.2 bar) $\mathrm{Ar/Xe}$ (95:5) mixture also yielded a gain of $\mathrm{10^4}$ \cite{detlab_12}. 1THGEM and 2THGEM configurations operated in 375 Torr (0.5 bar) Kr yielded gains of $\mathrm{10^4}$ and $\mathrm{4 \times 10^4}$, respectively \cite{310}. The measured energy resolution values were 25\% and 35\%, respectively. 

\subsection{Room temperature and high pressure}
\label{sec:RTHP}

High-pressure gases are used to enhance the probability of the traversing particle to interact with the medium. Examples of relevant applications are TPCs for rare events \cite{170,240} and x-ray imaging \cite{310}.

Avalanche multiplication in high-pressure gases requires very high fields. Thus, reaching high charge gains is challenging. By utilizing the fact that light emission starts at voltages below the charge multiplication threshold, studies of \acrshort{THGEM}-based detectors under such conditions focused mostly on light readout schemes using highly scintillating gases. Compared to charge readout, it was demonstrated that light readout allows reaching one or two orders of magnitude higher gains with fairly good energy resolutions \cite{170,240}. 

The majority of measurements were performed with soft x-rays. A 1THGEM detector operated in pure Xe and Ar achieved light yields of the order of $\mathrm{10^4}$ and several $\mathrm{10^3}$ photons per \acrshort{PE} at 1.5 and 2.5 bar pressures, respectively \cite{170,240}. The light yield measured with a \acrshort{FAT-GEM} (a 5 mm thick \acrshort{THGEM}) operated in 2-10 bar Xe ranged between 100-350 photons per electron per cm drift per bar, while the energy resolution ranged between 20-30\% \cite{248}.

Using charge readout, 0.8 mm thick 1THGEM and 2THGEM detectors operated in Xe yielded gains ranging from $\mathrm{3 \times 10^3}$ at 1 bar to 800 at 2 bar and from $\mathrm{10^4}$ at 1 bar to $\mathrm{10^3}$ at 2 bar, respectively \cite{detlab_12}. 1THGEM and 2THGEM detectors of 0.4 mm thicknesses were also investigated in Xe, yielding gains of 100 and 200 at 2.9 bar, respectively. Slightly higher gains, ranging from $\mathrm{3 \times 10^4}$ at 1 bar to $\mathrm{10^4}$ at 2 bar could be reached with a 2THGEM configuration operated in $\mathrm{Xe/Ar}$ (95:5) \cite{detlab_12}.

1THGEM and 2THGEM configurations operated in 1.9 bar Ar yielded gains of the order of
$\mathrm{2 \times 10^3}$ and $\mathrm{8 \times 10^3}$, respectively. While a slightly higher gain of $\mathrm{2 \times 10^4}$ could be reached with a 2THGEM configuration made of Kevlar, a lower gain of $\mathrm{2 \times 10^3}$ was recorded with a 2RETGEM configuration under the same conditions \cite{3}. Lower gains were reached with purified Ar \cite{239}.
1THGEM and 2THGEM detectors operated in 2-3 bar Kr reached gain values ranging between $\mathrm{10^3}$-300 and $\mathrm{10^4}$-$\mathrm{8 \times 10^3}$, respectively. The energy resolution of the two configurations was in the range of 20-30\%, where the poorer resolution was measured at the higher pressures \cite{310}. A THCOBRA detector operated in 1.7 bar Ne reached a gain of $\mathrm{10^5}$ \cite{51}.

\begin{figure}
\centering
\includegraphics[width=0.9\textwidth]{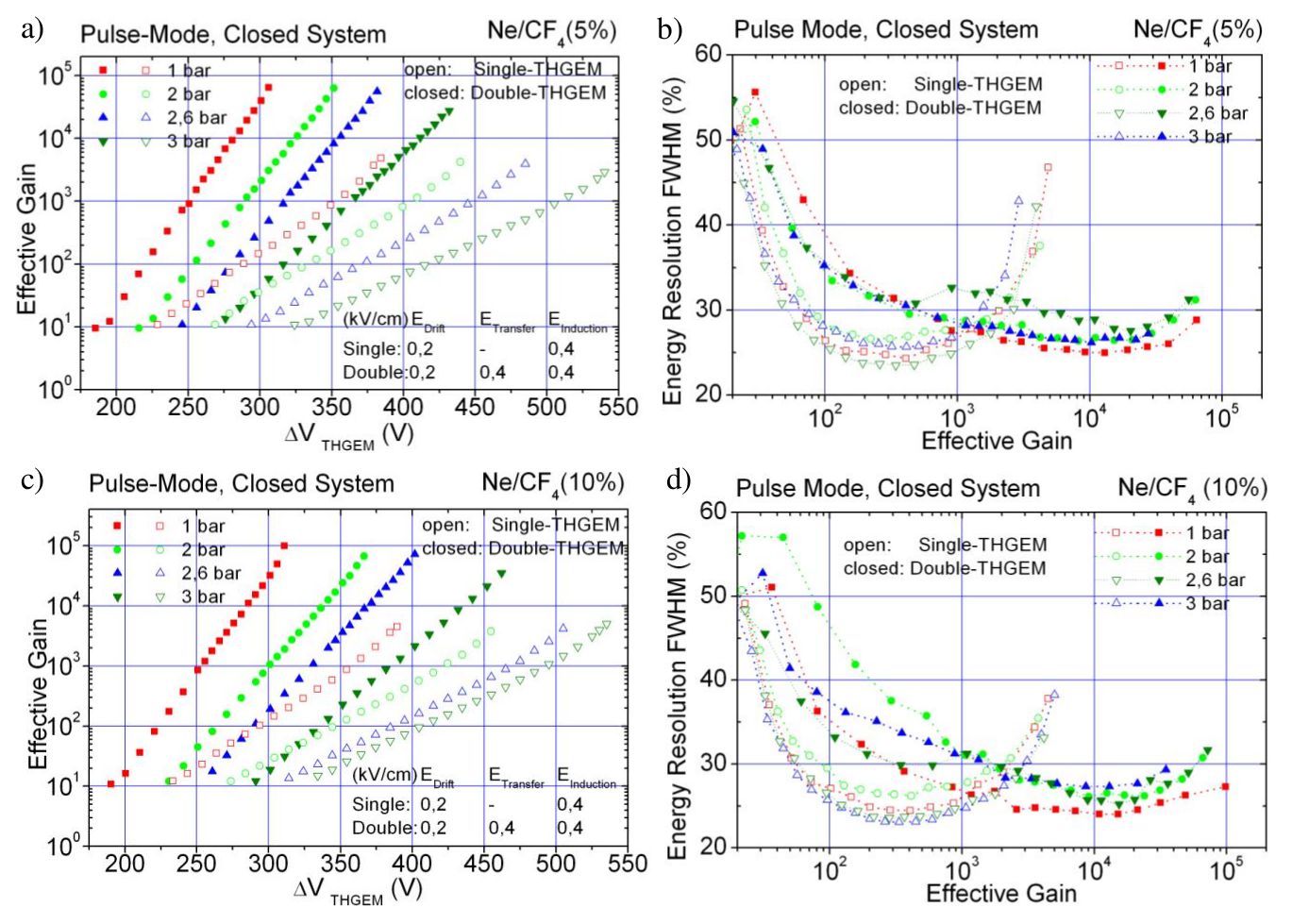}
    \caption{Effective single photoelectron gain curves and energy resolutions in 1–3 bar Ne/CF$_4$ (95:5) for 1THGEM and 2THGEM detector configurations. Figure taken from \cite{detlab_73}.}
    \label{fig:gain highP Ne/CF4}
\end{figure}

Gain and energy resolution curves measured in 1-3 bar $\mathrm{Ne/CF_4}$ (95:5) are presented in Figure \ref{fig:gain highP Ne/CF4}. This gas is interesting for \acrshort{GPM} applications since it provides high photoelectron extraction efficiencies. 1THGEM and 2THGEM configurations yielded gain values of 7-$\mathrm{4 \times 10^3}$ and 9-$\mathrm{3 \times 10^4}$, respectively. Gain values of 4-$\mathrm{3 \times 10^3}$ and 10-$\mathrm{7 \times 10^4}$ were reached with the two configurations operated in 1-3 bar $\mathrm{Ne/CF_4}$ (90:10). In all cases, the energy resolution was of the order of 25-30\% \cite{detlab_73}.

\subsection{Cryogenic temperatures}
\label{sec:CT}

\begin{figure}[htbp]
    \centering
    \subfloat[]{
    \includegraphics[scale=0.35]{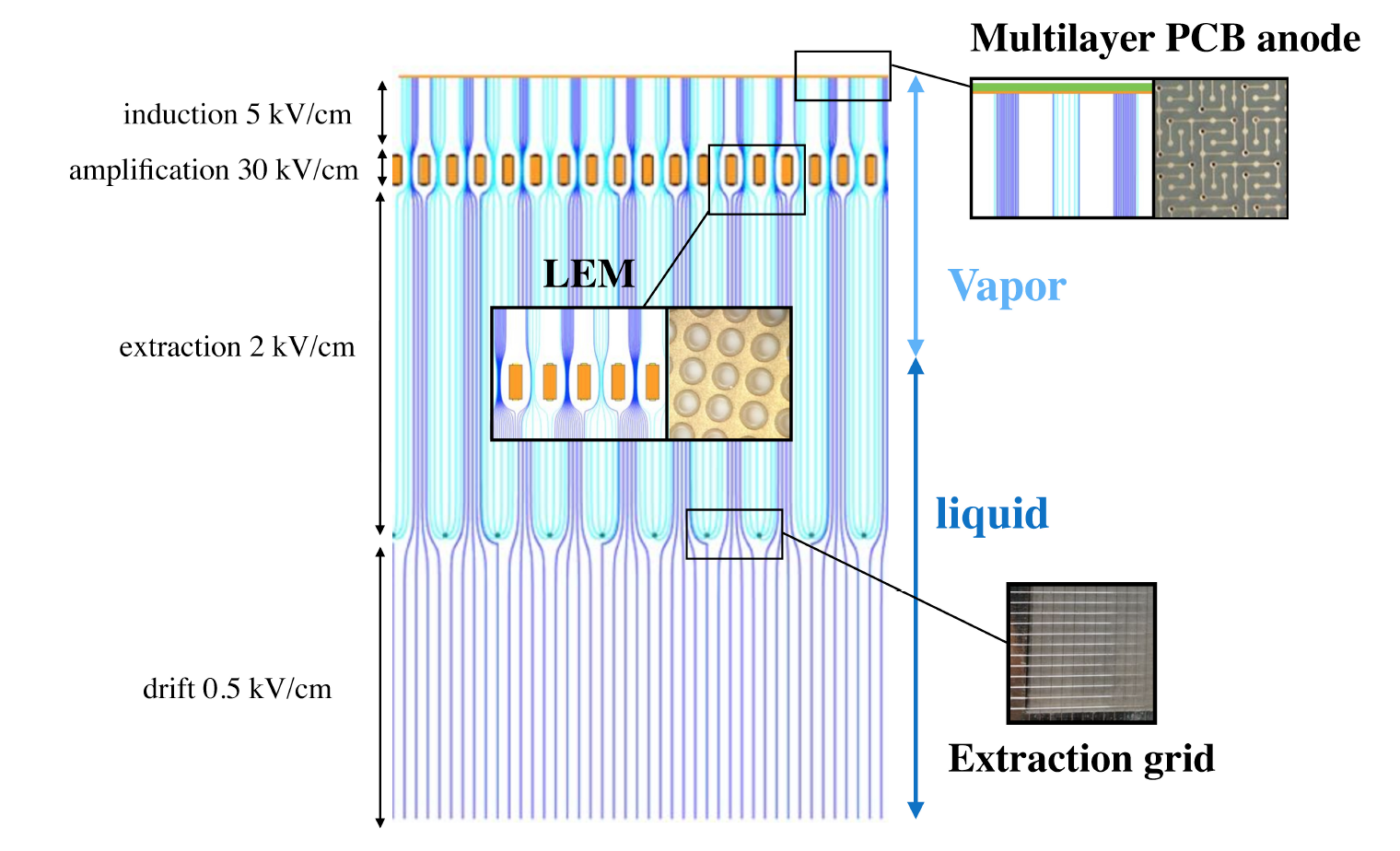}
\label{fig:CRAD}
    }\\
    \subfloat[]{
        \includegraphics[scale=0.45]{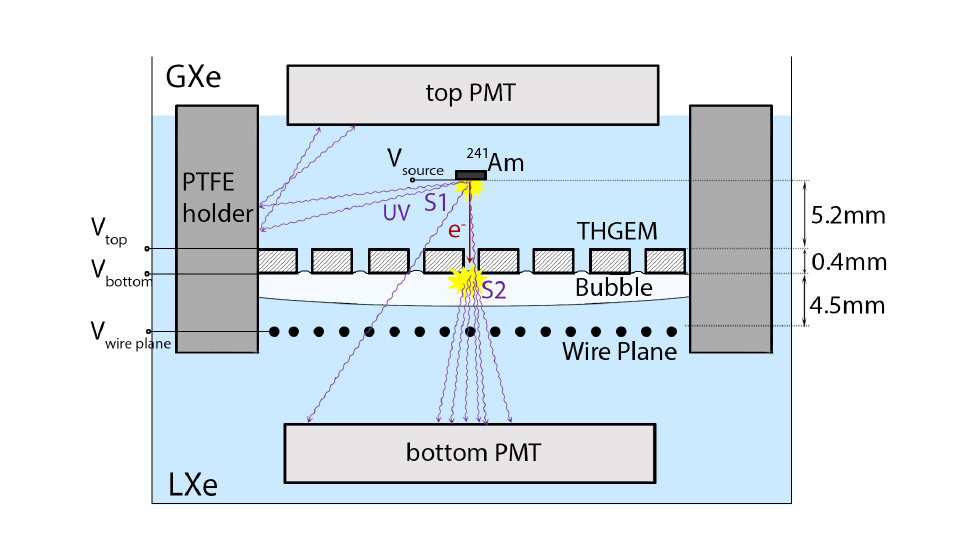}
        \label{fig:LHM}
    }\\
    \subfloat[]{
        \includegraphics[scale=0.30]{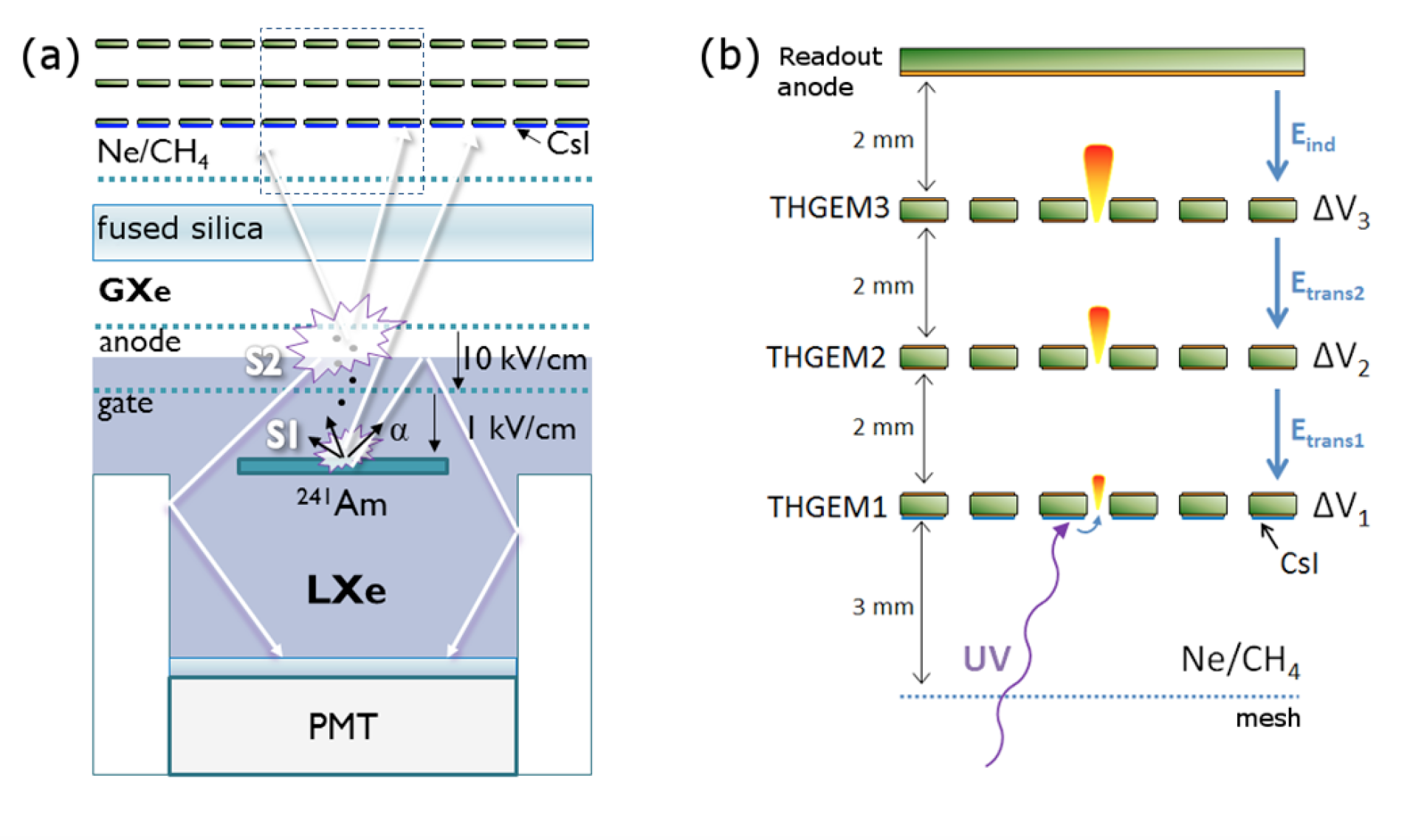}
        \label{fig:cryo GPM}
    }
\caption{Cryogenic \acrshort{THGEM} detector configurations: \protect\subref{fig:CRAD} \acrshort{CRAD}/\acrshort{LEM} (\cite{443}), \protect\subref{fig:LHM} \acrshort{LHM} (\cite{detlab_43}), \protect\subref{fig:cryo GPM} \acrshort{GPM} (\cite{detlab_44}).}
\label{fig: cryo THGEM}
\end{figure}

The performance of gaseous detectors operated in cryogenic conditions has been studied mostly in the context of \acrshort{TPC}s for rare-event experiments (Section \ref{sec:TPCforRareEvents}). \acrshort{THGEM} configurations have been developed for detecting both charge and scintillation light. The detector can be operated in the vapor of dual-phase noble liquid \acrshort{TPC}s (\acrshort{CRAD}/\acrshort{LEM} depicted in Figure \ref{fig:CRAD}), immersed in the liquid of single-phase \acrshort{TPC}s (\acrshort{LHM} depicted in Figure \ref{fig:LHM}), and as a \acrshort{GPM} coupled with the \acrshort{TPC} through a transparent window (as demonstrated in Figure \ref{fig:cryo GPM}). A detailed summary of the early studies of \acrshort{GEM} and \acrshort{THGEM}-based detectors operated in cryogenic conditions can be found in \cite{417,review_15}. For a recent and comprehensive review of the physics processes involved in charge and light multiplication in dual-phase detectors and related technologies, see \cite{427} and the recent book \cite{akimov2021two}. A summary of novel electron and photon readout concepts for noble liquid detectors is presented, together with some newly proposed ones \cite{detlab_79}.

In this section, we present the performances of \acrshort{THGEM}-based detectors operated in cryogenic conditions. Quantitative comparison of similar properties measured using different experimental setups should be made with care. The results could be very sensitive to the operation conditions, such as small differences in temperature, pressure, gas purity, etc.. The methodology used, i.e., gain stabilization time, irradiation source used, readout electronics, and others could also affect the results (see Table \ref{tab: CRAD} and \ref{tab: GPM} for \acrshort{CRAD}/\acrshort{LEM} and \acrshort{GPM}, respectively). 

\subsubsection{Dual-phase noble-liquid TPCs}
\label{ref:DoalPhaseNobleLiquidTPC}
\subsubsection*{Charge readout}
 
\acrshort{THGEM} configurations operated in the vapor phase of noble liquids are known as CRiogenic Avalanche Detectors (\acrshort{CRAD}) \cite{3} or Large Electron Multipliers (\acrshort{LEM}) \cite{431}. The \acrshort{CRAD}/\acrshort{LEM} scheme is depicted in Figure \ref{fig:CRAD}.
The operation of gaseous detectors in dense cryogenic noble gases is mainly hindered by electrical instabilities caused by the high fields required to obtain charge multiplication and by photon feedback effects. The latter is mitigated by exploiting multipliers of closed geometry, among which \acrshort{THGEM}-based configurations have exhibited excellent performances. 

Studies were conducted primarily in Ar. As part of the \acrshort{CRAD} R\&D \cite{3}, various configurations were characterized with an $\mathrm{^{241}Am}$ source emitting 60 keV photons and 5.49 MeV alpha particles. Operated at a liquid temperature of 84 K (the vapor temperature at the \acrshort{THGEM} proximity was not reported) and detecting a dense cloud of $\sim$1000 \acrshort{PE}s escaping recombination from the gamma photons' conversion in the liquid, charge gains of the order of 20 and 2000 were measured with 0.4 mm thick FR4 $\mathrm{25 \times 25 mm^2}$ 1THGEM and 2THGEM configurations, respectively. For the 2THGEM configuration, the energy resolution was approximately 20\% \acrshort{FWHM}. The dependency of the gain and energy resolutions on time was not reported. A Kevlar-2THGEM system reached a gain of 6000 but with a degraded energy resolution \cite{3}. Using $\mathrm{10\times10 ~cm^2}$ electrodes, a 2RETGEM \cite{156} reached a gain value of several hundred, whereas a double Polyimide-\acrshort{THGEM} was limited to a gain of $\sim$30 \cite{414}. The different performances could stem from the different substrate materials and geometries (electrode thickness, hole diameter) or due to vapor condensation within the holes \cite{414}. An effective gain of the order of 5000 was reached using a hybrid 2THGEM+\acrshort{GEM} structure \cite{414}. 
A gain of 100 was reported with a PTFE-\acrshort{THGEM}, operated at 117 K \cite{166}. Using PTFE-2THGEM in 99 K, a maximal achievable gain of 1500 was reported \cite{447}. 

In \cite{420,490,423} a 2THGEM \acrshort{CRAD} is used to measure the electrons produced by 2.45 MeV neutrons recoiling in \acrshort{LAr}. This kind of calibration can be useful for rare event searches (see Section \ref{sec:TPCforRareEvents}).

In \cite{3}, the \acrshort{CRAD} was also studied as a \acrshort{UV}-photon detector. Using the copper as a photon converter, an avalanche induced by tens of photoelectrons could be detected with the 2THGEM configuration. 

\acrshort{CRAD} were also investigated in Xe. 1THGEM and 2THGEM detector configurations were operated in cold, 167 K, gaseous Xe at 1 atm \cite{57}. Using x-ray photons in the range of 15-40 \kev, an average of $\sim$1500 \acrshort{PE}s reached the multiplication region. A maximal achievable gain of 600 was recorded with the 2THGEM configuration. A similar performance was measured in Xe vapor \cite{57}.  

Within the \acrshort{LEM} R\&D project \cite{431}, cosmic muon tracks were used to characterize various detector configurations in a dual-phase Ar \acrshort{TPC}. Operated at 87 K, approximately $\mathrm{5\times 10^4}$ electrons/cm were induced by a muon in the \acrshort{LAr}. Using a $\mathrm{100 \times 100 ~mm^2}$ 1 mm-thick segmented electrode, an effective charge gain of $\sim$30 was reported after \acrshort{THGEM} substrate charge-up \cite{432,443}. An effective gain of $\sim$90 was reported in \cite{459}, under similar conditions. 

Studies comparing the dependency of the performance on different \acrshort{THGEM} parameters (thickness, hole diameter, hole patterns, rim size) are reported in \cite{443}. An induction field of 5 \kvcm{} was set, most probably extending the avalanche formation to outside the holes. A maximal stable effective gain in the range 20-30 was measured with 0.6-1 mm thick electrodes perforated with a 0.5 mm-diameter hole with 80 \um{} rim after charging-up. In all configurations, the gain after charging-up was $\sim$3 times lower than the original value, similar to the observation made at standard temperature and pressure (see Section \ref{sec:THGEM}). The \acrshort{LEM} performance was simulated in \cite{442}. Studies focused on gain evaluation, \acrshort{IBF} and \acrshort{EL} light yield.

Due to unavoidable defects in the \acrshort{PCB}, the probability of electrical instabilities was related to the multiplier surface area. As a result, lower maximal gains could be reached with larger electrodes. A factor of two loss in the gain was reported in the transition from a $\mathrm{25 \times 25 ~mm^2}$ to a $\mathrm{100 \times 100 ~mm^2}$ 2THGEM configuration \cite{414}. A similar drop was also reported with 1THGEM in the transition from $\mathrm{100 \times 100 ~mm^2}$ to $\mathrm{400 \times 760 ~mm^2}$ large electrodes \cite{}. 

\subsubsection*{Optical readout}

The optical readout scheme under cryogenic conditions was studied in \cite{239,412,413,415,418,422,424,425,435,385,50,449,416,460}. In this scheme, \acrshort{EL} \acrshort{UV} photons are emitted in the avalanche process in the holes \cite{239}. In some cases, \acrshort{IR} photons are emitted as well \cite{50,384,484}. Imaging is performed with various photosensors, including cameras. Thick electrodes ($\sim$1.5 mm) are potentially preferable for having a longer electron path in the high-field region, resulting in a higher light yield \cite{239}.  

Most of the measurements were performed in Ar \acrshort{TPC}s \cite{50,413,435,385} and some in gaseous Xe \cite{449}. Operated in the \acrshort{IR} range \cite{416,460}, a photon yield of the order of 10 photoelectrons was recorded in Ar over 4$\mathrm{\pi}$ per unit gain and \kev{} of deposited energy \cite{50}. The position resolution in the mm range was demonstrated \cite{413,435}. When combining proportional \acrshort{EL} in the drift gap with avalanche \acrshort{EL} (at a gain of the order of 40) in the \acrshort{THGEM} holes, 0.7 detected photons/electron, and a spatial resolution of 26 mm/$\mathrm{\sqrt(N_{electrons})}$ were recorded in Ar \cite{435}. In the \acrshort{UV} range in Ar, a light yield of $\sim$200 photons/electron was estimated using a 65 mm diameter electrode \cite{239}. A large $\mathrm{500 \times 500 ~mm^2}$ \acrshort{THGEM} was operated in Ar over a wide range of fields, from the linear regime to the beginning of the exponential one (24 \kvcm), resulting in 10-fold larger \acrshort{EL}-photon yield \cite{385}.  

\subsubsection{Single-phase noble liquid TPCs}
\label{ref:SinglePhaseNobleLiquidTPC}

Charge and light multiplications in noble liquids require a very high field (of the order of hundreds of kV/cm) for \acrshort{EL} and an even higher field for charge multiplication (see \cite{Aprile_2014,DOKE198287} in \acrshort{LXe} and references therein), which are beyond the reach of most experiments. However, based on early observations \cite{239,detlab_28}, it was shown that both multiplications can occur under low field values suggesting uncontrolled spontaneous gas bubbles trapped below the perforated electrode. The phenomenon led to the development of bubble-assisted liquid hole multiplier (\acrshort{LHM}) \cite{detlab_45,detlab_43,detlab_47,detlab_57}.

The \acrshort{LHM} can be considered as a “local dual-phase multiplier” operating in a single-phase medium. Shown in Figure \ref{fig:LHM}, it consists of a perforated electrode (\acrshort{GEM} or \acrshort{THGEM}) immersed in the noble liquid with a vapor bubble trapped underneath. \acrshort{PE}s deposited in the liquid drift into the multiplier holes, traverse the liquid-to-gas interface, and emit \acrshort{EL} under high field across the bubble.  The light is measured using photosensors (e.g., \acrshort{PMT} or \acrshort{SiPM}). Coating the \acrshort{THGEM}/\acrshort{GEM} electrode with a photoconverter (e.g., CsI) allows the detection of \acrshort{UV} photons through photoelectrons collected into the holes and transferred into the bubble.

Studies were performed in \acrshort{LXe} \cite{detlab_52,detlab_57} with \acrshort{THGEM}, \acrshort{GEM}, and double conical GEM electrodes. Approximately 7000 \acrshort{PE}s were induced in the \acrshort{LXe} by an $^{241}$Am alpha source. The best performance was obtained with a single conical GEM, attributed to deeper penetration of the bubble into the high field region (see \cite{detlab_57} for a detailed summary). Using a 0.4 mm-thick \acrshort{THGEM} electrode perforated with 0.3 mm diameter holes, approximately 200 photons/electron were measured with an energy resolution of $\sim$6\% \cite{detlab_57} and a position resolution of the order of 200 \um{} \acrshort{RMS} \cite{detlab_60}. Enhanced \acrshort{EL} could be obtained by increasing the transfer field across the bubble at the expense of some degradation in energy \cite{detlab_57} and position \cite{detlab_60} resolutions. However, the low (3-5\%) \acrshort{PDE} reached so far, suggests upon electron losses during their transfer through the liquid-gas interface. The \acrshort{LHM} concept was also demonstrated in \acrshort{LAr} \cite{detlab_58}. The dynamics of a bubble in a \acrshort{THGEM}-like hole was studied in \cite{detlab_77}.

\subsubsection{Cryogenic GPM}
\label{sec:CryoGPM}

Cryogenic \acrshort{GPM}s have been developed in the context of rare event experiments (see Section \ref{sec:TPCforRareEvents}) as a potentially cost-effective, large-area photon-imaging solution for substituting \acrshort{PMT} and \acrshort{SiPM} arrays.  In a cryogenic \acrshort{GPM}, the gaseous detector is separated from the noble-liquid scintillator by a \acrshort{UV} transparent window. This allows operating the detector in quenched gas mixtures, reducing photon-feedback effects. The \acrshort{GPM} scheme is illustrated in Figure \ref{fig:LHM}.

A 2THGEM \acrshort{GPM} was operated in $\mathrm{Ne/CH_4}$ (95:5) and $\mathrm{Ne/CF_4}$ (95:5) at 
173 K \cite{detlab_23}. A gain of the order of $\mathrm{10^4}$ was measured with 5.9 keV x-rays \cite{detlab_23} and \acrshort{UV} photons \cite{detlab_24}. In a similar experiment, a hybrid structure of \acrshort{THGEM}+double grid+\acrshort{MM} showed an unprecedented gain of 10$^6$ at liquid Xe temperature while measuring charge from 5.9 keV x-rays \cite{detlab_27}.

The operation of a 10 cm in diameter 3THGEM \acrshort{GPM} was demonstrated at 180 K $\mathrm{Ne/CH_4}$ (5\%, 10\%, 20\%) at 0.7 bar, corresponding to a room temperature density of 1.1 bar \cite{detlab_44}. The \acrshort{GPM} coupled with an \acrshort{LXe} could measure both primary and secondary scintillation from an $^{241}$Am alpha-gamma source. This type of operation aimed to test the large dynamic range, e.g., required in dark matter experiments.  The maximum gain for single photoelectrons was $\mathrm{8\times 10^5}$ for Ne/CH$_4$ (95:5) and $\mathrm{3\times 10^5}$ for Ne/CH$_4$ (80:20) in an asymmetric bias scheme with a higher voltage on the first THGEM layer to get high photoelectron extraction efficiency from its aurface-coated CsI photocathode. The maximum achievable gain was reduced by a factor of 2-3 when the alpha source was turned on. The detector operated stably for two months in sealed mode. An excellent time resolution of $\sim$1 ns was obtained for the secondary scintillation signal yielding $\sim$200 photoelectrons per alpha particle. A considerable degradation of this performance is expected when measuring single photoelectrons, mainly due to the worse signal-to-noise ratio. A 9\% \acrshort{RMS} energy resolution was achieved. A 2THGEM \acrshort{GPM} detector was operated also in $\mathrm{He/CH_4}$ (92.5:7.5) \cite{detlab_23}.

Detectors with resistive electrodes can be used to effectively quench electrical instabilities at high gains (See Section \ref{Sec:SurfaceCoatings} for details regarding resistive materials for cryogenic temperatures). In \cite{detlab_59},  a $\mathrm{3 \times 3 ~cm^2}$ \acrshort{RPWELL} \acrshort{GPM} was operated both as a single element and in a cascade configuration with a THGEM. 
Gain values of $\mathrm{10^4}$ and $\mathrm{10^5}$ were measured with 5.9 \kev{} x-rays and single \acrshort{UV} photons, respectively, in $\mathrm{Ne/CH_4}$ (95:5)
at 163 K. Some small electrical instabilities were observed above these values. 
Energy resolutions of the order of $\sim$20\% were measured with the single and double-stage configurations. The double structure granted a 10-fold higher maximum achievable gain with respect to the single \acrshort{RPWELL}, both for x-rays and \acrshort{UV} photons. The energy resolution for x-rays improved by 5\% with respect to the single \acrshort{RPWELL} case. The single photon spectrum at the highest gain presented a well-defined Polya shape, suggesting high detection efficiency. No degradation in performance was observed when coating the \acrshort{THGEM}-top electrode with CsI.

\section{Applications}
\label{sec:Applications}

\acrshort{THGEM} and its derivatives derivatives have been investigated as potential multipliers for a large variety of applications. In this section, we summarize some of the leading efforts. 

\subsection{RICH detectors}
\label{sec:RICH}

Ring Imaging CHerenkov (\acrshort{RICH}) counters with solid or gaseous radiators are core components in experiments requiring particle identification (\acrshort{PID}) of protons, pions, and kaons with momenta up to $\sim$50 GeV. Examples include \acrshort{COMPASS} \cite{ABBON2007455}, \acrshort{ALICE} \cite{Vercellin:2008zz}, the Hadron Blind detector of Phenix \cite{FRAENKEL2005466}, the foreseen detector for the future Electron-Ion Collider (\acrshort{EIC}) \cite{accardi2016electron}, and future super tau-charm facility (\acrshort{STCF}) \cite{487,495}.
 
Cherenkov radiation is emitted by charged particles traversing through a transparent dielectric radiator at a velocity greater than the speed of light in the medium. A few tens of photons are typically emitted per particle, at an angle relative to its velocity, creating a ring-shaped image on the photo-sensor. High-quality ring imaging allows for precise velocity measurements. It relies on photon detectors with high \acrshort{PDE} and adequate position resolutions. \acrshort{PID} is obtained by combining the velocity measurement with an independent momentum measurement to extract the particle mass.   

Having low material budget and being cost-effective make gas-avalanche-\acrshort{RICH} photo-sensors an attractive solution for applications requiring large area coverage. Various photon detector technologies have been developed for gas-avalanche-\acrshort{RICH} photo-sensors over the years (see \cite{TESSAROTTO2018278} and references therein); the ones employed in recent experiments are based on \acrshort{MPGD} technologies with (unlike wire chambers, e.g., \cite{Vercellin:2008zz}) a closed geometry, e.g., cascaded \acrshort{GEM} \cite{FRAENKEL2005466} and \acrshort{THGEM} or hybrid configurations \cite{87}.  When coated with a photocathode, these operate stably with high \acrshort{PDE} and low \acrshort{IBF}; the latter reduces the aging effects of the photocathode (see, for example, \cite{Milov:2006ew}).  In what follows, we focus on \acrshort{THGEM}-based \acrshort{RICH} counters.

\subsubsection{COMPASS RICH} 
\label{subsubsec: COMPASS RICH}

The COmmon Muon Proton Apparatus for Structure and Spectroscopy (\acrshort{COMPASS}) experiment at the CERN Super Proton Synchrotron (\acrshort{SPS}) aims at studying the hadron structure and spectroscopy with high-intensity muon and hadron beams \cite{ABBON2007455}. 

The \acrshort{COMPASS} \acrshort{RICH} counter \cite{Albrecht:2002ty} provides pion-kaon separation within the momentum range of 3-55 GeV over $\pm$200 mrad angular acceptance. It consists of a 3 m long $\mathrm{C_4 F_{10}}$ gaseous radiator and 21 $\mathrm{m^2}$ \acrshort{VUV} spherical mirrors for focusing the photons onto a 5.5 $\mathrm{m^2}$ detection surface sensitive to single photons. In its phase I, the photon detectors incorporated \acrshort{MWPC} with reflective CsI photocathodes \cite{Tessarotto_2014}. Despite their good performance, \acrshort{MWPC}s have limitations in terms of spatial resolution, maximum achievable gain ($\sim$$\mathrm{10^4}$), time response, rate capability, and CsI photocathode aging due to the high \acrshort{IBF}. 

To overcome the high particle flux of the experiment's central region, the central \acrshort{MWPC}-based photon detectors were replaced with Multi Anode Photo-Multiplier Tubes (\acrshort{MAPMT}s) coupled to individual fused silica lens telescopes. In parallel, an extensive R\&D program aiming to develop \acrshort{MPGD}-based photon detectors \cite{detlab_1} was established. One proposed technology was a hybrid configuration combining \acrshort{THGEM} and an asymmetric \acrshort{MWPC} \cite{120} operated in $\mathrm{CH_4}$. Gains as high as $\mathrm{10^5}$ were reached with reduced $\sim$$\mathrm{25\%}$ \acrshort{IBF}.  Photon detectors based on cascades \acrshort{THGEM} with reflective CsI phocathodes \cite{288} were characterized with 3THGEM photo-sensors and performances compared with a CsI-\acrshort{MWPC} ones, in \nech, \necf, \ch, and \cf{}  \cite{detlab_18}. By staggering the holes in the three layers and optimizing the transfer field configuration, an \acrshort{IBF} of 3$\%$ at a gain of $\sim$$\mathrm{2 \times 10 ^5}$ per single photoelectron could be reached, albeit at relatively high absolute voltages \cite{19}. 

\begin{figure}[htbp]
    \centering
    \subfloat[]{
        \includegraphics[width=0.60\textwidth]{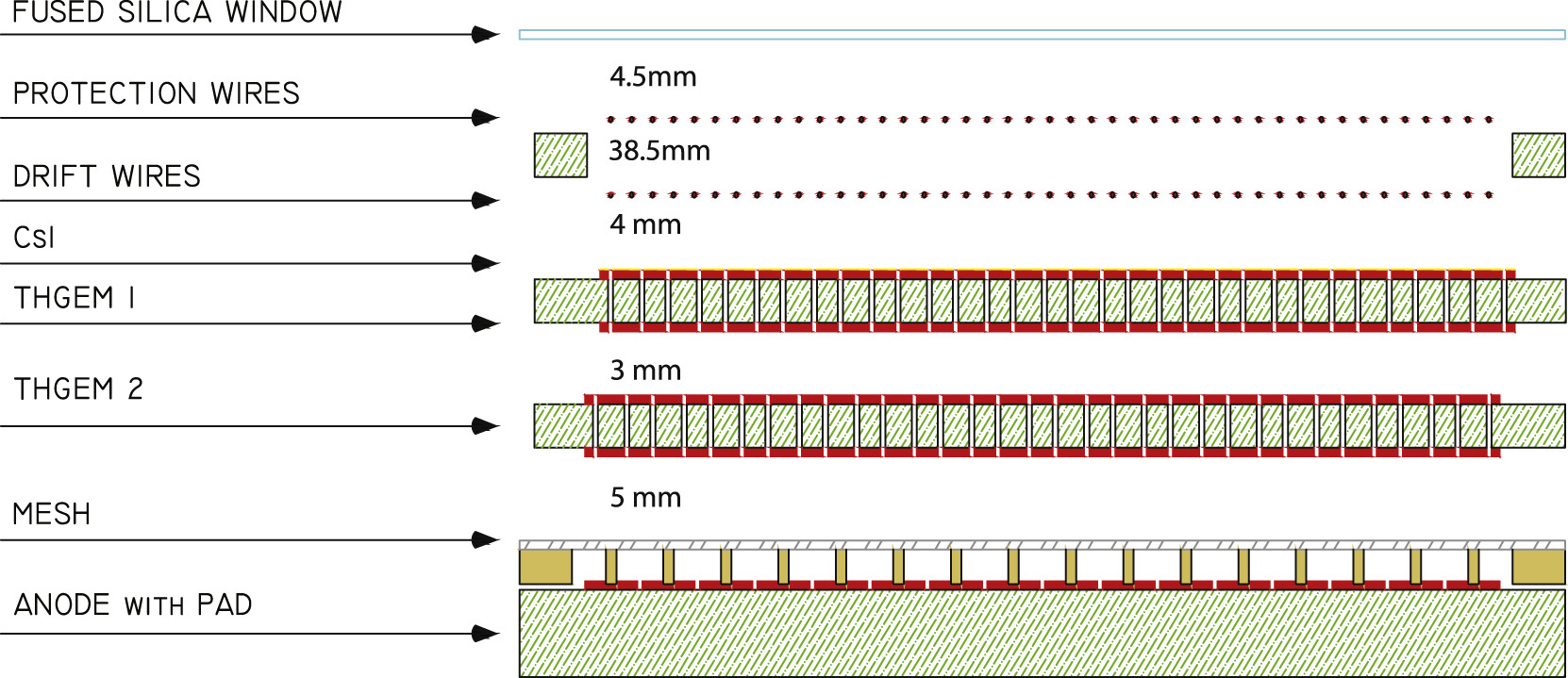}
        \label{fig:COMPASSREACHSchema}
    }
    \subfloat[]{
        \includegraphics[width=0.35\textwidth]{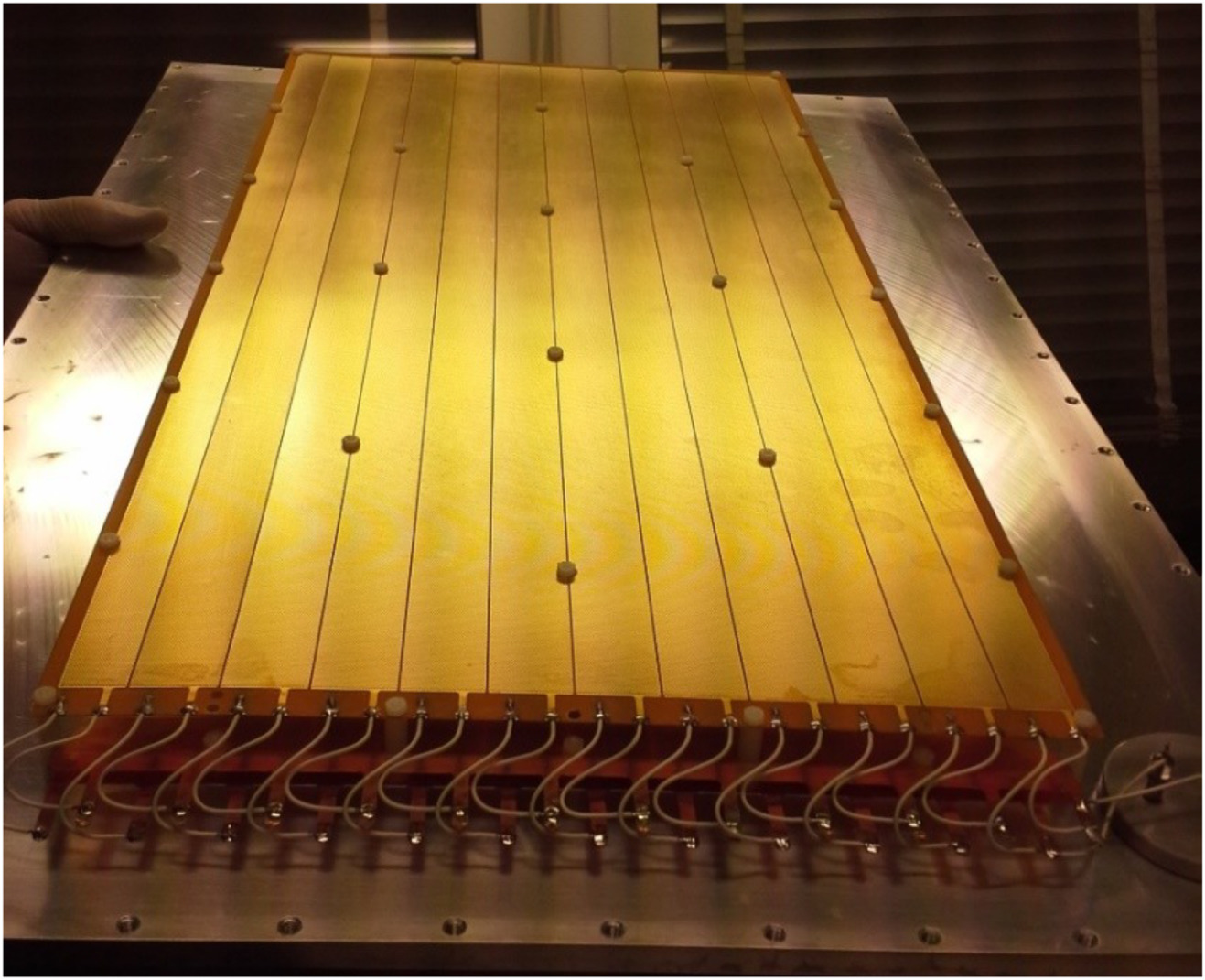}
        \label{fig:COMPASSTHGEM}
    }
    \caption{\protect\subref{fig:COMPASSREACHSchema} Schematics of the \acrshort{COMPASS} \acrshort{THGEM}-based \acrshort{RICH} detector. \protect\subref{fig:COMPASSTHGEM} A \acrshort{COMPASS} CsI-coated \acrshort{THGEM} electrode showing the individual electrode segments. The figures were obtained from \cite{87}.}
\label{fig:COMPASSRICH}
\end{figure}

A schematic description of \acrshort{COMPASS} hybrid \acrshort{MPGD} detector is provided in Figure \ref{fig:COMPASSREACHSchema}, together with a picture of the CsI coated \acrshort{THGEM} electrode \ref{fig:COMPASSTHGEM}. It consists of two \acrshort{THGEM} electrodes proceeded by a \acrshort{MM}. The latter was shown to have an intrinsically low \acrshort{IBF} \cite{Bhattacharya_2015,COLAS2004226}. The two 470 \um{} thick \acrshort{THGEM} electrodes have 0.4 mm diameter holes drilled with a hexagonal pattern with a 0.8 mm pitch. The first \acrshort{THGEM} electrode is coated with a thin CsI reflective photocathode. No rims are etched around the holes to maximize the photon conversion surface and minimize charging-up effects. To reduce the intensity of occasional discharges and provide the possibility of tuning the voltage of unstable areas, the \acrshort{THGEM}-top and bottom electrodes are segmented and electrically decoupled. Larger-diameter holes, 500 \um{}, along the external borders prevent an increased electric field in the periphery.

A total sensitive area of approximately 1.5 $\mathrm{m^2}$ was assembled by merging module pairs of $\mathrm{300\times 600 ~mm^2}$, for better thickness-uniformity control \cite{105}. The detector was operated with $\mathrm{Ar/CH_4}$ (50:50). The relatively large $\mathrm{CH_4}$ fraction is favorable to achieve a good photoelectron extraction from the photocathode: it both reduces photoelectron backscattering \cite{detlab_21} and it allows for high enough dipole electric field at the THGEM surface. The drift voltage was optimized to reach, at the same time, low \acrshort{IBF} \cite{38}. In the optimal field configuration, with an \acrshort{IBF} equals 3\%, the gain values of the three layers were estimated to be $\sim$13, 9, and 120 for THGEM1 THGEM2 and the \acrshort{MM}, respectively \cite{87}.

\paragraph{Performance} 

The \acrshort{COMPASS} \acrshort{RICH} hybrid \acrshort{MPGD} detector has been operating successfully since 2016. The detectors were successfully commissioned, and they have been operating ever since. A comprehensive description of the detectors' setup and performance can be found in \cite{93,86,37,66}.

During 12 months of operation at the nominal beam rates, no \acrshort{HV} trips have been recorded. The recovery time after an occasional discharge was 10 seconds; the discharge rate was typically 1/h/detector, imposing a negligible dead time on the measurement. 

\begin{figure}[htbp]
    \centering
    \includegraphics[scale=0.4]{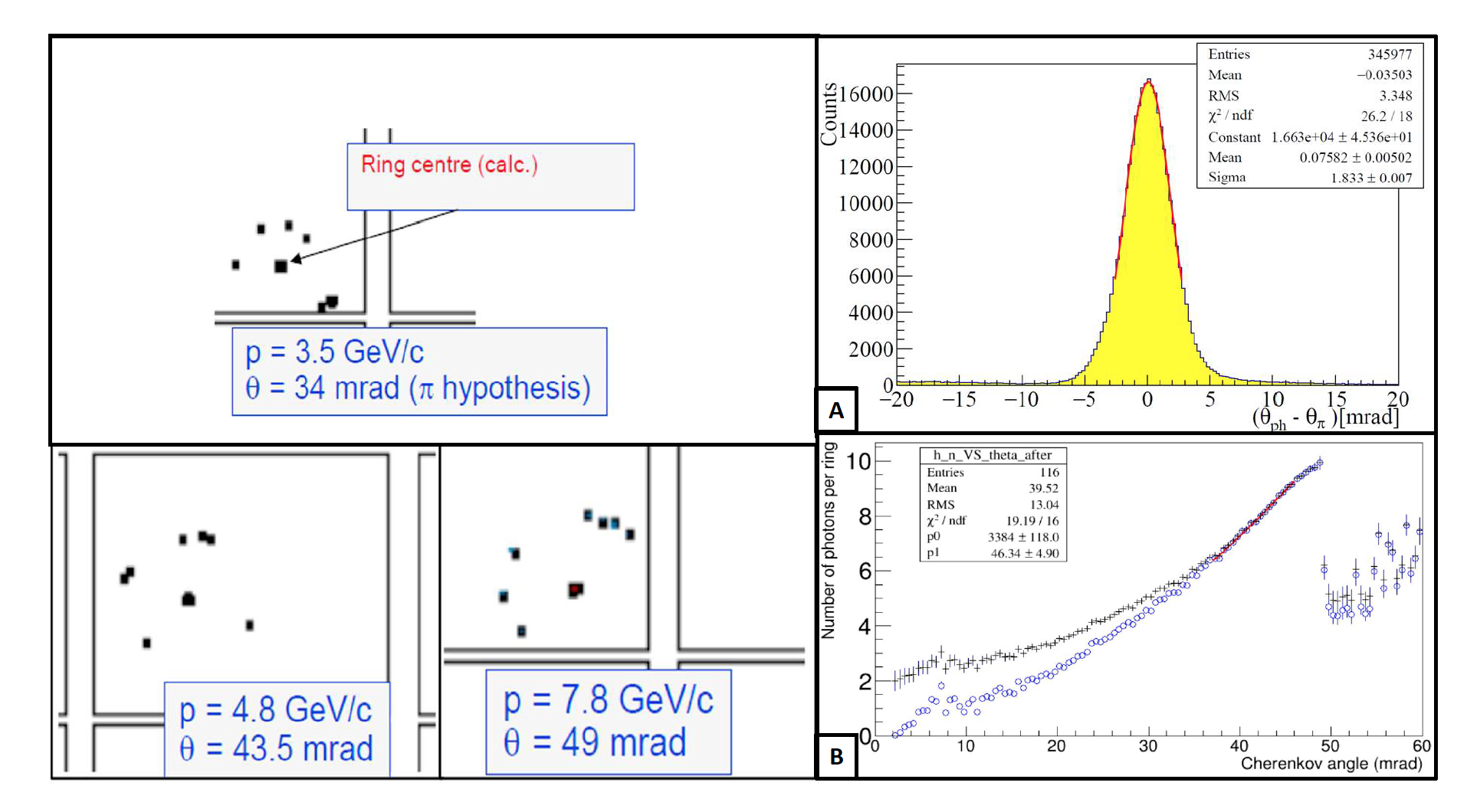}
    \caption{Left: Typical Ring images measured with the \acrshort{COMPASS} \acrshort{RICH} detector. The center of the expected ring patterns is obtained from the reconstructed particle trajectories; the particle momentum and the expected Cherenkov angle are also shown. Upper right: the angular resolution of the \acrshort{RICH} detector. Bottom right: number of detected photon-electrons as a function of the Cherenkov angle. The figure is taken from \cite{66}.}
    \label{fig:CopassRing}
\end{figure}

A typical ring image is illustrated in Figure \ref{fig:CopassRing}; 11 photoelectrons per particle were measured on average. The single photo-electron gain is of the order of $1.4 \times 10^4$. A \acrshort{PDE} greater than 80\% was estimated from the gain and readout electronics threshold. The Cherenkov angle was measured with a precision of 1.7 to 1.8 mrad \acrshort{RMS}, in full agreement with the expected values. 

\subsubsection{HMPID in the electron-ion collider}
\label{subsubsec: HMPID at the Electron-Ion collider}

The future electron-ion collider (\acrshort{EIC}) BNL aims at precision measurements of the quark-gluon plasma properties with unprecedented luminosity and energy range. Beams of polarized electrons will collide with that of either polarized nucleons or nuclei. \acrshort{PID} is crucial for the \acrshort{EIC} experiments, in which the \acrshort{RICH} systems will play a key role \cite{He:2020oyy}. In the hadron-going direction (i.e., proton or ion beam), the final-state hadrons can have momenta up to 50 GeV. \acrshort{PID} capability in this region with continuous momentum coverage will be achieved by a dual radiator (gas+aerogel) \acrshort{RICH} detector (d-RICH) \cite{cisbani2020ai}.  

The d-RICH radiators are planned to be approximately half the length of the \acrshort{COMPASS} ones, thus posing stringent requirements on the photon detector; at a shorter length, the number of Cherenkov photons produced is smaller, requiring higher \acrshort{PDE} for maximizing the average number of photons per ring - for precise ring reconstruction. Furthermore, the shorter radiator length imposing a shorter focal length, requires a better spatial resolution for separating single photons. 

A windowless configuration with $\mathrm{CF_4}$ gas used for a dual purpose, as radiating material and amplification medium \cite{FRAENKEL2005466}, was proposed for optimizing the number of Cherenkov photons \cite{7349014}. The low refractive index of the $\mathrm{CF_4}$ allows obtaining far \acrshort{UV} photons ($\sim$120 nm), where the Cherenkov photon yield increases. The concept was demonstrated in a d-RICH-like prototype using a quintuple \acrshort{GEM}. Test beam results showed efficient separation of proton-kaon-pion up to momenta of 32 GeV.  Alternatively, a \acrshort{COMPASS} \acrshort{RICH}-like photon detector with an improved spatial resolution (obtained by employing smaller readout pads) was also proposed \cite{236,77}. Complete Cherenkov rings could be reconstructed in the area of a single detector module operated in $\mathrm{Ar/CH_4}$ (50:50) and pure $\mathrm{CH_4}$. The development phase is still ongoing.

\subsubsection{ALICE Upgrade}
\label{subsubsec:AliceUpgrade}

The \acrshort{ALICE} (A Large Ion Collider Experiment) at CERN is designed to detect heavy ion collisions, aiming to study quark-gluon plasma physics. \acrshort{PID} in \acrshort{ALICE} is provided by combining velocity measurements in the High Momentum Particle Identification Detector (\acrshort{HMPID}) \cite{MOLNAR200827} with dedicated energy loss and time-of-flight measurements. The \acrshort{HMPID} detector has been designed to identify p-kaon and kaon-p on a track-by-track basis for particles with energies of up to 3 GeV and 5 GeV, respectively. 
It is the largest \acrshort{RICH} detector in High Energy Physics using CsI photosensitive elements. The photon detectors are based on \acrshort{MWPC}, operated with $\mathrm{CH_4}$, covering an area of 11 $\mathrm{m^2}$ in total. 

The very high momentum PID (\acrshort{VHMPID}) \acrshort{RICH} counter could extend the track-by-track \acrshort{PID} capabilities of \acrshort{ALICE} to energies up to 30 GeV and enhance its discovery reach \cite{334,472}. Several alternative \acrshort{RICH} counter schemes were proposed in this context, some employing \acrshort{THGEM}-based photon counters: 1) a threshold imaging Cherenkov detector with \acrshort{RETGEM} coated by CsI photocathodes as photon counters \cite{216}. 2) An \acrshort{HMPID}-like detector with $\mathrm{C_4F_{10}}$ as the gaseous radiator and CsI-coated 3THGEM or \acrshort{RETGEM} photon counters operated in $\mathrm{Ne/CH_4}$ (90:10) or $\mathrm{Ne/CF_4}$ (90:10) \cite{14}. 3) A windowless $\mathrm{CF_4}$-based radiator with a 1THGEM operated with a weak reverse transfer field for hadron blockage \cite{detlab_18}.

\subsection{Noble liquid TPCs}
\label{sec:TPCforRareEvents}

In recent years, massive noble-liquid (\acrshort{LAr} and \acrshort{LXe}) \acrshort{TPC}s have become a tool of choice in experiments targeting the detection of weakly interacting particles \cite{Aprile:2008bga,Cavanna:2018yfk}. Some examples are neutrino detectors, such as ICARUS \cite{ICARUS:2004wqc}, MicroBooNE \cite{MicroBooNE:2016pwy}, and \acrshort{DUNE} \cite{DUNE:2018tke}, and dark matter detectors, such as XENON \cite{XENON:2017lvq}, LUX \cite{LUX:2012kmp}, and \acrshort{ArDM} \cite{ArDM:2010rnc}. 

\acrshort{LAr} and \acrshort{LXe} are excellent target materials due to their high density; they allow for conceiving massive target detectors required for efficient detection of low-interaction incident particles. Moreover, they provide high radiation-induced scintillation and ionization yields; measuring both permits effective discrimination between signals of interest and various background sources. In single-phase \acrshort{TPC}s \cite{Aprile:2008bga}, radiation-induced charges and scintillation photons in the liquid are recorded respectively with charge-sensing electrodes (typically wire grids and recently also with \acrshort{THGEM}-like multi-layer electrodes \cite{review_17}) or photon detectors (e.g., photomultipliers, avalanche photodiodes, etc.). In dual-phase (liquid and vapor) \acrshort{TPC}s \cite{Aprile:2008bga}, in addition to the prompt scintillation signal in the liquid, electrons extracted into the vapor phase are detected either after moderate charge-avalanche multiplication or through \acrshort{EL}. 

Being robust and self-supported and having high gain and potentially low photon feedback has made \acrshort{THGEM}- (a.k.a. \acrshort{LEM} in this context) based readout elements an attractive solution for large area noble liquid \acrshort{TPC}s. A \acrshort{THGEM}-based readout for dual-phase Ar \acrshort{TPC} was first considered for an Ar Dark Matter (\acrshort{ArDM}) direct search experiment \cite{441,463,457} and for the LBNO-GLACIER neutrino experiment \cite{438,464,477}. Since then, several \acrshort{THGEM} related R\&D projects have been carried out: 1) Charge readout in dual- and single-phase TPC modules was studied for the \acrshort{DUNE} experiment (Section \ref{sec:DUNE}). 2) Light readout with \acrshort{GPM}s based on low radioactivity PTFE-\acrshort{THGEM} was considered for the \acrshort{LAr} veto volume in the CDEX experiment \cite{166,447}. 3) Charge and light readouts in \acrshort{LXe} were studied with \acrshort{THGEM}-based \acrshort{LHM} in \acrshort{LXe} \cite{detlab_57} in the context of the \acrshort{DARWIN} experiment (Section \ref{sec:DARWIN}). 

\acrshort{THGEM} electrodes were also studied for \acrshort{EL} production  in their holes for optical readout  \cite{239,329,53,451,319,448,446}, leading to the \acrshort{ARIADNE} concept (Section \ref{sec:ARIADNE}), considered also in the context of the \acrshort{DUNE} experiment.

\subsubsection{Charge readout in DUNE}
\label{sec:DUNE}

The Deep Underground Neutrino Experiment (\acrshort{DUNE}) \cite{DUNE} is designed as a long-baseline neutrino beam experiment. It will consist of two neutrino detectors placed in the beam. One detector will record particle interactions near the beam source at the Fermi National Accelerator Laboratory in Batavia, Illinois. A second, much larger detector will be installed more than a kilometer underground at the Sanford Underground Research Laboratory in Lead, South Dakota, 1,300 kilometers downstream of the source. These detectors will enable searching for new subatomic phenomena and potentially transform the understanding of neutrinos and their role in the universe.

\begin{figure}[htbp]
    \centering
    \subfloat[]{
        \includegraphics[width=0.7\textwidth]{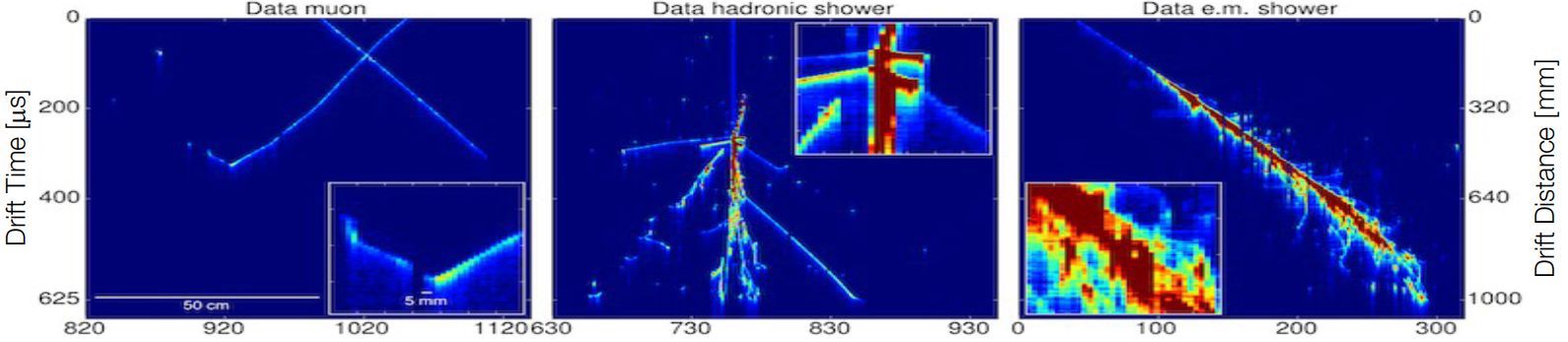}
        \label{fig:WA105muon}
    }\\
    \subfloat[]{
        \includegraphics[width=0.6\textwidth]{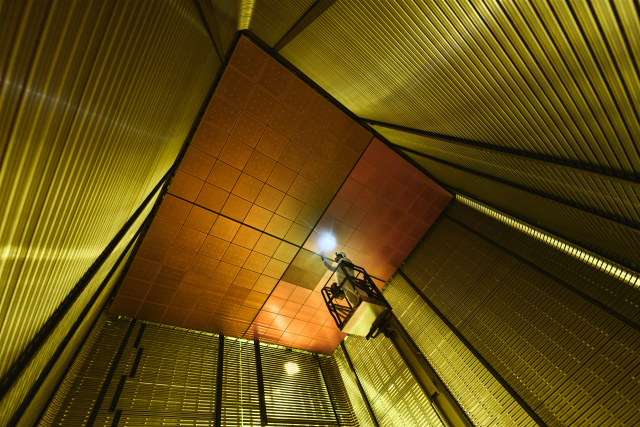}
        \label{fig:LEMatDUNE}
    }
    \caption{\protect\subref{fig:WA105muon} An example of particle tracks recorded with the WA105 experiment. The figure was obtained from \cite{475}. \protect\subref{fig:LEMatDUNE} A $\mathrm{50 \times 50 cm^2}$ \acrshort{LEM}-based charge readout assembled at the proto-\acrshort{DUNE} experiment in CERN. Figure obtained from CERN repository.}
\label{fig:MuonInLArTPC}
\end{figure}

The \acrshort{DUNE} far detector will consist of four $\mathrm{14 \times 14 \times 62~ m^3}$ \acrshort{LAr} \acrshort{TPC} modules \cite{DUNE}. The first one will consist of a single-phase \acrshort{TPC} with wire readout planes. Aiming at charge multiplication for better signal to noise separation, a dual-phase \acrshort{TPC} with a \acrshort{LEM}-based charge-readout has been considered as a second module. Double- \cite{428,430} and single- \cite{431,432,459,443} $\mathrm{10 \times 10 ~cm^2}$ \acrshort{LEM}-based charge-readout elements were studied in the context of a 3 lt dual-phase \acrshort{TPC}. They demonstrated a stable gain of the order of 30 \cite{443}; this motivated the construction of a larger module, a $\mathrm{40 \times 76 ~cm^2}$ readout plane deployed in a 200 lt volume, exhibiting a slightly degraded performance \cite{434,433,477}.  These studies were followed by a demonstration in the WA105 setup; a $\mathrm{3 \times 1 \times 1 ~m^3}$ dual-phase \acrshort{LAr} \acrshort{TPC} with $\mathrm{50 \times 50 ~cm^2}$ \acrshort{LEM} electrodes. Examples of events recorded by the WA105 experiment are presented in Figure \ref{fig:WA105muon}. Measurements and simulation studies of the primary and secondary scintillation in WA105 were carried out in \cite{478}. The targeted \acrshort{LEM}-gain of 20 could not be reached due to technical limitations \cite{444,475}. Studies at proto-\acrshort{DUNE} - two large volumes of $\mathrm{8 \times 8 \times 8 ~m^3}$ \acrshort{TPC}s - were conducted at the CERN neutrino platform. A fully assembled \acrshort{LEM}-based charge readout is shown in Figure \ref{fig:LEMatDUNE}. The targeted \acrshort{LEM}-gain could also not be reached in this setup\footnote{Private communications}.

The dual-phase proto-\acrshort{DUNE} presented additional significant difficulties \cite{DUNE_2021}. The very high voltages (600 kV) required across the \acrshort{TPC} were hard to obtain. In addition, it was challenging to provide a stable gas-liquid interface essential for uniform electron transfer efficiency over a large surface, inducing even more instabilities in the detector. 

Given the substantial difficulty posed by the dual-phase concept, \acrshort{DUNE} is focusing on single-phase \acrshort{TPC}s. In line with this decision, segmented readout planes made of multi-layer \acrshort{THGEM}-like electrodes immersed in the liquid are being studied as an alternative to wires \cite{review_17,426,480}. Relative to wires, such electrodes can be industrially mass-produced at relatively low costs. They grant large flexibility in several aspects, such as optimizing the geometrical parameters (strip orientation and size, holes, etc.), modularity, shielding, signal readout scheme, and more \cite{review_17}. Further, the mechanical robustness greatly reduces the chances of failure and renders the structure easier to support. 

\subsubsection{ARIADNE}
\label{sec:ARIADNE}

The ARgon ImAging DetectioN chambEr (\acrshort{ARIADNE}) concept was proposed as an optical readout for \acrshort{THGEM}-based dual-phase Ar \acrshort{TPC}s \cite{454}. It exploits the emission of secondary scintillation \acrshort{EL} \acrshort{UV} photons within the \acrshort{THGEM} holes in the vapor phase. The \acrshort{UV} light is wavelength-shifted and recorded by highly-pixelated sensors. For details about the physics of scintillation-light production by \acrshort{THGEM}s, see \cite{akimov2021two,50,416}.  

Relative to a charge readout, the optical readout relies on a smaller number of readout channels, making it potentially simpler and cheaper when scaled to large volume \acrshort{TPC}s. The \acrshort{ARIADNE} \acrshort{TPC} has a fiducial volume of $\mathrm{54 \times 54 \times 80}$ $\mathrm{cm^3}$ instrumented with 16 \acrshort{THGEM} elements. Viewports on the top of the cryostat enable light readout by four Electron-Multiplying CCD or Timepix3 cameras. 

The \acrshort{THGEM} electrodes used in \acrshort{ARIADNE} are 1 mm thick, with an effective area of $\mathrm{53 \times 53 ~cm^2}$. A one-sided segmentation into 16 square regions has also been tested. The holes of 0.5 mm diameter have an 800 \um{} pitch and 50 \um{} rim. The charge induced on the segmented side is read out by charge-sensitive pre-amplifiers. \acrshort{ARIADNE} was tested at the CERN \acrshort{PS} test beam as well as with cosmic muons \cite{454}.  The \acrshort{THGEM}s were operated in a proportional \acrshort{EL} regime (no charge multiplication). A track resolution of 1 mm/pixel was measured, comparable to the electrons' transverse diffusion and well below the $\sim$$\mathrm{4 ~mm}$ diffusion foreseen for a 12 m drift in \acrshort{DUNE}. 

The 3D event reconstruction capabilities of \acrshort{ARIADNE} were demonstrated by employing Timepix3 ASIC \cite{385}. In this configuration, the \acrshort{THGEM} was operated in the linear (charge-drift) and exponential (some charge-avalanche multiplication) light production regimes, which allowed recording of high-quality muon \acrshort{MIP} and decay tracks. Following these results, preparations are currently being made for a larger volume \acrshort{TPC} at the CERN neutrino platform, instrumented with 4 Timepix3 cameras reading out a $\mathrm{2 \times 2 ~m^2}$ area. The large demonstrator will include newly developed glass-\acrshort{THGEM} electrodes \cite{476} which could be made of radio pure materials.

\subsubsection{DARWIN}
\label{sec:DARWIN}

The DARk matter WImp search with liquid xenoN (\acrshort{DARWIN}) observatory \cite{479} is foreseen as the next large volume \acrshort{LXe} experiment for direct dark matter and other searches. Several novel detector configurations were proposed over the years to tackle the challenges posed by such multi-ton experiments. Among them are the \acrshort{GEM} and \acrshort{THGEM}-based \acrshort{GPM} and \acrshort{LHM} (Section \ref{sec:CryoGPM} and references therein). The \acrshort{THGEM}-\acrshort{GPM} concept was demonstrated in \acrshort{LXe} on small-scale prototypes \cite{detlab_44,detlab_64}. It could provide better spatial resolution and be more cost-effective relative to state-of-the-art \acrshort{PMT}s. The \acrshort{LHM} concept provides controlled, dual-phase regions in a large single-phase volume, such that the difficulties attributed to the large liquid-gas interfaces might be avoided.

\subsubsection{Novel concepts}
\label{sec:appCrypTPCNovel}

In order to solve the substantial problem of liquid-to-vapor interface instabilities in large volume dual phase \acrshort{TPC}s \cite{DUNE_2021}, two novel charge and \acrshort{UV}-photons detection concepts have been recently proposed \cite{detlab_81}.

In the bubble-free \acrshort{LHM} (bf-\acrshort{LHM}), the gas bubble is replaced by a liquid-to-vapor interface located in between two perforated (e.g. \acrshort{THGEM}) electrodes. The bottom one, fully immersed in the liquid, has a CsI \acrshort{VUV} photocathode underneath; the top one is located in the vapor phase, with photo-sensors above it. Ionization electrons and photoelectrons emitted from the photocathode are focused into the holes of the immersed electrode, transferred through them, extracted into the gas volume, and induce fast \acrshort{EL} signals within the top \acrshort{THGEM} holes. The concept was validated in \acrshort{LXe}\footnote{\url{https://events.camk.edu.pl/event/47/contributions/377/attachments/126/281/LIDINE-2022}}. 
The concept of a Floating Hole Multiplier (\acrshort{FHM}) is similar to that of a \acrshort{LHM}, where a \acrshort{THGEM}) is freely floating on the surface of the liquid\footnote{\url{https://events.camk.edu.pl/event/47/contributions/379/attachments/122/275/LIDINE2022_Chepel_presentation\%20.pdf}}.

Some recent single-phase detector concepts encompassing \acrshort{THGEM}-like multipliers were proposed in \cite{detlab_81}. They rely on \acrshort{THGEM}-like electrodes immersed in the liquid phase and coated with \acrshort{VUV} photocathodes. High field can be created either by a \acrshort{THCOBRA} structure of by properly tailored nanostructured surfaces. The \acrshort{EL} (and possibly small charge multiplication) in liquid, would result in fast \acrshort{UV}-photon flashes - detected by nearby photo sensor arrays.

\subsection{Calorimetry}
\label{sec:AppCalorimetry}

Particle flow \cite{Thomson:2009rp} is a leading approach toward reaching the challenging jet energy resolution ($\frac{\sigma}{E}\leq 30\%/\sqrt{E[\text{GeV}]}$) required in future collider experiments \cite{Behnke:2013xla,CEPCStudyGroup:2018rmc,FCC:2018evy,Linssen:2012hp}. It is based on the observation that, on average, over 60\% of the particles in a jet are charged hadrons. Hence, their energy can be measured by the tracking system with higher precision than a traditional measurement based on the Hadronic Calorimeter (\acrshort{HCAL}). In this approach, only the energy of the neutral hadrons ($\sim$10\% of the jet energy) is measured in the \acrshort{HCAL}. Particle-flow calorimeters are designed to allow associating the energy deposits with individual particles, ignoring the ones deposited by charged particles. This requires high transverse and longitudinal granularity and, thus, many readout channels. In this respect, Digital and Semi-Digital Hadronic Calorimeters ((S)\acrshort{DHCAL}) \cite{Sefkow:2015hna} with a 1-2 bit Analog to Digital Converter readout are appealing; they offer a cost-effective solution for reading out a large number of channels.

A typical configuration consists of alternating layers of absorbers, where the hadronic shower develops, and sampling elements with a pad-readout. The absorbers' thickness (several cm) and the pads' size (order of cm$^2$) define the longitudinal and transverse granularity, respectively. The energy of a single hadron is reconstructed from the number and pattern of all fired pads (hits). The performance of a sampling element is characterized in terms of \acrshort{MIP} detection efficiency and average pad-multiplicity, i.e., the number of pads firing per impinging \acrshort{MIP}. A low detection efficiency reduces the total number of hits, and large average pad-multiplicities increase the probability of overestimating the number of hits fired by a single particle. Both parameters may degrade the energy resolution. 

Although \acrshort{RPC} is the most studied technology for \acrshort{DHCAL}, \acrshort{MPGD} sampling elements have some advantages; they demonstrate lower average pad-multiplicities for similar \acrshort{MIP} detection efficiencies and are operated in environmentally-friendly gas mixtures. The various technologies that have been studied in test beams with large prototypes, include 1 $\mathrm{m^2}$ glass-\acrshort{RPC} \cite{CALICE:2019rct}, 1 $\mathrm{m^2}$ \acrshort{MM} \cite{1sqmMM, Adloff:2014qea}, $\mathrm{30 \times 30}$ $\mathrm{cm^2}$ double \acrshort{GEM} \cite{Hong:2020hwx}, $\mathrm{50 \times 50}$ $\mathrm{cm^2}$ \acrshort{RWELL}, and up to $\mathrm{50 \times 50}$ $\mathrm{cm^2}$ \acrshort{RPWELL} \cite{detlab_48,detlab_49,detlab_53,detlab_63}. Their measured performances are summarized in Table \ref{tab:DHCAL}.

\begin{table}[htbp]
\centering
\caption{\label{tab:DHCAL} A summary of the average pad-multiplicity and \acrshort{MIP} detection efficiency measured with sampling elements of different technologies.}
\smallskip
\begin{tabular}{lcc}
\hline
& \begin{tabular}{c}
     Average  \\
     Pad-Multiplicity 
\end{tabular}&\begin{tabular}{c}
     MIP detection  \\
     Efficiency 
\end{tabular}\\
\hline
Glass RPC \cite{CALICE:2019rct} & 1.6 & 98\%\\
MM \cite{1sqmMM} & 1.1 & 98\%\\
Resistive MM \cite{Geralis:2018fmn} & $\sim$1.1 & 95\%\\
Double GEM \cite{Hong:2020hwx} & $\sim$1.2 & 98\%\\
RWELL  & Not Reported & $\sim$96\%\\
RPWELL \cite{detlab_48} & 1.2 & 98\%\\
\hline
\end{tabular}
\end{table}

\begin{figure}[htbp]
    \centering
    \subfloat[]{
        \includegraphics[width=0.47\textwidth]{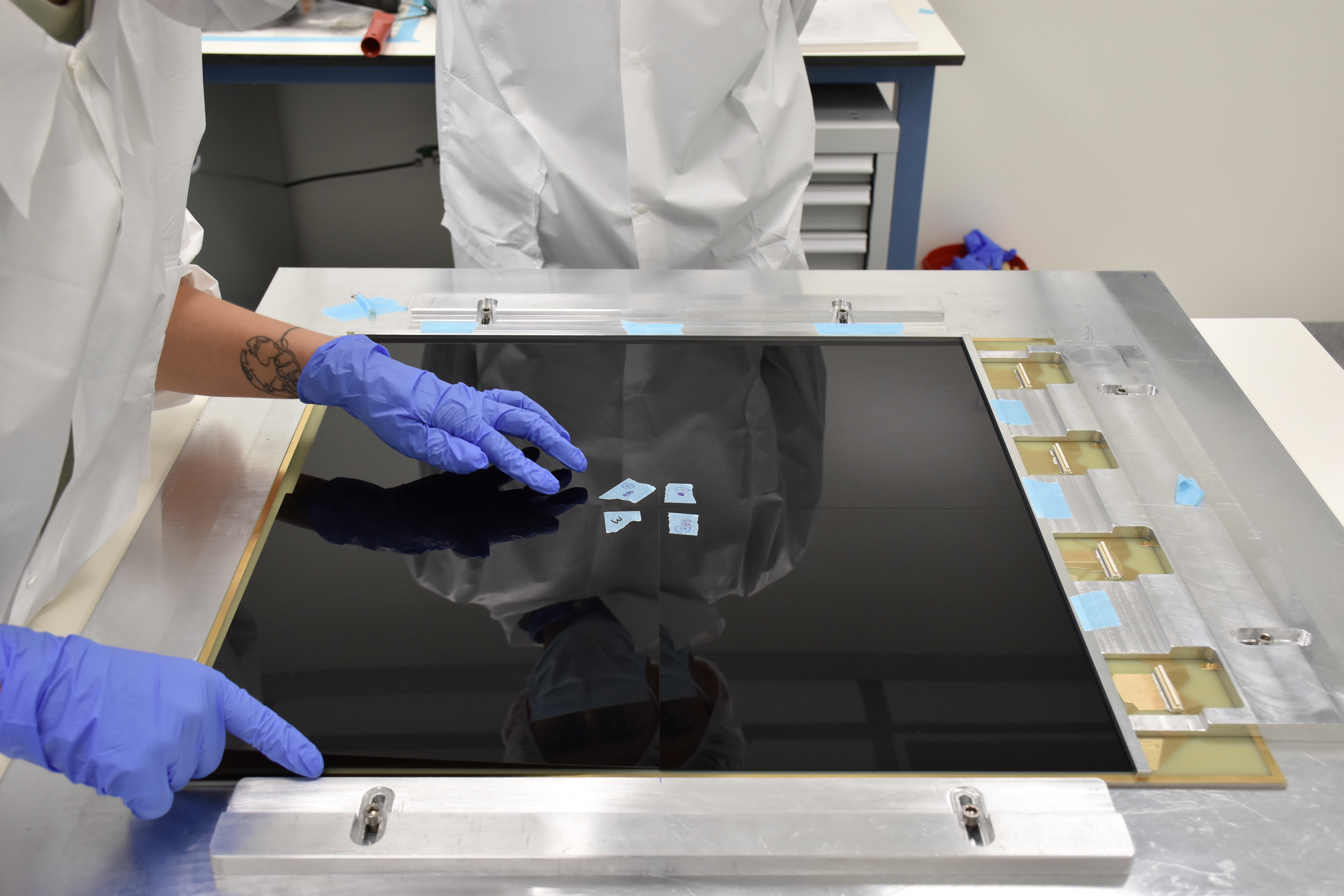}
        \label{fig:RPWELLGlass}
    }
    \subfloat[]{
        \includegraphics[width=0.47\textwidth]{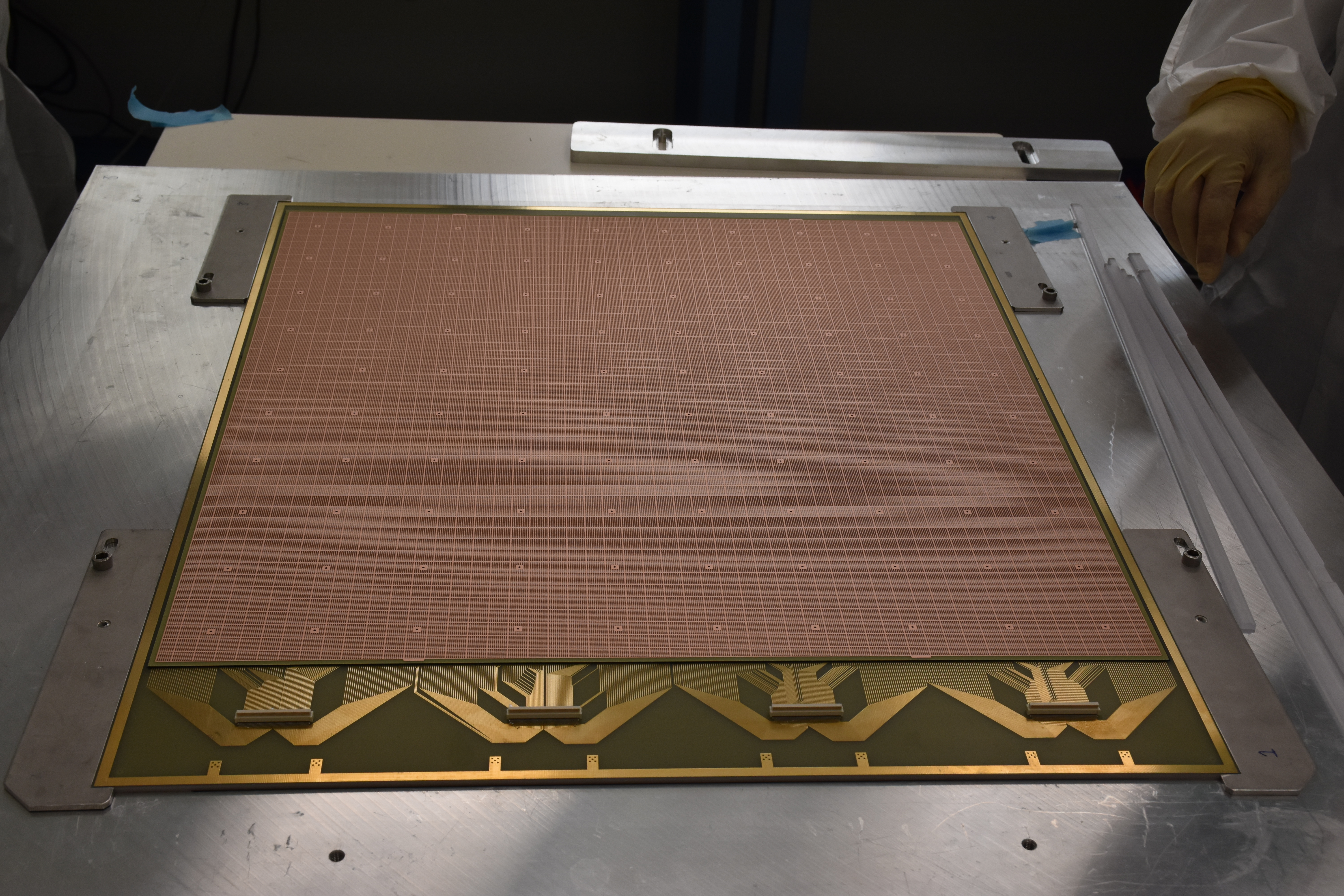}
        \label{fig:RPWELLElectrode}
    }
    \caption{The assembly procedure of a large, $\mathrm{50 \times 50~cm^2}$, \acrshort{RPWELL} sampling element prototype. \protect\subref{fig:RPWELLGlass}: the four Fe-doped glass tiles glued on top of the anode. \protect\subref{fig:RPWELLElectrode}: the large WELL electrode.}
\label{fig:largeRPWELL}
\end{figure}

The assembly procedure of a large, $\mathrm{50 \times 50 ~cm^2}$, \acrshort{RPWELL} \acrshort{DHCAL} sampling element prototype is shown in Figure \ref{fig:largeRPWELL}. Based on data collected with a small \acrshort{MPGD}-based \acrshort{DHCAL} (several \acrshort{MM} and \acrshort{RPWELL} layers), simulation studies have shown that a full-size fully equipped \acrshort{RPWELL}-based \acrshort{DHCAL} could reach the desired hadron energy resolution \cite{detlab_80}. 

\subsection{Negative Ion TPCs}
\label{sec:IonTPC}

Negative Ion \acrshort{TPC}s \cite{MARTOFF2000355} are proposed as a solution for applications requiring high precision tracking of radiation depositing small energy, such as searches of directional dark matter \cite{101,411,76}, neutrinoless double beta decay \cite{462}, etc. Operated in a highly electronegative gas, negative ions are formed by the attachment of the \acrshort{PE}s to the gas molecules. The ions drift towards the high field region; there, with a certain probability, the extra electrons are stripped from the negative ions and undergo a standard avalanche multiplication. This concept conjugates the advantages of the ions' small diffusion values with the large gains of electron multiplication \cite{462}.

In \cite{462}, the concept of a single negative ion counting \acrshort{TPC} based on an $\mathrm{O_2}$-doped gas mixture of $\mathrm{Ar/CO_2/O_2}$ (66:30:4) at 0.25 bar is demonstrated. \acrshort{PE}s are captured over a $\sim$cm distance by $\mathrm{O_2}$ molecules, allowing to count them separately, resulting in an improved energy resolution. The high field region, where the primary electrons are stripped from the ion and multiplied, is defined by a \acrshort{THGEM} (\acrshort{LEM}). $\mathrm{SF_{6}}$ gas is also considered for such applications \cite{101}. It has excellent electron attachment features and is rich in Fluorine, which is important for the search of spin-dependent interactions. It was demonstrated that \acrshort{THGEM} electrodes could operate at high gains in this gas at low pressures (see Section \ref{sec:RTLP}).

In \cite{101}, a negative ion \acrshort{TPC} based on single- and cascaded-\acrshort{THGEM} with light readout was developed for low pressure (20-100 Torr) $\mathrm{SF_6}$; gain of $\sim$$\mathrm{10^3}$ was reached. A similar concept is explored in \cite{76} with a \acrshort{THGEM} capacitively coupled to a multiwire plane in 20 Torr $\mathrm{SF_6}$.
\subsection{Tracking systems for high energy physics and nuclear physics}
\label{sec:tracking and nuclear}

\subsubsection{Magnetic spectrometers}
\label{sec: focal plane tracker}

Magnetic spectrometers employ focal plane trackers (\acrshort{FPT}s) for track reconstruction, momentum measurement, and \acrshort{PID}. A common choice for trackers are low-pressure \acrshort{TPC}s (also called drift chambers in this context). 

The MAGNEX magnetic spectrometer is the \acrshort{FPT} deployed in the Nuclear Matrix Elements (NUMEN) experiment \cite{345}. The experiment aims at measuring the half-life of neutrinoless double $\beta$ decay via double charge exchange in the nuclear reactions of ions on target. MAGNEX is a gaseous hybrid detector followed by a wall of 60 silicon detectors. It measures the horizontal and vertical positions of each highly ionizing ion at four sequential points along its trajectory in four low-pressure (10-15 mbar) \acrshort{TPC}s. The \acrshort{TPC}s are separated two by two by a large proportional chamber. 

\acrshort{THGEM}-based detectors were proposed as an alternative to the wire-based proportional chambers to improve the spectrometer's rate capabilities \cite{345}. The final choice was made for \acrshort{M-THGEM} because of their lower voltage operation in low-pressure gases \cite{491,496,503,527} (Section \ref{sec:RTLP}).

The \acrshort{FPT} of the S800 magnetic experiment at the National Superconducting Cyclotron Laboratory (NSCL) incorporates two drift chambers with a wire readout. To improve the rate capabilities of the experiment, the wire readout will be replaced by an \acrshort{M-THGEM}+\acrshort{MM} readout \cite{74}. Tests have been carried out with low-energy alpha particles, high-energy heavy (100 MeV) ion beams, and reaction fragments. The same detector will be implemented at the focal planes of the HRS spectrometer at the Facility for Rare-Isotope Beams (FRIB) in the Michigan State University \cite{524}.

The Cooling Storage Ring (\acrshort{CSR}) External-target Experiment (\acrshort{CEE}) is the first multi-purpose nuclear physics experimental device to operate in the GeV energy range at the Heavy-Ion Research Facility (HIRFL-CSR) in Lanzhou. The primary goals of the \acrshort{CEE} are to study the bulk properties of dense matter and understand the quantum chromo-dynamic (QCD) phase diagram by measuring the charged particles produced in heavy-ion collisions at the target region with a large acceptance. A core component in the \acrshort{CEE} experiment is a \acrshort{THGEM}-based \acrshort{TPC}. High energy and position resolutions were demonstrated with a prototype chamber operated in 50 - 101 kPa $\mathrm{Ar/CH_4/CF_4}$ (90:7:3) \cite{36}.

Similar systems were also used for other purposes. A THGEM detector array (ELITHGEM) is proposed to measure the angular distribution of photofission fragments at the Extreme Light Infrastructure-Nuclear Physics (ELI-NP) facility \cite{494} at an expected fission event rate of $\sim$$\mathrm{10^3}$ Hz. The detector, operated with 5 mbar $\mathrm{iC_4H_{10}}$, is designed to cover $\sim$$\mathrm{80 \%}$ of 4$\mathrm{\pi}$ with a 5$^{\circ}$ resolution.

\subsubsection{Low-pressure TPCs for nuclear experiments}
\label{sec: AT TPC}

\acrshort{AT} \acrshort{TPC}s, in which the operation gas is also used as a target, are a common tool used in nuclear physics experiments to measure unstable isotopes and cross-sections of low-energy nuclear reactions in astrophysical sources. A broad range of measurements can be performed by tuning the gas mixture (He based mixtures are commonly used) and pressure (Section \ref{sec:RTLP}). For a recent review of \acrshort{AT} \acrshort{TPC} experiments see \cite{344}.

The \acrshort{GEM}-MSTPC \acrshort{AT} \acrshort{TPC} was operated with various \acrshort{THGEM}-based readout elements. A stable operation was recorded with heavy ions at $\mathrm{10^5}$ Hz at 0.16 atm in $\mathrm{He/CO_2}$ (90:10) \cite{281}. Studies were conducted also in low-pressure He \cite{110,108}, $\mathrm{H_2}$ \cite{109,108}, $\mathrm{D_2}$ \cite{109,160,180,342,343}, and $\mathrm{CO_2}$ \cite{97} (Section \ref{sec:RTLP}). In \cite{97}, images of alpha particle tracks in 50 Torr $\mathrm{CO_2}$ with 2THGEM and a \acrshort{MM} were recorded. The latter configuration was implemented in the PAT-TPC \cite{268}, operated in 60 Torr $\mathrm{^4He/CO_2}$ (90:10) to measure the $\mathrm{^{11}Be}$ decay emission spectrum and, specifically, its beta delayed proton emission \cite{346}.  A \acrshort{THGEM} readout for the ACTAR TPC \cite{271} was tested in $\mathrm{iC_4H_{10}}$ at 25-75 mbar, providing a high energy resolution. A 2M-THGEM was tested within the Notre Dame Cube (ND-Cube) \acrshort{AT} \acrshort{TPC} in Ne/H$_2$ (95:5) and it will be implemented in a hybrid 2M-THGEM+\acrshort{MM} configuration \cite{511}. A 2THGEM configuration in the Compact \acrshort{AT}-\acrshort{TPC} operated in He/CO$_2$ (96:4) demonstrated good performance in terms of time and spatial resolution when measuring alpha particles \cite{513}. A \acrshort{THGEM}+\acrshort{MM} configuration was implemented in the Texas \acrshort{AT} (TexAT) \acrshort{TPC} \cite{520,521}. It allowed measuring beta-delayed particle decays in CO$_2$ at 20 Torr \cite{521} and the $^{12}$C3$\alpha$ cross section in 50 Torr CO$_2$ \cite{520}.

In \cite{509} a low-pressure \acrshort{TPC} was developed to measure $^{12}$C and $^{16}$O nuclear reactions, which are fundamental for stellar evolution and nucleosynthesis. The system is operated in 35-100 mbar iC$_4$H$_{10}$ and in 90, 160 Torr Ar/CH$_4$ (90:10) with 2THGEM charge readout configuration. Tracks from products of $^{12}$C+$^{12}$C reactions were successfully recorded \cite{519}. 

\subsubsection{Low pressure TPCs for low energy tracking}

Low pressure \acrshort{TPC}s are important tools for x-ray polarimetry and directional dark-matter experiments, which require reconstructing small tracks generated by low energy depositions. They can work either with light or charge readout.

The \acrshort{ARIADNE} detector concept developed in the context of light readout in cryogenic systems (Section \ref{sec:ARIADNE}) is also suitable for room temperature and low-pressure \acrshort{TPC}s. Prototypes employing 2THGEM with either a Timepix3 camera or Linearly Graded-SiPM as photo sensors were tested in 100 mbar $\mathrm{CF_4}$, demonstrating track reconstruction of alpha particles and cosmic muons \cite{456,64}. 1THGEM and 2THGEM operated in 25-50 Torr CF$_4$ with a CCD light readout was used to record electron tracks from \fe{} x-rays, while operated at a gain of about 10$^5$ and energy resolution down to 30\% \cite{410}.

Optical \acrshort{TPC}s operated with Oxygen-rich gas mixtures (suited for the study of nuclear reactions of astrophysical interest) were tested with 1THGEM and 2THGEM detectors in 75 Torr $\mathrm{CO_2/N_2}$ (90:10) \cite{40}. In \cite{507,510} a 1THGEM detector operated in 0.8 atm Ne/DME (80:20) coupled to a Si sensor is used to detect electron tracks from x-ray photons. 

\begin{figure}[htbp]
    \centering
    \subfloat[]{
        \includegraphics[width=0.55\textwidth]{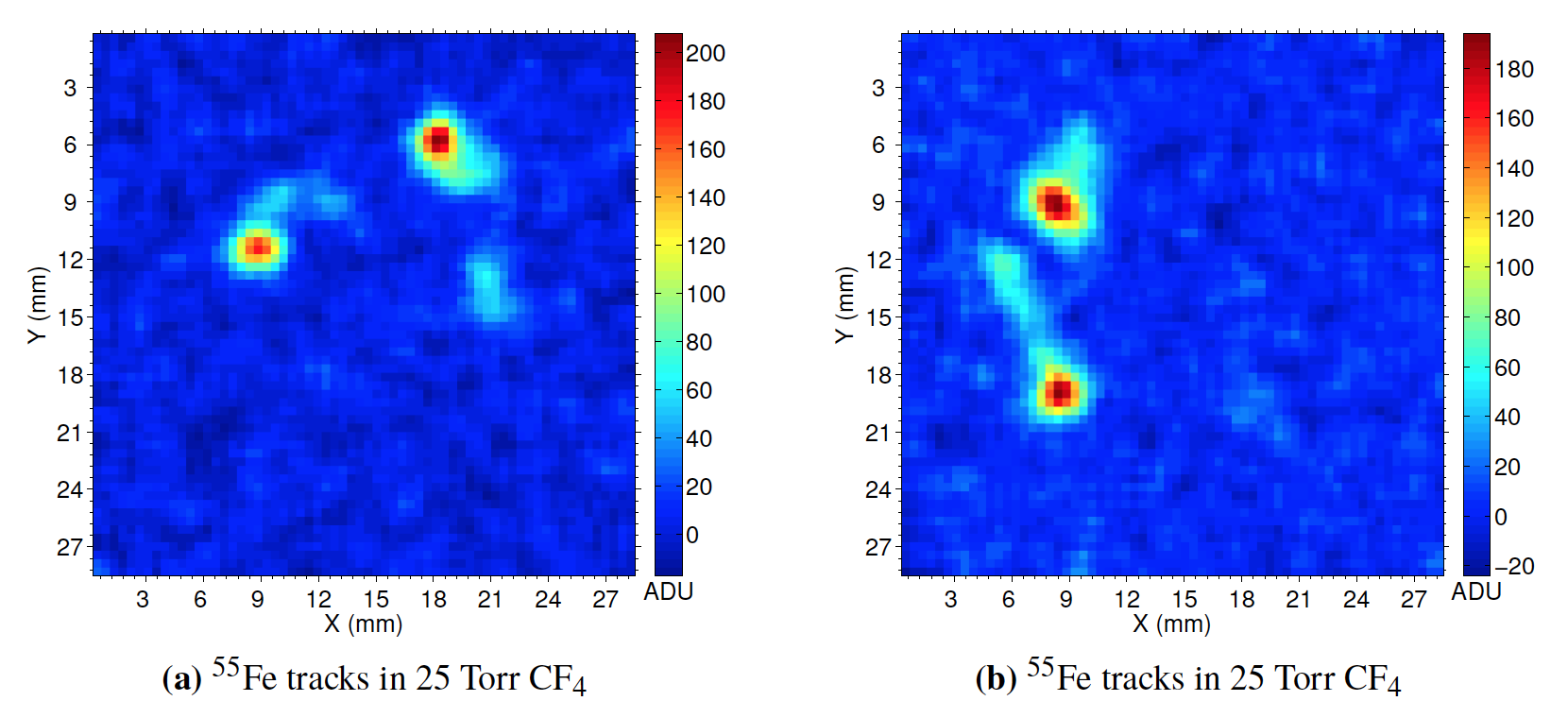}
        \label{fig:lowP x-rays}
    }
    \subfloat[]{
        \includegraphics[width=0.4\textwidth]{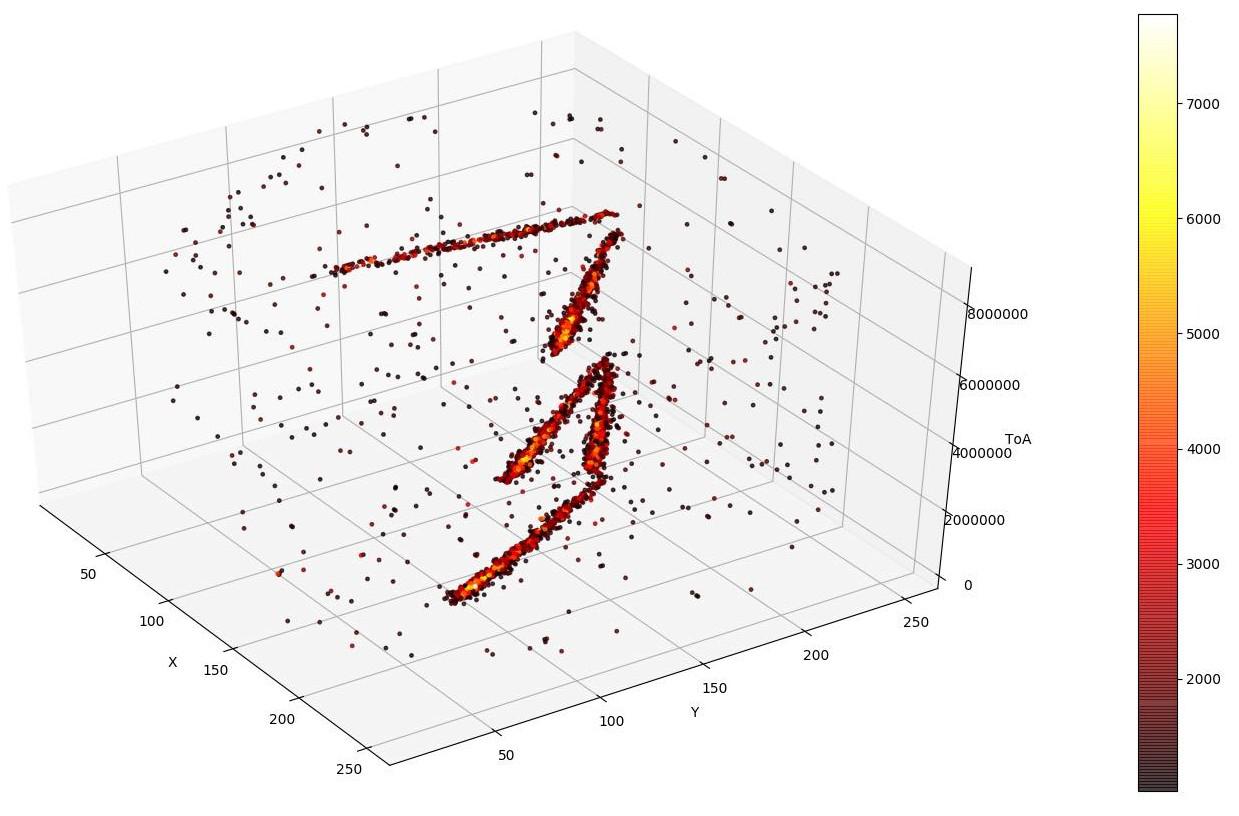}
        \label{fig:lowP alpha}
    }
    \caption{\protect\subref{fig:lowP x-rays} Optical readout of electron tracks from \fe{} x-rays in 25 Torr $\mathrm{CF_4}$ using a 2TGHEM detector. The figure was taken from \cite{410}. \protect\subref{fig:lowP alpha} Optical readout of alpha particle tracks in 75 Torr $\mathrm{CF_4}$ using a 1TGHEM detector. Figure obtained from \cite{456}.}
\label{fig:lowP imaging}
\end{figure}

Example images of electrons from x-ray photons and alpha particle tracks acquired with optical readout in low-pressure $\mathrm{CF_4}$ with \acrshort{THGEM} detectors are presented in Figure \ref{fig:lowP imaging}.

In \cite{529} a glass-\acrshort{THGEM} was installed in a Bragg curve counter instead of a Frisch grid. Alpha particles were measure in Ar/CH$_4$ and Ar/CO$_2$ at low pressures. 

\subsubsection{Other proposed systems}

\subsubsection*{ALICE VHMPID trigger}

\acrshort{THGEM}-based detectors were considered for triggering rare high \pt{} events in \acrshort{ALICE} high \pt{} Trigger Detector (\acrshort{HPTD}) \cite{5,472}. In the proposed scheme, four planes of 2THGEM with digital pad readouts are used to sample \acrshort{MIP} tracks. A test chamber was built and tested in a pion beam. Using the \acrshort{ALICE} electronic chain, full efficiency was achieved in different Ar-based gas mixtures. The rate of occasional discharges remained compatible for efficient operation in \acrshort{ALICE}. A simulation was performed to optimize the pad segmentation and the trigger logic with respect to the trigger event rate. It was demonstrated that the concept could work in the experiment \cite{472}. The project was not pursued further\footnote{Private communications}. 

\subsubsection*{CBM}

The Compressed Baryonic Matter (\acrshort{CBM}) experiment at the FAIR accelerator facility at GSI is an ion-on-target experiment aiming at studying nuclear matter at $\sim$10 times its normal density (therefore 'compressed'). Targeting tracking at a flux of $\mathrm{10^7 Hz/cm^2}$, a muon spectrometer based on various \acrshort{MPGD} technologies was proposed. The 3rd and 4th stations could be based on \acrshort{THGEM}. This possibility was studied in proton and pion beams \cite{333,291,332} with a 2THGEM configuration, a 4M-THGEM in a WELL configuration, and a hybrid \acrshort{THGEM}+\acrshort{MM}.

\subsubsection*{J-PARC E15}

The J-PARC E15 experiment studies the kaon nuclear-bound states via the $\mathrm{^3 He(K^-,n)}$ reaction. \acrshort{THGEM} detectors were studied for upgrading the E15 inner tracker \cite{405}. The proposed technology was a \acrshort{TPC} with a \acrshort{THGEM} readout operating in $\mathrm{Ar/CH_4}$ (90:10) at atmospheric pressure. Different \acrshort{THGEM} electrodes with either Cu, graphite or Cu+graphite with different geometrical parameters were studied in 1THGEM and 3THGEM configurations \cite{16,317}. The project was not pursued to the end for lack of resources\footnote{Private communications}.
\subsection{Other scientific and civil applications}
\label{sec:civil}

\acrshort{THGEM}-based detectors are robust and cost-effective. They can be industrially produced over large areas and operated under harsh conditions conditions with non-flammable gaseous mixtures. Thus, they are  candidates of choice for a wide variety of scientific and civil applications.

\subsubsection{Safety}
\label{sec:appsSafety}

2RETGEM electrodes in a \acrshort{RICH}-like configuration (see Section \ref{sec:RICH})
were studied as flame detectors by visualizing \acrshort{UV} photons in daylight conditions \cite{401,7}. Operated in a gas mixture of Ar and a photosensitive vapor like EF or TMAE, and even in air, 1-dimensional images of a small flame were acquired in daylight at a distance of tens of meters, outperforming the sensitivity of a commercial \acrshort{UV}-based flame sensor by two orders of magnitude. In the same works, a similar detector was proposed for hyperspectroscopy in the \acrshort{UV} range. The concept was further developed in \cite{225} using Ne mixtures.  The authors addressed challenges of outdoor operation, e.g., temperature variations, visible light background, and operation in sealed mode. \acrshort{RETGEM} detectors were also considered for the detection of dangerous vapors in the air \cite{224}.  

A \acrshort{THGEM}-based electron mobility spectrometer was studied in the context of tritium monitoring in the working environment of heavy water power reactors \cite{106}. \acrshort{THGEM} detectors operated in $\mathrm{Ar/iC_4H_{10}}$ (97:3) in sealed mode with a \um{} thick window on the cathode side were used in an alpha-particle surface 2D contamination monitor \cite{182}. Detection efficiency of the order of 50\% of alpha particles emitted from $^{238}$Pu was measured with $\sim$3 mm position resolution. A \acrshort{UV} detector based on \acrshort{THGEM} was used to identify contaminants in water by their different absorption spectra \cite{522}.

\subsubsection{Non-invasive material imaging}
\label{sec:non invasive imaging}

Non-invasive material imaging techniques are frequently used in medical applications, material sciences, archaeology, and artistic heritage.

\subsubsection*{X-ray imaging}

\begin{figure}[htbp]
    \centering
    \subfloat[]{
        \includegraphics[width=0.5\textwidth]{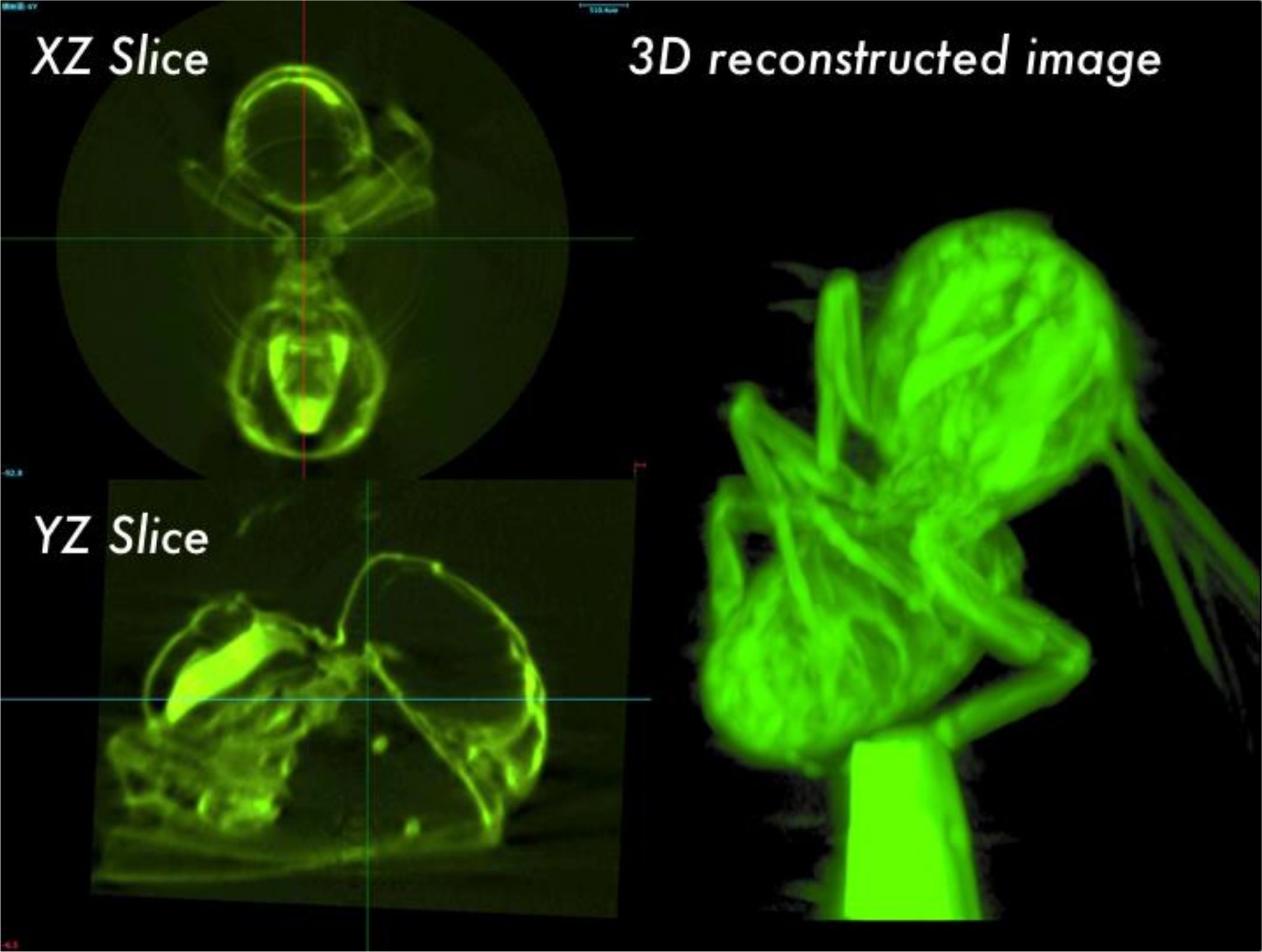}
        \label{fig:GTHGEM-CT}
    }\\
    \subfloat[]{
        \includegraphics[width=0.60\textwidth]{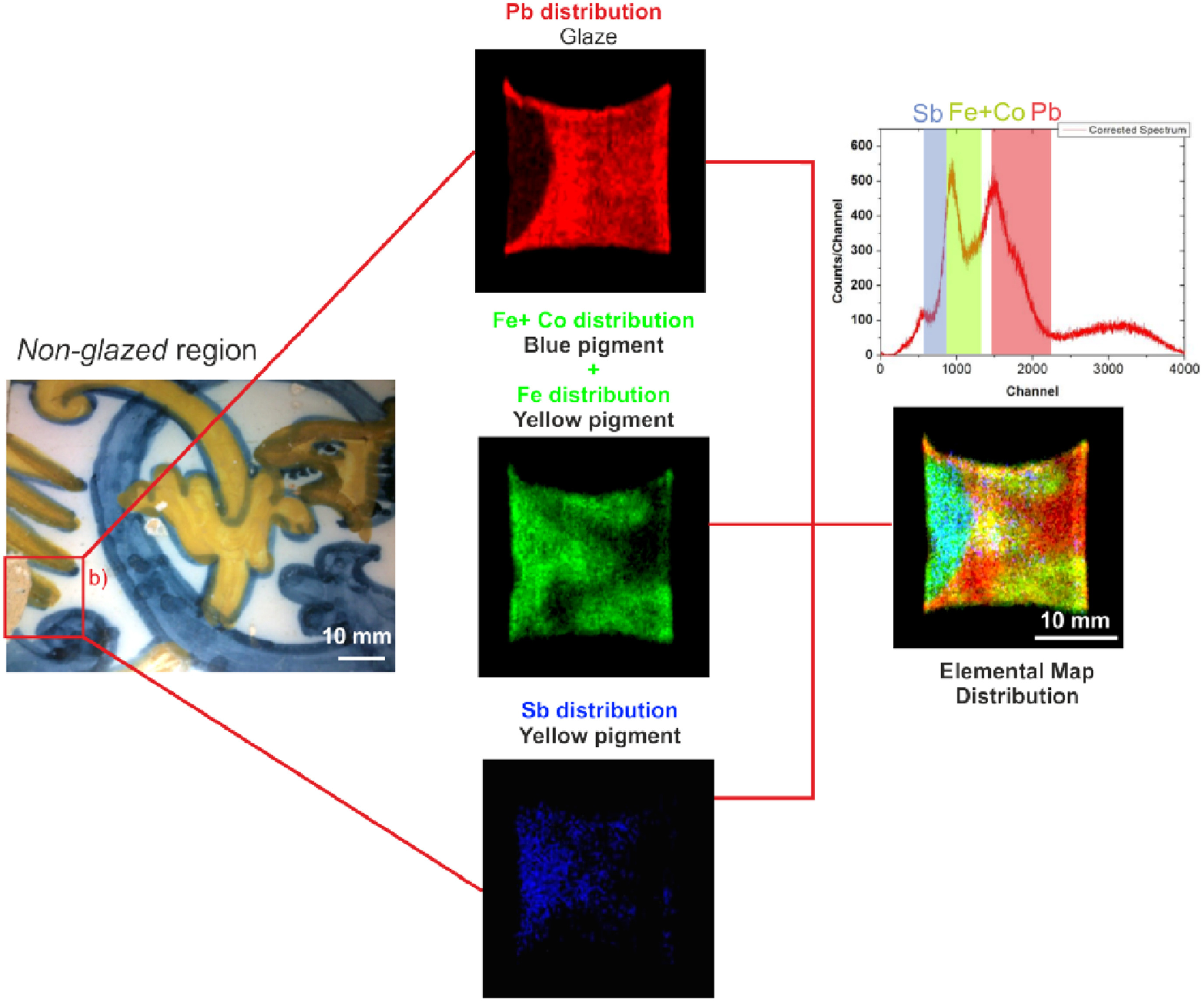}
        \label{fig:THCOBRA-elements}
    }
    \caption{\protect\subref{fig:GTHGEM-CT} An x-ray \acrshort{CT} scan obtained with the optical readout of a 1THGEM detector in $\mathrm{Ar/CF_4}$ (90:10). The figure was obtained from \cite{305}. \protect\subref{fig:THCOBRA-elements} An \acrshort{EDXRF} image acquired with a \acrshort{THCOBRA}, operated in $\mathrm{Ne/CH_4}$ (95:5) with charge readout of resistive lines. The figure was taken from \cite{383}.}
\label{fig: x-ray imaging}
\end{figure}

Several optical-readout x-ray imaging concepts were studied. An instrument made of a glass-\acrshort{THGEM} operated in an $\mathrm{Ar/CF_4}$ scintillating gas with a CCD light readout was studied in \cite{251,292}. A similar concept employing a photodiode array panel as a photo-sensor was studied in \cite{311}. The same instrument was also used to produce images by normal and magnified transmission \cite{256}, as well as for 3D computed tomography (\acrshort{CT}) \cite{305}.  High gain and light yields allow for high sensitivity for low-energy x-rays and fast image acquisition. An example image is presented in Figure \ref{fig:GTHGEM-CT}. In \cite{241}, the same concept was demonstrated using charge readout. In \cite{250}, the system was scaled up in size to $\mathrm{280 \times 280 ~mm^2}$. Similar setups employing a hybrid of 2GEM+\acrshort{THCOBRA} and a Kapton-2THGEM are described in \cite{135} and \cite{265}, respectively.

X-ray imaging with charge readout was obtained using a \acrshort{THCOBRA} with resistive lines of strips operated in pure Kr \cite{136,471} and in $\mathrm{Ne/CH_4}$ \cite{194}. It resulted in a position resolution of $\sim$2.5 mm and 0.6 mm (FWHM) for $\mathrm{Ne/CH_4}$ \cite{194} and Kr \cite{136,471}, respectively. The former is limited by the large photoelectron range in the gas. The energy resolution measured was $\sim$20\% in both cases. The detector was employed in an energy-dispersive x-ray Fluorescence (\acrshort{EDXRF}) system (Figure \ref{fig:THCOBRA-elements}), allowing the detection of materials' local compositions in multi-component samples, including biological ones. The detector performance was demonstrated in $\mathrm{Ne/CH_4}$ \cite{382,383,386,497,525} and in Kr \cite{351}.

In \cite{229,264}, the imaging capabilities of a small \acrshort{CT} scanner based on a \acrshort{THCOBRA} with and without a \acrshort{THGEM} in cascade were demonstrated in $\mathrm{Ne/CH_4}$. These (0.2 mm thick) electrodes were used in 1THGEM and 2THGEM configurations to detect x-rays in a diffractometer. They were mounted in a curved geometry to avoid parallax effects \cite{132,21,340}.

\subsubsection*{Neutron imaging}
An advantage of gaseous detectors for neutron detection is their intrinsically low sensitivity to gamma photons, which constitute the main background in neutron measurements.

In \cite{390}, it is demonstrated that a glass-\acrshort{THGEM} coupled to a $^{10}$B thermal neutron converter cathode and a resonance filter is able to identify different materials by their resonance energy. As an alternative to resonant filters, the \acrshort{THGEM} detector can be synchronized with a \acrshort{ToF} system for energy measurement, resulting in energy-resolved neutron imaging \cite{514}.

\begin{figure}[htbp]
    \centering
        \includegraphics[width=0.45\textwidth]{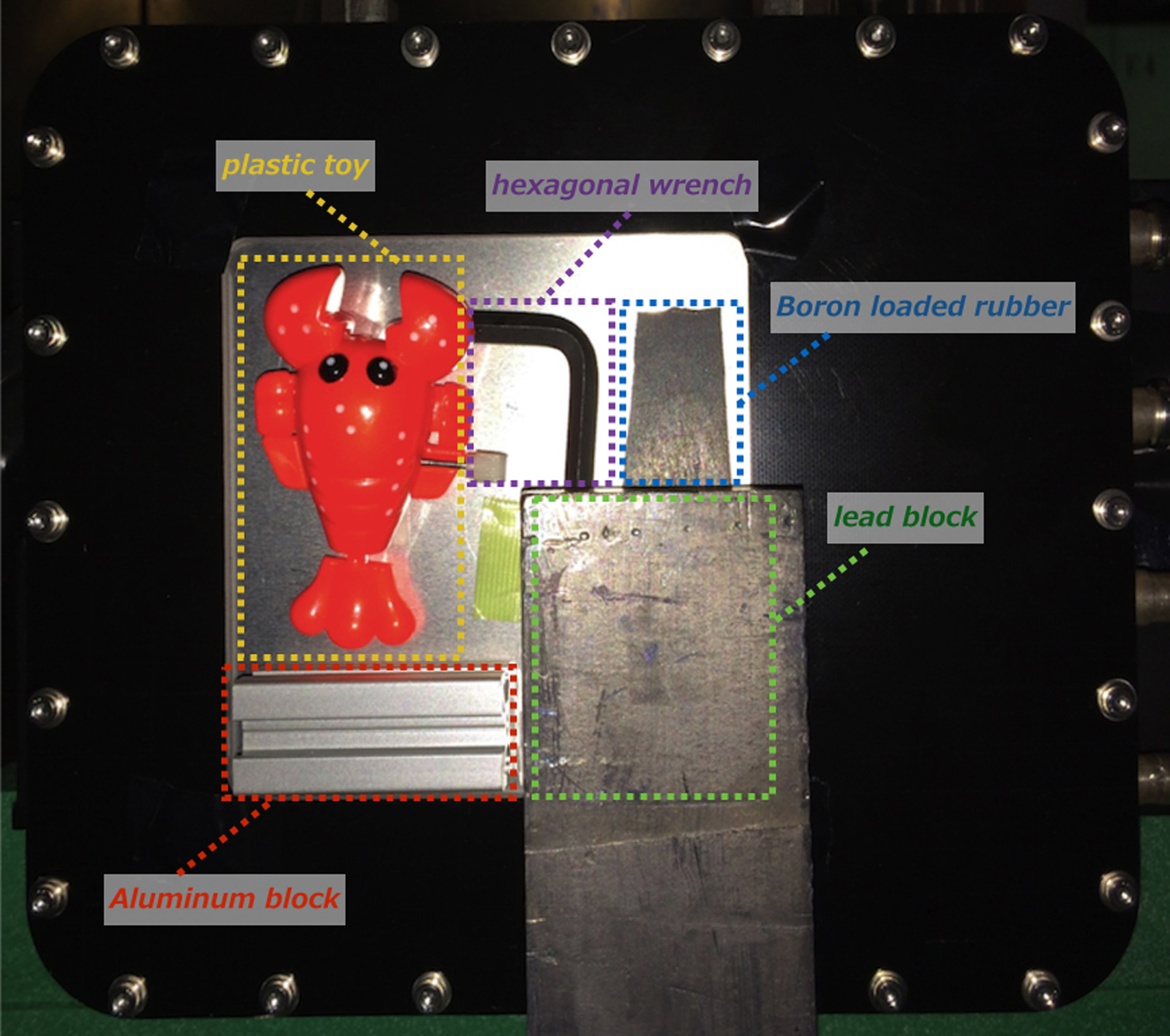}
        \includegraphics[width=0.48\textwidth]{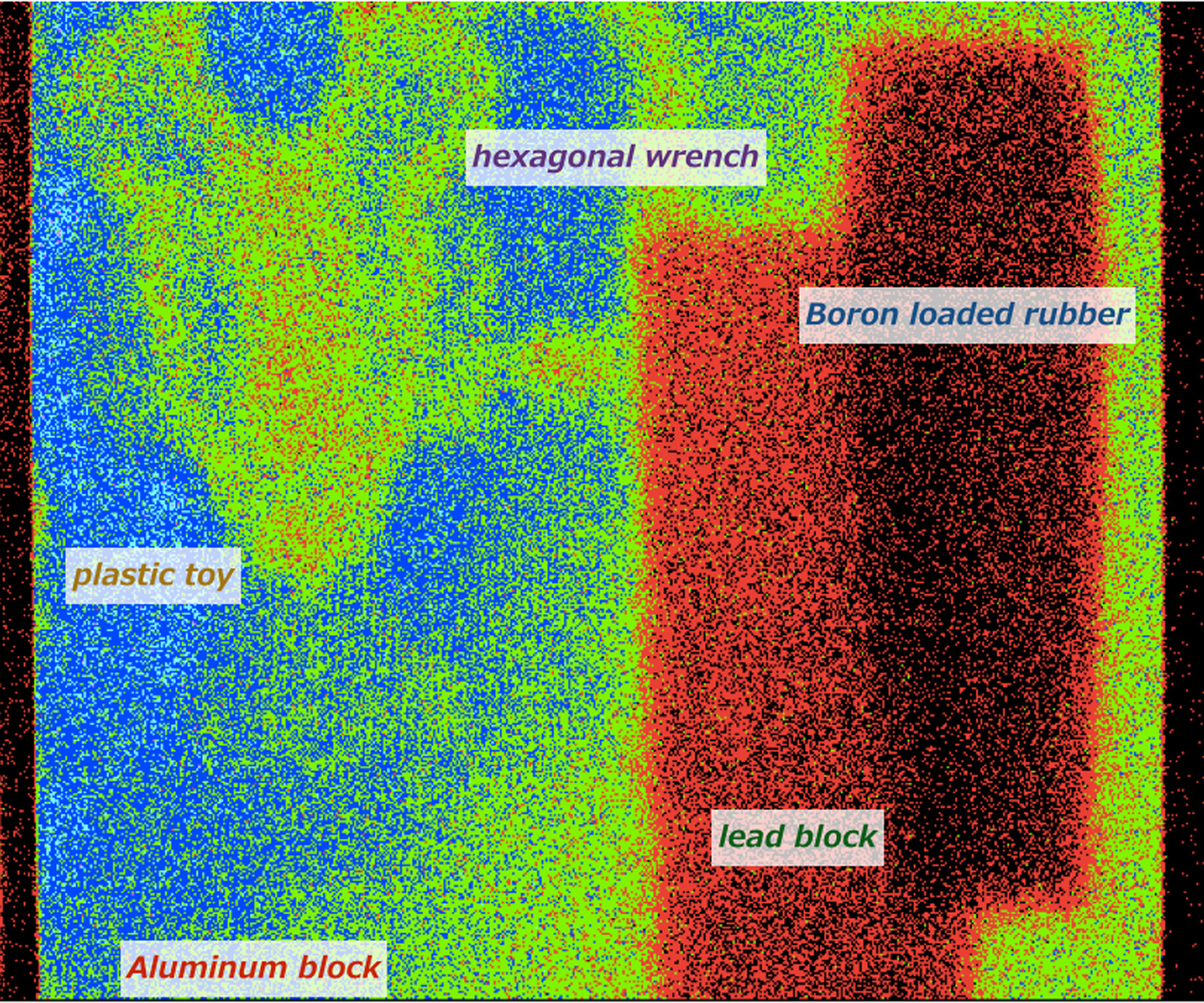}
        \label{fig:GTHGEM-neutrons}
\caption{Neutron imaging with a glass-\acrshort{THGEM} detector. Figures taken from \cite{290}.}
\label{fig:neutron imaging}
\end{figure}

A similar detector operated in an $\mathrm{Ar/CF_4}$ (90:10) scintillating gas mixture can substitute scintillators with image intensifiers \cite{290}. A CCD camera is used to acquire images of the light produced in the holes of a glass-\acrshort{THGEM} with 100 \um{} diameter holes and a small pitch to maximize the filling factor \cite{233_capillary}. A typical neutron imaging picture is presented in Figure \ref{fig:neutron imaging}. A fast-neutron imaging detector, for contraband detection, comprising a \acrshort{LXe} capillary converter coupled to a 3THGEM \acrshort{GPM} is described in  \cite{detlab_50}.
In \cite{506}, a detector including a ceramic-2THGEM configuration is implemented as a thermal neutron imager for on the VESUVIO spectrometer at the ISIS neutron and muon source. This instrument allowed measuring the cross section of cold neutrons onto amminoacids, which is of particular relevance for medical and biological applications \cite{523}. A ceramic-1THGEM operated in Ne/CO$_2$ (90:10) in a sealed mode is chosen instead for the China Spallation Neutron Source (CSNC) \cite{518,517}. 

Imaging of fast neutrons (the order of MeV) can be applied for non-destructive testing of nuclear waste, detection of explosives or drugs, and investigation of thermal hydraulics phenomena. In \cite{133,128}, a system of \acrshort{THGEM} electrodes coupled to slabs of neutron absorbers is proposed and developed for fan-beam tomography. The absorber's structure can also be tailored to select specific neutron energies, for which neutron spectroscopy is possible \cite{350}. This technology can be complemented by a similar system optimized for cold neutron imaging, where the absorber slabs are substituted by a $^{10}$B neutron converter \cite{188,117,29,266,296,299}. In \cite{35}, cascaded \acrshort{THGEM} electrodes coated with $^{10}$B and operated at a gain of 1 serve as neutron converters. Such techniques are applicable for imaging of nuclear fuel \cite{349,350}.

In \cite{detlab_65,detlab_46,detlab_64,detlab_50}, a setup including a \acrshort{THGEM}-based \acrshort{GPM} coupled to a \acrshort{LXe} converter/scintillator is developed to image fast neutrons and gamma rays. This could allow identifying hidden explosives and fissile materials in cargo containers.

\subsubsection*{Imaging with alpha particles}

In \cite{81,90,504}, the light produced by discharges in a 3THGEM electrode is recorded with a CMOS camera to produce radiographic images with alpha particles. A similar system comprising 1THGEM configuration was used to measure the concentration of radioactive Rn in air \cite{501}.

\subsubsection{Medical applications}
\label{sec:medicalApps}

\subsubsection*{Gamma-ray imaging}

A 3THGEM \acrshort{GPM} was proposed for scintillation-photons recording in a large-size \acrshort{LXe} \acrshort{TPC} Compton Camera. It was developed within a small-animal imaging concept of a 3-gamma imager incorporating a COMPTON camera and PET scanner, for small-animal medical imaging \cite{detlab_13}. A Compton camera based on a photosensitive \acrshort{THCOBRA} operated in a high pressure (10-20 bar) scintillating gas (Ne, Xe, or Ar) was shown to be more sensitive than the standard devices in use \cite{302}.

\subsubsection*{Dosimetry, microdosimetry and nanodosimetry}  

\paragraph{Dosimetry}
\label{sec:dosimetry}
Dose imaging in radiation therapy is of crucial importance for a successful treatment with minimal collateral damage. \acrshort{MPGD} detectors are well-suited for this kind of application due to their large active area, radiation hardness, linear response, and high position resolution.

Preliminary studies towards real-time dosimetry during radiation treatments were carried out in \cite{254}. They employed a glass-\acrshort{THGEM} detector operated in Ar/CF$_4$ (90:10) with the optical readout presented in Section \ref{sec:non invasive imaging}. The detector was used for measuring a 6 MeV photon beam and demonstrated the proportionality between the light emitted by the glass-\acrshort{THGEM} and the dose rate.  In \cite{251}, the same detector was used to obtain a 70 MeV proton radiography and a 160 MeV proton energy loss profile. The detector was optimized in \cite{515,516} to obtain improved proportionality between dose and light response despite the high linear energy-transfer of a carbon beam. The improvements aimed at minimizing the production of Cherenkov light, in particular using the glass-\acrshort{THGEM} in a \acrshort{THWELL} configuration with a thin anode.

\paragraph{Microdosimetry}

Microdosimetry aims to measure the energy deposition from a microscopic (cellular or sub-cellular) volume of biological tissue. The goal is to understand the effect of radiation on biological cells.

Tissue-equivalent gaseous microdosimeters were studied in \cite{142,146,147,183,187}. The neutron spectra obtained with the \acrshort{THGEM} device operated in propane (a favorable choice for mimicking tissue) were similar to those acquired with a standard spherical tissue equivalent proportional counter \cite{138}. To increase the neutron detection efficiency and obtain 2D information, structures made by arrays of \acrshort{THGEM} microdosimeters were studied. In \cite{145}, 1THGEM and 2THGEM were coupled to a 3$\times$3 array of separate gas volumes. As a further development, a stack of arrays was designed and simulated \cite{148,144,303}. The concept was demonstrated with a prototype made of 21 cells coupled to 3THGEM configuration \cite{177}. In this case, the achieved sensitivity was three times higher than that of a standard tissue equivalent neutron dosimeter. Preliminary studies were carried out with a similar prototype, including ceramic-\acrshort{THGEM} with low intrinsic radioactivity \cite{304}.

\paragraph{Nanodosimetry}

Hadron therapy for cancer is based on clustered DNA damages to malignant cells. The treatment can also damage healthy cells. Thus,  the study of the quantity, quality, and topology of absorbed doses of radiation at the nm scale (DNA scale) can help improving the precision and efficacy of such treatments \cite{Bellinzona2021}. Energy depositions by ionizing radiation in low-pressure gases can be related to those in the equivalent nanometric volumes of water. This is the basic principle of nanodosimeters, which consist of mm scale low-pressure gas volumes (see \cite{gartincl515} and references therein). For a recent review on this field of research see \cite{AGOSTEO2022106807}.

\acrshort{THGEM}-like structures have been implemented as 3D ion counters in nanodosimeters. The detector concept is demonstrated in \cite{298}. The configuration used is a 3 mm-thick \acrshort{RPWELL} with 1 mm holes operated with reverse bias in 0.1-10 Torr propane (a favorable choice for mimicking tissue), air, or water vapor. When colliding with a gas molecule, an ion reaching a hole can extract one electron and yield an avalanche towards the \acrshort{THGEM}-top electrode. The concept was further optimized in \cite{203,321,100,322,325,323}. The efficiency of the detector is estimated to be of the order of several \%; it increases with the electrode thickness (up to 10 mm) due to higher probability of primary ion to extract an electron. The concept was implemented in a single-hole configuration in \cite{396,489}. In \cite{483}, the study was complemented with Monte Carlo simulations, which include optimizations for a multi-hole setup.

\subsubsection{Other applications}  
\label{sec:Geology}

In \cite{67,71}, a \acrshort{THGEM}-based low-pressure \acrshort{TPC} was operated in pure $\mathrm{iC_4H_{10}}$ for ion track reconstruction aimed at dating geological objects. Tests were conducted with alpha particles of different energies. Track ranges were measured with 2\% accuracy and signals from various sources could be distinguished by track length discrimination.

A system, similar to that employed in \cite{81}, was used to estimate the level of radioactivity in biological and archaeological samples, allowing for $^{14}$C dating \cite{498}.

\acrshort{THGEM}-based ion counters similar to the ones proposed for nanodosimetry are compact, making them portable and therefore usable in a clinical environment, e.g., for cancer detection. In \cite{324}, it is suggested as a rough but fast and cheap alternative solution to standard chromatography-mass spectroscopy for recognizing volatile organic compounds emitted by healthy or malignant cells. The measurement is performed by recording basic signal features like amplitude, rise and fall time, or more complex ones like energy resolution, pulse area, ionization cluster size, and ion drift time.

\section{Outlook}

In the last two decades, significant effort has been made to understand the physics of the \acrshort{THGEM} detector, to overcome its limitations and improve its performance. This effort resulted in the development of many novel detector concepts. Some of them have been successfully deployed in recent particle- and nuclear-physics experiments, as well as in civil applications, and others are likely to be used in the future. Yet, there is room for additional studies and probably many more ideas to come.

When preparing this review, we scanned an enormous number of publications and tried to refer to all the relevant studies. We regret if other \acrshort{THGEM}-based concepts and results might have been, mistakenly, missed. 

\section{Acknowledgements}

This work was partly supported by the Israel Science Foundation, grant 3177/19, the Pazy foundation, Nella and Leon Benoziyo Center for High Energy Physics, and Sir Charles Clore Prize. We would like to give special thanks for Martin Kushner Schnur for supporting this research. We thanks our close and dearest collaborators at the Weizmann Institute of Science, and in particular Prof. Amos Breskin, Dr. David Vartsky and Dr. Dan Shaked-Renous.

\appendix

\newpage
\section{Summary tables}
\label{sec: summary tables}

\begin{longtable}{|l|l|p{4cm}|p{2cm}|p{1.5cm}|p{1.5cm}| }
\caption{Summary of references to studies at standard temperature and pressure.}
\label{tab:STTP}
\\\hline
 Derivative        & Gas & x-ray  & UV & MIP & $\alpha$   \\
\hline
 \multirow{4}{*}{\acrshort{THGEM}}  	& Ne & \cite{detlab_18,detlab_14,detlab_41,detlab_55,detlab_73,detlab_66,detlab_31,21,116,124,230,316,283,250,256,311,251} & \cite{detlab_18,detlab_21,detlab_14} & \cite{193,detlab_26} & \cite{detlab_20} \\
\cline{2-6}
			& Ar & \cite{detlab_55,detlab_61,detlab_12,detlab_4,detlab_66,57,29,166, 117,27,58,79,112,116,123,124,173,181,197,205,278,293,316,458,196,158,309,223,240,49,131,261,204,245,254,283,288,241,254,292,242,201,387,266,296} & \cite{detlab_4,detlab_1,detlab_14,detlab_69,20,339,88} & \cite{detlab_61,115}  & \cite{detlab_20,106,282,294,476}  \\
\cline{2-6}
			& Xe & \cite{detlab_12,57} &  &  & \cite{detlab_20,57} \\
\cline{2-6} 
            & Others & \cite{310} & \cite{detlab_1,detlab_18,detlab_4,detlab_75} &  &  \\
\hline
  \multirow{4}{*}{Cascade}            & Ne & \cite{detlab_16,detlab_20,detlab_23,detlab_73,detlab_14,detlab_18,detlab_34,470,112,124,173,199,230,265} & \cite{detlab_24,215,470,235} & & \cite{detlab_20,285} \\
\cline{2-6}
			& Ar & \cite{detlab_16,detlab_6,detlab_12,16,119,112,124,173,262,267,286,317} & \cite{detlab_1,detlab_4,detlab_16,19,228,131,114,19} & \cite{25} & \cite{285} \\
\cline{2-6}
			& Xe & \cite{detlab_16,detlab_12} & - & - & - \\
\cline{2-6}
			& Others & \cite{detlab_16,310} & - & - & \cite{102} \\
\hline
  \multirow{3}{*}{\acrshort{RETGEM}}  & Ne & \cite{1,4,13,45,156,157,287,2,285,216,162,7} & \cite{4,287,162,216,162,7} & - & \cite{1,2,4,156,157,285} \\
\cline{2-6}
			& Ar & \cite{1,4,13,45,156,157,259,287,285,249,2,16,162,216,249,7} & \cite{4,216} & - & \cite{2,4,157,285} \\
\cline{2-6}
			& Others & \cite{45,7} & \cite{7} & - & \cite{2,157} \\
\hline
  \multirow{3}{*}{\acrshort{THCOBRA}} & Ne & \cite{252,194,119} & - & - & - \\
\cline{2-6}
			& Ar & \cite{51,301} & \cite{51} & - & - \\
\cline{2-6}
			& Kr & \cite{136} & - & - & - \\
\hline
   \multirow{3}{*}{\acrshort{THWELL}}  & Ne & \cite{detlab_41,detlab_30,detlab_40,detlab_78,detlab_26,detlab_54} & \cite{detlab_68} & \cite{detlab_26,detlab_48,detlab_53,detlab_49,detlab_51,detlab_32,detlab_33} & -  \\
\cline{2-6}
			& Ar & \cite{detlab_61,206,65,249} & - & \cite{detlab_61,detlab_48,detlab_53} & - \\
\cline{2-6}
			& He & - & \cite{110} & - & - \\
\hline
   \multirow{2}{*}{\acrshort{M-THGEM}}           
			& Ar & \cite{258} & - & - & - \\
\cline{2-6}
			& He & - & \cite{102} & - & \cite{102} \\
\hline
   \multirow{2}{*}{Exotics}             
			& Ar & - & \cite{82} & - & - \\
\cline{2-6}
			& He & - & \cite{82} & - & - \\
\hline
   \multirow{3}{*}{Hybrids}             & Ne & \cite{detlab_27,199,220,207,119} & \cite{detlab_68,60,225,191} & \cite{detlab_26,detlab_32,detlab_33} &  \\
\cline{2-6}
			& Ar & \cite{291,231,220,207,119} & \cite{detlab_68,105,82} & \cite{333} & - \\
\cline{2-6}
			& Others & \cite{291,110} & \cite{120,110,82,108,82} & \cite{333} & - \\
\hline
\end{longtable}

\begin{landscape}
\begin{table}[ht]
\caption{Main results with cryogenic GPM detectors.}
\label{tab: GPM}
\centering
\begin{tabular}{|c|c|c|c|c|c|}
\hline
Ref & Gas & T[K], p[bar] &Detector & Radiation & notes \\ [0.5ex] 
\hline

\multirow{5}{*}{\cite{detlab_23}} & $\mathrm{Ne/CH_4}$ (95:5) & \multirow{5}{*}{173, 1.1}   & \multirow{4}{*}{2 THGEM}& \multirow{3}{*}{$\mathrm{^{55}Fe}$}  &\multirow{3}{*}{gain}  \\\cline{2-2}
&$\mathrm{Ne/CF_4}$ (95:5) &&&&\\\cline{2-2}
&$\mathrm{Ne/CF_4/CH_4}$ (90:5:5) &&&&\\\cline{2-2}\cline{5-6}
&$\mathrm{He/CH_4}$ (92.5:7.5)&&&\multirow{2}{*}{$\mathrm{^{238}Pu}$ scintillation in \acrshort{LXe}}&\multirow{2}{*}{S$_1$ signals}\\\cline{2-2}\cline{4-4}
&$\mathrm{Ne/CH_4}$ (90:10)&&THGEM+PIM+MM&&\\\hline

\multirow{2}{*}{\cite{detlab_24}}&$\mathrm{Ne/CH_4}$ (95:5)&150-242, 1&\multirow{2}{*}{2 THGEM}&\multirow{2}{*}{UV}&\multirow{2}{*}{gain}\\\cline{2-3}
&$\mathrm{Ne/CF_4}$ (95:5)&168, 1&&&\\\hline

\multirow{2}{*}{\cite{detlab_27}}&\multirow{2}{*}{$\mathrm{Ne/CF_4}$ (90:10)}&\multirow{2}{*}{171, 1.1}&THGEM&\multirow{2}{*}{$\mathrm{^{55}Fe}$}&gain\\\cline{4-4}\cline{6-6}
&&&THGEM+PIM+MM&&gain, energy resolution\\\hline

\cite{detlab_44}&$\mathrm{Ne/CH_4}$ (95:5)-(90:10)-(80:20)&180/190, 0.7&3 THGEM&$\mathrm{^{241}Am}$ scintillation in \acrshort{LXe}&S$_1$/S$_2$ signals, gain, energy/time resolution\\\hline

\multirow{2}{*}{\cite{447}}&\multirow{2}{*}{Ar}&99-153, 1.1&\multirow{2}{*}{2 PTFE THGEM}&Cu K$_{\alpha}$&gain\\\cline{3-3}\cline{5-6}
&&113, 1.1&&Cu K$_{\alpha}$, $\mathrm{^{241}Am}$ scintillation in \acrshort{LXe} &efficiency, gain, energy resolution\\\hline

\multirow{3}{*}{\cite{detlab_59}}&\multirow{3}{*}{$\mathrm{Ne/CH_4}$ (95:5)}&\multirow{3}{*}{163, 0.8}&WELL&\multirow{3}{*}{UV, $\mathrm{^{55}Fe}$}&gain, energy resolution\\\cline{4-4}
&&&RPWELL&&gain, energy resolution \\\cline{4-4}
&&&THGEM+RPWELL&&gain, energy resolution \\\hline

\cite{detlab_50}&$\mathrm{Ne/CH_4}$ (95:5)&210, 475-647&3 THGEM&$\mathrm{^{60}Co}$/AmBe scintillation in \acrshort{LXe}& energy/position resolution\\
 [1ex]
\hline
\end{tabular}
\end{table}
\end{landscape}


\newpage

\begin{landscape}
\begin{table}[ht]
\caption{Main results with CRAD/LEM detectors. $\mathrm{T_{eq}}$ is the equilibrium temperature at the given pressure. The "Ph" column indicates operation in single gaseous (S) or double (D) phase.}
\label{tab: CRAD}
\centering
\begin{tabular}{|c|c|c|c|c|c|c|c|}
\hline
Ref & Gas & T[K], p[bar] & Ph &Detector & Readout & Radiation & notes \\ [0.5ex] 
\hline
\multirow{5}{*}{\cite{3}} & \multirow{5}{*}{Ar} & \multirow{5}{*}{84, 1} & \multirow{5}{*}{D} & THGEM & \multirow{5}{*}{charge} & \multirow{5}{*}{$\mathrm{^{241}Am}$} & gain\\\cline{5-5}\cline{8-8}
 & &  && 2 THGEM &  &  & gain, energy resolution \\\cline{5-5}\cline{8-8}
& & && 2 Kevlar THGEM &  & &  gain, charging up\\\cline{5-5}\cline{8-8}
 & &  && RETGEM &  &  & condensation \\\cline{5-5}\cline{8-8}
& &  && 2 RETGEM &  &  & condensation\\\hline 

\multirow{4}{*}{\cite{156}} & \multirow{4}{*}{Ar} & 100, 1 &S& \multirow{2}{*}{RETGEM} & \multirow{4}{*}{charge} & \multirow{4}{*}{$\mathrm{^{55}Fe}$} & gain \\\cline{3-3}\cline{8-8}
&  & 89, 1.1 &D&  &  &  &gain \\\cline{5-5}\cline{3-3}\cline{8-8}
 &  & 100 &S&  \multirow{2}{*}{2 RETGEM} &  &  &gain \\\cline{3-3}\cline{8-8}
 &  & 89, 1.1 &D&  &  &  & gain\\\hline

\multirow{4}{*}{\cite{414}} & \multirow{4}{*}{Ar} & \multirow{3}{*}{87, 1} &\multirow{3}{*}{D}&  THGEM & \multirow{4}{*}{charge} & \multirow{4}{*}{$\mathrm{^{241}Am}$} &gain \\\cline{5-5}\cline{8-8}
 &  & && 2 THGEM & &  &gain, also N$_2$(0.3\%) impurity  \\\cline{5-5}\cline{8-8}
 &  & && 2 THGEM + GEM & &  &gain, also N$_2$(0.3\%) impurity\\\cline{5-5}\cline{8-8}
 &  & 123, 1.42 / 87, 1 &S/D& Polymide 2 THGEM &  &  & gain\\\hline


\cite{432,434} & Ar & 87, 1 & D & THGEM & charge & cosmics & TPC operation\\\hline

\cite{459} & Ar & 87, 1 & D& THGEM & charge & cosmics & gain stabilization, TPC operation\\\hline

\cite{443} & Ar & 87, 1 & D & THGEM & charge & cosmics & gain stabilization, THGEM parameters \\\hline

\cite{444,475} & Ar & 87, 1 &D& THGEM & charge & cosmics & WA105 TPC operation\\\hline

\cite{239} & Ar & 87, 1 &D& THGEM & charge,light & $\mathrm{^{55}Fe}$  & gain, light yield\\\hline

\cite{50} & Ar & 87, 1 &D& 2 THGEM & charge,light & $\mathrm{^{241}Am}$  & gain, light yield\\\hline


\cite{435} & Ar & 87, 1 &D& THGEM & charge,light & $\mathrm{^{109}Cd}$, Mo $\mathrm{K_\alpha}$ &gain, light yield, position resolution\\\hline

\cite{448} & Ar & 87, 1 &D& 2 THGEM & light & $\mathrm{^{241}Am}$ &imaging\\\hline

\cite{446} & Ar & 87, 1 &D& THGEM & light & cosmics &TPC operation\\\hline

\cite{454} & Ar & $\mathrm{T_{eq}}$, 1.2 / $\mathrm{T_{eq}}$, 1.08 &D& THGEM & light & cosmics/beam  & ARIADNE TPC\\\hline

\cite{385} & Ar & $\mathrm{T_{eq}}$, 1.04 &D& THGEM & light & cosmics & ARIADNE TPC \\\hline

\cite{476} & Ar & $\mathrm{T_{eq}}$, 1.04 &D& Glass THGEM & light &  $\mathrm{^{241}Am}$ & imaging \\\hline

\cite{57} & Xe & 191, 1.1 / 178, 1 / 165, 1  &S/S/D& 2 THGEM & charge, light & $\mathrm{^{241}Am}$, Mo $\mathrm{K_\alpha}$, $\mathrm{^{22}Na}$ &gain, light yield\\
 [1ex]
\hline
\end{tabular}
\end{table}
\end{landscape}

\newpage


\bibliographystyle{elsarticle-num} 
\bibliography{cas-refs}





\end{document}